\documentclass[11pt,letterpaper]{article}

\usepackage{jheppub_mod}
\usepackage{slashed}
\usepackage{graphicx}

\newcommand{\vev}{\Delta}

\newcommand{\ud}{\mathrm{d}}
\newcommand{\uD}{\mathrm{D}}
\newcommand{\ue}{e}
\newcommand{\ui}{i}
\newcommand{\us}{\mathrm{s}}
\newcommand{\uS}{\mathrm{S}}
\newcommand{\uw}{\mathrm{w}}
\newcommand{\uT}{\mathrm{T}}
\newcommand{\vol}{\mathrm{vol}}
\newcommand{\Str}{\mathrm{Str}}
\newcommand{\tr}{\mathrm{tr}}
\newcommand{\cA}{\mathcal{A}}
\newcommand{\cB}{\mathcal{B}}
\newcommand{\cD}{\mathcal{D}}
\newcommand{\cF}{\mathcal{F}}
\newcommand{\cK}{\mathcal{K}}
\newcommand{\cN}{\mathcal{N}}
\newcommand{\cO}{\mathcal{O}}
\newcommand{\cS}{\mathcal{S}}
\newcommand{\cV}{\mathcal{V}}
\newcommand{\cW}{\mathcal{W}}

\newcommand{\vp}{\varphi}

\newcommand{\ul}{\underline}
\newcommand{\ovl}{\overline}
\newcommand{\bs}{\boldsymbol}

\newcommand{\ep}{\epsilon}
\newcommand{\kap}{\kappa}
\newcommand{\lam}{\lambda}
\newcommand{\sig}{\sigma}
\newcommand{\im}[1]{\mathrm{Im}\,#1}
\newcommand{\re}[1]{\mathrm{Re}\,#1}
\newcommand{\SU}[1]{\mathrm{SU}\left(#1\right)}
\newcommand{\SO}[1]{\mathrm{SO}\left(#1\right)}
\newcommand{\U}[1]{\mathrm{U}\left(#1\right)}
\newcommand{\abs}[1]{\bigl\vert #1 \bigr\vert}
\newcommand{\absb}[1]{\left\vert#1\right\vert}

\title{Chiral matter wavefunctions in warped compactifications}

\author[a]{Fernando Marchesano,}
\author[b]{Paul McGuirk,}
\author[b]{and Gary Shiu}

\affiliation[a]{Instituto de F\'{\i}sica Te\'orica UAM/CSIC, Cantoblanco,
  28049 Madrid, Spain}
\affiliation[b]{Department of Physics, University of Wisconsin-Madison, USA}

\emailAdd{fernando.marchesano@csic.es}
\emailAdd{mcguirk@physics.wisc.edu}
\emailAdd{shiu@physics.wisc.edu}

\date{\today}

\preprint{IFT-UAM/CSIC-10-89\\MAD-TH-10-05}

\abstract{We analyze the wavefunctions for open strings stretching
  between intersecting 7-branes in type IIB/F-theory warped
  compactifications, as a first step in understanding the warped
  effective field theory of 4d chiral fermions.  While in general the
  equations of motion do not seem to admit a simple analytic solution,
  we provide a method for solving the wavefunctions in the case of
  weak warping. The method describes warped zero modes as a
  perturbative expansion in the unwarped spectrum, the coefficients of
  the expansion depending on the warping. We perform our analysis with
  and without the presence of worldvolume fluxes, illustrating the
  procedure with some examples. Finally, we comment on the warped
  effective field theory for the modes at the intersection.}

\begin{document}

\maketitle

\newpage 

\section{Introduction}

Although string theory is the leading candidate for a quantum theory
of gravity, finding realistic models in a string framework is a
difficult task.  Among the challenges faced by such constructions, as
well as by any candidate ultraviolet completion of the Standard Model,
is an explanation of the electroweak hierarchy.  A virtue of string
models is that they typically contain extra dimensions, the existence
of which potentially allows the hierarchy problem to be translated
into a question of geometry.  For example, if the degrees of freedom
in the visible sector are realized by open strings localized at
D-branes and their intersections, then the hierarchy can in principle
result from extra dimensions that are large with respect to the string
length \cite{ArkaniHamed:1998rs,*Antoniadis:1998ig,Shiu:1998pa} (see
also \cite{Antoniadis:1990ew,Lykken:1996fj}). In practice, however, it
remains challenging to find compactifications with stabilized moduli
that admit such a low string scale and yet are phenomenologically
viable (see for
example~\cite{Balasubramanian:2005zx,*Conlon:2005ki,*Conlon:2006gv}).

Another related approach to translating the hierarchy problem into a
geometrical problem is warping.  In type II string theories and
F-theory, D-branes and other extended objects are necessary for the
cancellation of tadpoles and providing the open strings required to
produce realistic models.  The gravitational influence of these
ingredients results in a spacetime that is warped in the sense that it cannot be
described as a direct product.  Additionally, stabilization of the
internal geometry requires the addition of fluxes and the
back-reaction of such fluxes also leads to warping.  If the warping is
strong, then the gravitational redshift provides a mechanism for the
exponential suppression of the electroweak scale.  Although this
method of generating the hierarchy was first introduced from a 5d
point of view~\cite{Randall:1999ee}, one may also implement this
scheme in the context of string
theory~\cite{Verlinde:1999fy,*Dasgupta:1999ss,*Greene:2000gh,*Becker:1996gj,
  *Becker:2000rz,*Giddings:2001yu}
(see~\cite{Douglas:2006es,*Blumenhagen:2006ci,*Grana:2005jc} for
reviews). In addition to providing phenomenologically attractive
constructs from the point of view of particle physics, warped
geometries have played an important role in string cosmology by
providing a framework to describe either
inflation~\cite{Kachru:2003sx} (for reviews
see~\cite{Linde:2005dd,*Cline:2006hu,*Kallosh:2007ig,*Burgess:2007pz,
  *McAllister:2007bg,*Baumann:2009ni}) or late time
acceleration~\cite{Kachru:2003aw}.  Finally, warped geometries are
also key to the understanding of strongly-coupled gauge theories by
way of the gauge/gravity
correspondence~\cite{Maldacena:1997re,*Gubser:1998bc,*Witten:1998qj}.

Due to the many applications of warped compactifications in string
theory, it is of clear value to understand their dynamics.  Although
in principle such dynamics follow from worldsheet methods, warped
compactification of type II theories necessarily include Ramond-Ramond
fluxes and, except in special
cases~\cite{Metsaev:1998it,*Metsaev:2001bj}, it is challenging to
quantize string theory in such backgrounds. An alternative method to
describe the low energy dynamics is to consider the effective action
resulting from a dimensional reduction of the supergravity description
of these geometries.  However, even when considering only the fields
in the 4d supergravity multiplet, deducing such an action has proven
to be a subtle problem~\cite{Giddings:2005ff, *Frey:2006wv,
  *Burgess:2006mn,*Douglas:2008jx,*Shiu:2008ry,*Frey:2008xw,*Martucci:2009sf,*Underwood:2010pm}. The
problem becomes even more involved if one considers compactifications
with a realistic gauge sector, as such sectors are localized on the
worldvolumes of D-branes, the light degrees of freedom of which are
also be affected by the presence of
warping~\cite{Marchesano:2008rg,Chen:2009zi}.  An understanding of the
warped effective field theory of these open string modes is thus a
crucial ingredient in any detailed phenomenological study of warped
compactifications.

In~\cite{Marchesano:2008rg}, we considered warped type IIB
compactifications and studied the wavefunctions for open strings
beginning and ending on the same $\uD 7$-brane. Such a $\uD 7$-brane
fills the large, non-compact four dimensions and wraps a 4-cycle of
the internal geometry.  Since the low-energy effective action follows
from dimensional reduction to 4d, almost any quantity arises as an
overlapping integral of warped wavefunctions, which are in turn
computed by solving a warped Dirac or Laplace equation. From our
analysis in~\cite{Marchesano:2008rg}, we found that the wavefunctions
for the bosonic degrees of freedom remain unmodified by the presence
of warping, while the wavefunctions associated with the fermions are
modified in a way that is consistent with supersymmetry. In
particular, the effect of warping on the fermionic degrees of freedom
depends on the chirality of such fermions in the internal $\uD
7$-brane dimensions. The behavior of the wavefunctions and 4d
effective action can then be deduced by the 8d Dirac-Born-Infeld and
Chern-Simons action describing the bosonic degrees of freedom,
together with the 8d action of~\cite{Martucci:2005rb} describing the
fermionic degrees of freedom.

In this work, we extend our analysis in~\cite{Marchesano:2008rg} by
analyzing the wavefunctions for open strings stretching between
intersecting $\uD 7$-branes in warped compactifications.  Such strings
generically give rise to {\it chiral} bifundamental fields and are
thus of obvious phenomenological interest. The strategy that can be
followed to describe such an intersection is to consider the
non-Abelian generalization of the $\uD 7$-brane action describing
the low-energy dynamics in the limit where $N$ branes are coincident.
The non-trivial intersection can then be described by a varying
background profile for the transverse deformations field $\Phi$ of the
non-Abelian $\uD 7$-brane theory, in such a way that the initial gauge
group is broken as $\U{N} \rightarrow \U{N_a} \times \U{N_b}$ by the
presence of $ \Phi $.  Since the energy of a string is proportional to
its length, one expects that the massless strings stretching between
the intersecting $\uD 7$-branes are localized at the intersection
locus. Indeed, in the unwarped case it is known that the corresponding
wavefunctions are exponentially peaked there~\cite{Katz:1996xe,
  Hashimoto:2003xz, *Nagaoka:2003zn}.

Although an intersection of $\uD 7$-branes is sufficient to obtain
bifundamental fields, this does not automatically yield a 4d chiral
spectrum. In order to obtain 4d chirality one must either place the
intersection at a singularity or to consider intersections that
support a non-trivial worldvolume flux $ F $. In the latter, more
generic case, the Laplace and Dirac equations are modified by the
presence of a non-vanishing vector potential $A$, requiring that the
wavefunctions at the intersection be modified as well. For instance,
if the intersection is a flat two-torus, one can show that the
unwarped wavefunctions are constant in the unmagnetized case, while
they are described by Riemann $\vartheta$-functions as soon as $F \neq
0$~\cite{Cremades:2004wa}.

For the adjoint fields studied in~\cite{Marchesano:2008rg}, the
warping modification of the open strings wavefunctions could be simply
expressed in terms of the warp factor, as these fields have a
well-defined chirality in the internal $\uD 7$-brane dimensions. As
the massless fields at the intersection do not have a well-defined
internal chirality, the warped wavefunctions no longer take such a
simple expression.  However, in the weak warping case (i.e., a slowly
varying warp factor), the effect of warping can be treated as a
perturbation. The wavefunction can then be expanded in terms of the
massive modes of the unwarped geometry, and the coefficients
characterizing the expansion can be determined using perturbation
theory.

Our paper is organized as follows.  In section~\ref{sec:setup}, we
consider some generalities of intersecting $7$-branes in warped
compactifications.  Drawing on~\cite{Butti:2007aq}
and~\cite{Myers:1999ps}, we propose a non-Abelian generalization of
the superpotential and $D$-terms of~\cite{Jockers:2005zy,
  Martucci:2006ij} which allow us to extend the supersymmetry
conditions of~\cite{Marino:1999af,Gomis:2005wc,Martucci:2005ht} 
to intersecting $\uD 7$-branes in
warped backgrounds.  In section~\ref{sec:nonchiral}, we consider the
fluctuations about unmagnetized intersections, as a warm-up for the
more involved, magnetized case. The equations of motion for these
fluctuations follow again by considering the $F$- and $D$-flatness
conditions as well as a non-Abelian generalization of the fermionic
action of~\cite{Martucci:2005rb}.  The massive spectrum in the
unwarped case is determined in subsection~\ref{subsec:nonchiralMM} and
the expansion of the warped zero mode in terms of the unwarped
spectrum is presented in subsection~\ref{sec:unmagnetizedMM} with some
simple examples worked out in
subsection~\ref{sec:unmagnetized_examples}.  We then extend our
analysis to the magnetized case in section~\ref{sec:chiral}.
These results lead us in section~\ref{sec:eft}, to address some issues
regarding the 4d warped effective field theory for the chiral modes at
the intersection.  We draw our conclusions in
section~\ref{sec:conclusion}, while our conventions, some technical
details, and a discussion on corrections to the $D$-flatness
conditions are left for the appendices.

The effective action for bifundamental fields arising from
intersecting $\uD$-branes have been considered in many other places in
the literature, though the effects of warping, which is our focus
here, has not yet been widely explored.  Such modes are often
considered via the worldsheet as reviewed
in~\cite{Blumenhagen:2005mu,Blumenhagen:2006ci}. Field theory
treatments include~\cite{Hashimoto:2003xz, Nagaoka:2003zn} in the
context of brane recombination and~\cite{Cremades:2004wa} for the
purpose of calculating Yukawa couplings.  The intersections of general
$7$-branes were considered in~\cite{Beasley:2008dc,Cecotti:2009zf,
  Donagi:2008ca,Font:2009gq,Conlon:2009qq,Leontaris:2010zd} (see also 
  \cite{Conlon:2008qi}), though
again in the absence of warping.  Finally, in addition to the
consideration of warped effective actions referenced above, background fluxes
which give rise to warping can have additional influence on the
wavefunctions; such effects were considered
in~\cite{Camara:2009xy,*Camara:2009zz} for the open string sector.

\section{\label{sec:setup}Intersecting D7s in warped compactifications}

Let us consider type IIB superstring on the warped product of
$\mathbb{R}^{1,3} \times_\omega X_{6}$, where $X_6$ a compact
six-dimensional manifold.  That is, we consider the Einstein frame 10d
background metric
\begin{equation}
  \label{eq:warpedmetric}
  \ud s_{10}^{2}=e^{2a}\eta_{\mu\nu}\ud x^{\mu}\ud x^{\nu}+
  \ue^{-2a}\ud\tilde{s}_{6}^{2},\quad \quad
  \ud\tilde{s}_{6}^{2}=\tilde{g}_{mn}\ud y^{m}\ud y^{n},
\end{equation}
where the warp factor $\ue^{-4a}$ varies over $X_{6}$.  Such a
geometry is supported by the RR $5$-form field
strength~\cite{Verlinde:1999fy, Giddings:2001yu}
\begin{equation}
  \label{eq:5form}
  F_{5}=\bigl(1+\ast_{10}\bigr)F_{5}^{\mathrm{ext}},\qquad
  F_{5}^{\mathrm{int}}=\tilde{\ast}_{6}\ud \ue^{-4a}
\end{equation}
where $\ud\vol_{\mathbb{R}^{1,3}}$ is the volume element of
$\mathbb{R}^{1,3}$ and $\ast_{10}$ is the Hodge-$\ast$ built from the
warped metric~\eqref{eq:warpedmetric} and $\tilde{\ast}_{6}$ is the
Hodge-$\ast$ built from $\tilde{g}$.  Such $5$-form flux is sourced by
objects with finite $\uD 3$-brane charge such as $\uD 3$-branes,
$\mathrm{O}3$-planes, magnetized $7$-branes and $3$-form flux $G_{3}$.
Focusing on supersymmetric warped compactifications requires that
$G_{3}$ is a primitive $\left(2,1\right)$-form, $\tilde{g}$ is
K\"ahler and the axio-dilaton $\tau$ is a holomorphic function on
$X_{6}$~\cite{Grana:2000jj, *Gubser:2000vg, *Grana:2001xn}, so that the
elliptic fibration over $X_{6}$ specified by $\tau$ is a Calabi-Yau
four-fold.  The divisors $\cS \subset X_{6}$ on which the fiber
degenerates correspond to the location of $7$-branes with the
corresponding gauge group $G_{\cS}$~\cite{Vafa:1996xn,*Morrison:1996na,
  *Morrison:1996pp}.

Our primary interest in this paper will be on the intersection of two
of these divisors where the symmetry further enhances.  Localized
along this matter curve are additional degrees of freedom that are
charged under $G_{\cS} \times G_{\cS'}$ and generalize the well-known
bifundamental fields appearing in the low energy spectrum of
intersecting $\uD 7$-branes~\cite{Bershadsky:1996nh, Katz:1996xe}.
For a single stack of $\left(p,q\right)$ $7$-branes, the effective
action is given by an $\mathrm{SL}\bigl(2,\mathbb{Z}\bigr)$ rotation
of the usual Dirac-Born-Infeld and Chern-Simons actions, as such
branes are simply Dirichlet branes for $\left(p,q\right)$-strings.  An
intersection of two $\left(p,q\right)$ $7$-branes can then be
described by Higgsing this low energy theory, just like the
intersection of two $\uD 7$-branes. Finally, the intersection of a
stack of $\left(p,q\right)$ with $\left(p',q'\right)$ $7$-branes with
$\left(p,q\right)\neq\left(p',q'\right)$ can be treated,
following~\cite{Beasley:2008dc}, by means of a topologically twisted
YM 8d action with an exceptional gauge group, also Higgsed down to
describe the massless modes on a matter curve.

While the latter strategy allows to describe the fields as the
$7$-brane intersection in terms of wavefunctions, it is a priori not
obvious how to include the effects of warping in this topologically
twisted 8d action. In this sense, it seems more reliable to consider
an intersection of two $\uD 7$-branes and make use of the non-Abelian
DBI and CS actions, as well as their fermionic counterpart, in order to
derive the warped equations of motion for bosonic and fermionic
degrees of freedom. Such computations (performed in appendix
\ref{app:bosonic_eom} and next section, respectively), will however not
be our main strategy to derive the warped equations of motion. Instead, we will take
 a different approach based on the supersymmetry conditions for a stack of 
 $\uD 7$-branes in a general type IIB background, conditions which 
 we will derive in the remainder of this section. 

As we will see, this last approach allows to consider general closed string backgrounds
in a rather simple way. Indeed, while we turn off background $3$-form fluxes in our computation, 
we allow for a varying dilaton and hence a non-Calabi-Yau geometry for the 
internal space $X_6$. This, together with the fact that the BPS equations for 
a $\uD 7$-brane and a $(p,q)$ $7$-brane are identical, leads us to believe
that our warped equations of motion apply to the more general
$\left(p,q\right)$-$\left(p',q'\right)$ $7$-brane intersection that
are of main interest in local F-theory GUT models. It would be interesting to
check from first principles if this is indeed the case.

In order to derive the non-Abelian BPS equations, let us first
consider a single $\uD 7$-brane wrapping a $4$-cycle
$\mathcal{S}_{4}\subset X_{6}$.  The massless open string excitations
of this D-brane consist of a gauge field $A$ living on the 8d
worldvolume and its transverse fluctuations of its embedding
$\Phi^{i}=\lam^{-1}X^{i}$, where $\lambda=2\pi\alpha'$. The $\uD 7$
will be supersymmetric if~\cite{Marino:1999af,Gomis:2005wc,Martucci:2005ht}
\begin{enumerate}

\item $\mathcal{S}_{4}$ is holomorphically embedded into $X_{6}$, and

\item the worldvolume field strength $F_{2}=\ud A$ satisfies the
  self-duality condition
  \begin{equation}
    \ast_{4}F_{2}=F_{2},
  \end{equation}
where $\ast_{4}$ is the Hodge-$\ast$ on $\mathcal{S}_{4}$ built from the induced metric.

\end{enumerate}

These conditions follow from consideration of an effective potential
resulting from the superpotential and $D$-term~\cite{Jockers:2005zy,
  Martucci:2006ij}.
\begin{subequations}
\label{BPSpure}
\begin{align}
  W=&\int_{\Sigma_{5}}\ue^{3\alpha}(\im\tau)^{-1/2}\,
  P\bigl[\Psi_{2}\wedge \ue^{B_{2}}\bigr]\wedge \ue^{\lambda F_{2}}, \\
  D=&\int_{\mathcal{S}_{4}}\ue^{2\alpha}\,
  P\bigl[\im\Psi_{1}\,
  \wedge \ue^{B_{2}}\bigr]\wedge\ue^{\lambda F_{2}},
\end{align}
\end{subequations}
in which $B_{2}$ is the NS-NS $2$-form, $\tau$ is the axio-dilaton,
$P$ indicates the pullback to $\mathcal{S}_{4}$, and $\Sigma_{5}$ is a
$5$-chain whose boundaries are $\mathcal{S}_{4}$ and its
deformation. Finally, $\alpha$ is related to the Einstein frame warp factor $a$ through
\begin{equation}
  \alpha=a+\frac{1}{4}\log\left(\im\tau\right).
\end{equation}
As it will turn out, the equations of motion will be written naturally
in terms of $\alpha$, and so for simplicity we will often refer to
$\ue^{-4\alpha}$ as the warp factor.  The pure spinors $\Psi_{1}$ and
$\Psi_{2}$ are given in terms of the warped K\"ahler form $J =
e^{-2\alpha}J_{X_6}$ and the (unwarped) fundamental $3$-form of
$X_{6}$ by $\Psi_{1}=\ue^{i J}$ and $\Psi_{2}=
\ue^{-3\alpha}\Omega_{X_6}$.  Demanding $F$-flatness implies that
$\mathcal{S}_{4}$ is holomorphic and that $F_{2}$ is
$\left(1,1\right)$ while demanding $D$-flatness to leading order in
$\alpha'$ implies that $F_{2}$ is a primitive in $\cS_4$; together,
these two conditions on $F_{2}$ imply that it is self-dual.

Interestingly, the expressions (\ref{BPSpure}) for $W$ and $D$ allow
for a simple generalization to the non-Abelian case, following some
observations made in~\cite{Butti:2007aq}.\footnote{We would like to
  thank L.~Martucci for discussions on this point.}  To this end, we
locally write\footnote{In general, this will be possible whenever
  $\left(\ud +
    H\wedge\right)\left(\ue^{3\alpha}\left(\im\tau\right)^{-1/2}\,\Psi_2\right)=0$,
  which in the language of~\cite{Lust:2008zd} is the BPS condition for
  domain walls.  Hence, in the $\mathcal{N}=0$ vacua
  of~\cite{Giddings:2001yu} and the DWSB vacua discussed
  in~\cite{Lust:2008zd}, this analysis should be reconsidered.}
\begin{equation}
  \ue^{3\alpha}\left(\im\tau\right)^{-1/2}\, \Psi_{2}\wedge \ue^{B_{2}} =\ud\gamma.
\end{equation}
The superpotential and $D$-term can then be expressed as
\begin{equation}
  W=\int_{\mathcal{S}_{4}}P\bigl[\gamma\bigr]\wedge
  \ue^{\lambda F_{2}},\quad
  D=\int_{\mathcal{S}_{4}}P
  \bigl[\im\eta\,\bigr]\wedge \ue^{\lambda F_{2}},
\end{equation}
where
\begin{equation}
  \eta=\ue^{2\alpha}\, \Psi_{1}\wedge e^{B_{2}}.
\end{equation}
Now, as observed in Appendix  A  of~\cite{Butti:2007aq}, these expressions take the same
form as the Chern-Simons action for a $\uD$-brane
\begin{equation}
\label{CSDp}
  S_{\uD p}^{\mathrm{CS}}=\int_{\mathcal{W}} P\bigl[
  \mathcal{C}\wedge \ue^{B_{2}}\bigr]
  \wedge \ue^{\lambda F_{2}},
\end{equation}
where $\mathcal{C}$ is the formal sum of R-R potentials and $\mathcal{W}$ is the worldvolume 
of the brane. Following~\cite{Myers:1999ps}, the non-Abelian generalization of (\ref{CSDp}) 
is then given by
\begin{equation}
  S_{\uD p}^{\mathrm{CS}}=\int_{\mathcal{W}}
  \Str\biggl\{P\bigl[\ue^{i\lambda\iota_{\Phi}\iota_{\Phi}}
  \mathcal{C}\wedge\ue^{B_{2}}\bigr]\wedge\ue^{\lambda F_{2}}\biggr\},
\end{equation}
where, as detailed in Appendix~\ref{app:bosonic_eom}, $\mathrm{Str}$
indicates a symmetrized trace and $\iota_{\Phi}$ stands for the interior product. 
The transverse fluctuations are then promoted to adjoint-valued scalars $\Phi$ and 
the field strength to $F_{2}=\bigl(\ud-\ui A\wedge\bigr)A$. Finally, the non-Abelian pullback
replaces derivatives with gauge covariant derivatives
\begin{equation}
  P\left[v\right]_{\alpha}=v_{\alpha}+\lambda v_{i}D_{\alpha}\Phi^{i},
\end{equation}
where $D_{\alpha}=\partial_{\alpha}-\ui\left[A_{\alpha},\cdot\right]$.

Making use of the fact that the pure spinors $\Psi_{1}$ and $\Psi_{2}$
transform under T-duality in a way that is analogous to $\mathcal{C}$,
one can then deduce that the non-Abelian superpotential and $D$-term are
given by~\cite{Butti:2007aq}
\begin{subequations}
\label{naBPS}
\begin{equation}
  \label{eq:na_superpotential}
  W=\int_{\mathcal{S}_{4}}\Str\biggl\{
  P\bigl[e^{i\lambda\iota_{\Phi}\iota_{\Phi}}\gamma\bigr]
  \wedge e^{\lambda F_{2}}\biggr\},
\end{equation}
and
\begin{equation}
  D=\int_{\mathcal{S}_{4}}\mathrm{S}\biggl\{
  P\bigl[e^{i\lambda\iota_{\Phi}\iota_{\Phi}}
  \im\eta\bigr]\wedge\ue^{\lambda F_{2}}\biggr\},
\end{equation}
\end{subequations}
where in the latter  $\mathrm{S}$ indicates the symmetrization prescription
of~\cite{Myers:1999ps} without taking the trace.

In order to extract the F-term and D-term conditions from
(\ref{naBPS}) it is useful to consider a neighborhood of the internal
space $X_6$ around $\cS_4$ such that $(\im\tau)^{-1/2}\Omega_{X_6}
=\ud z^{1}\wedge\ud z^{2}\wedge \ud z^{3}$ with
\begin{equation}
  \ud z^{m}=\ud y^{m}+\tau_{m}\ud y^{m+3},
\end{equation}
and the warped K\"ahler form is given by
\begin{equation}
  \label{eq:warped_kaehler_form}
  J=\ue^{-2\alpha}\frac{\ui\alpha'}{2}
  \sum_{m=1}^{3}\left(2\pi R_{m}\right)^{2}\ud z^{m}\wedge
  \ud\bar{z}^{\bar{m}}.
\end{equation}
Moreover, let us consider a local coordinate system such that the
complex 4-cycle $\cS_4$ is parameterized by $(z^1, z^2)$, as is usual
in the literature of local F-theory models. Then, in absence of
background 3-form fluxes we can take $\gamma=z^{3}\ud z^{1}\wedge \ud
z^{2}$, so that $\gamma$ is globally well-defined on $\mathcal{S}_{4}$
and satisfies $\iota_{\Phi}\gamma=0$.  The resulting superpotential
takes the form
\begin{equation}
  \label{eq:particular_W}
  W=-\lambda
  \int_{\mathcal{S}_{4}}\ud^{4}z\, \Str\biggl\{\Phi F_{\bar{1}\bar{2}}\biggr\},
\end{equation}
where $\Phi$ is the complexified transverse fluctuation.  Demanding
$F$-flatness in the $\Phi$ direction immediately gives
$F_{\bar{1}\bar{2}}=0$ implying
\begin{equation}
  \label{eq:f_flatness_1}
  F^{\left(0,2\right)}=F^{\left(2,0\right)}=0.
\end{equation}
Likewise, variation with respect to $A$ gives
\begin{equation}
  \label{eq:f_flatness_2}
  D_{\bar{m}}\Phi=0.
\end{equation}
Both of these $F$-flatness conditions are what one would expect from their 
Abelian counterparts and, while derived in the type IIB framework, 
they have a simple generalization to F-theory.

Consider now the $D$-flatness condition $D=0$.  First we note that
since the $\uD 7$-brane is a real codimension $2$ object,
$\iota_{\Phi}^{3}=0$ and so
\begin{equation}
  e^{i\lambda\iota_{\Phi}\iota_{\Phi}}=1+i\lambda\iota_{\Phi}\iota_{\Phi}.
\end{equation}
It follows then that the non-Abelian $D$-term reads
\begin{equation}
  \label{eq:D_term_expanded}
  D=\int_{\mathcal{S}_{4}}\mathrm{S}\biggl\{
  \ue^{2\alpha}\biggl(\lambda P\bigl[J\bigr]\wedge F_{2}
  -\frac{i\lambda}{6}P\bigl[\iota_{\Phi}\iota_{\Phi} J^{3}\bigr]
  +\frac{i\lambda^{3}}{2}P\bigl[\iota_{\Phi}\iota_{\Phi} J\bigr]
  \wedge F_{2}\wedge F_{2}\biggr)\biggr\}
\end{equation}
with $\alpha$ a modified warp factor defined as in (\ref{eq:warped_kaehler_form}).

In the next section and in Appendix \ref{app:bosonic_eom} we will compare the 
equations of motion that result from the above $F$-flatness and $D$-flatness 
conditions to those derived from a DBI$+$CS action and its fermionic counterpart 
valid to leading order in $\alpha'$. For such comparison we need to truncate $D$ 
at order $\lambda$
\begin{equation}
  \label{eq:truncated_d_term}
  D=\lambda\int_{\mathcal{S}_{4}}
  \mathrm{S}\biggl\{\ue^{2\alpha}\biggl(P\bigl[J\bigr]\wedge F_{2}
  -\frac{\ui}{6}P\bigl[\iota_{\Phi}\iota_{\Phi}J^{3}\bigr]\biggr)\biggr\}.
\end{equation}
Then, defining the warped fundamental form on $\cS_4$ as
\begin{equation}
  \label{eq:warped_kahler_form_restricted}
  \mathfrak{J}=J\mid_{\mathcal{S}_{4}}=\ue^{-2\alpha}
  \frac{\ui\alpha'}{2}\sum_{m=1}^{2}\left(2\pi R_{m}\right)^{2}
  \ud z^{m}\wedge\ud\bar{z}^{\bar{m}},
\end{equation}
we have that $P\left[J\right]=\mathfrak{J}+\mathcal{O}\left(\lambda^{2}\right)$.
Finally, it is also straightforward to show that
\begin{equation}
  \frac{1}{6}\iota_{\Phi}\iota_{\Phi}J^{3}=\ue^{-2\alpha}\frac{\ui\lambda}{4}
  \left(2\pi R_{3}^{2}\right)\bigl[\Phi,\bar{\Phi}\bigr]\mathfrak{J}^{2},
\end{equation}
and so
\begin{equation}
  \label{eq:truncated_dterm_2}
  D=\lambda\int_{\mathcal{S}_{4}}\mathrm{S}\biggl\{
  \ue^{2\alpha}\mathfrak{J}\wedge F_{2}
  +\frac{\lambda}{4}\left(2\pi R_{3}^{2}\right)\mathfrak{J}^{2}
  \bigl[\Phi,\bar{\Phi}\bigr]\biggr\}.
\end{equation}
The symmetrization in this case is trivial and so the $D$-flatness
condition is
\begin{equation}
  \label{eq:d_flatness}
  \ue^{2\alpha}\mathfrak{J}\wedge F_{2}
  +\frac{\lambda}{4}\left(2\pi R_{3}^{2}\right)\mathfrak{J}^{2}
  \bigl[\Phi,\bar{\Phi}\bigr]=0,
\end{equation}
which is the expression we will work with from now on. The effect of
higher $\alpha'$ terms can be included as discussed in
Appendix~\ref{app:corrections}.  Note that the second term
in~\eqref{eq:d_flatness} is not to be interpreted as an
$\alpha'$-correction to the primitivity condition but is instead a
modification resulting from taking into account the non-Abelian
effects; the factor of $\lambda$ appears because
in~\eqref{eq:warped_kaehler_form}, we have taken $R_{3}$ and $z^{n}$
to be dimensionless.  When going from~(\ref{eq:truncated_dterm_2})
to~(\ref{eq:d_flatness}), we have ignored the non-Abelian nature of
the warping $\alpha$.  That is, by the general prescription of
non-Abelian pull-backs on $\cS_4$, $\alpha$ and other closed-string
fields should be interpreted as a functional of the non-Abelian field
$\Phi$. As we discuss below, for the case of intersecting $\uD
7$-branes, treating $\alpha$ as proportional to the identity
corresponds to taking the limit of small intersecting angles, and 
the case of arbitrary angles amounts to a redefinition of $\alpha$.

\begin{figure}[t]
  \begin{center}
  \includegraphics[scale=.475]{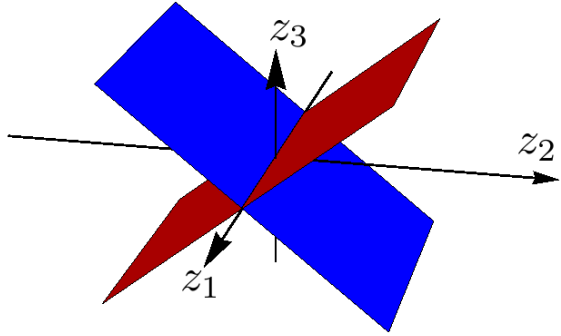}
  \caption{\label{fig:intersection}Local view of the intersection of
    $\uD 7$-branes described by~\eqref{eq:angle}.}
\end{center}
\end{figure}

Let us now consider the intersection of two stacks of $\uD 7$-branes.
Although our expressions for the superpotential and $D$-term are in
principle defined as an integral over a single 4-cycle
$\mathcal{S}_{4}$, we can consider $\uD 7$-branes wrapping different
cycles by giving a non-constant vev to the transverse scalar $\Phi$,
such that the initial worldvolume gauge theory is Higgsed as $\U{N}
\rightarrow \U{N_a} \times \U{N_b}$ by the presence of $\langle \Phi
\rangle$. This vev can be taken to be
\begin{equation}
\label{Sigma}
  \left\langle\Phi\right\rangle=\vev=
  \begin{pmatrix}
    \vev_{a}\mathbb{I}_{N_{a}} & \\
    & \vev_{b}\mathbb{I}_{N_{b}}
  \end{pmatrix},
\end{equation}
where $\vev_{a,b}$ are holomorphic functions of $z^{1}, z^{2}$, so that the 
F-flatness condition~\eqref{eq:f_flatness_2} is satisfied at the level of the background.
Note that this choice satisfies $\bigl[\vev,\bar{\vev}\bigr]=0$, so setting $F_{2} = 0$ 
is consistent with supersymmetry.
Geometrically, (\ref{Sigma}) describes a stack of $N_{a}$ $\uD 7$-branes wrapping
the $4$-cycle specified by $z^{3}=\lambda\vev_{a}$ and a stack of $N_{b}$ 
$\uD 7$-branes wrapping the $4$-cycle $z^{3}=\lambda\vev_{b}$, thus intersecting
at the complex curve $\Sigma = \vev_a \cap \vev_b \subset \cS_4$. 

Given this background for $\Phi$, the spectrum of open string modes arises from fluctuations around it such as
\begin{equation}
  \label{eq:flucs}
  \delta\Phi=\begin{pmatrix}
    \phi^{a} & \phi^{-} \\ \phi^{+} & \phi^{b}
  \end{pmatrix}.
\end{equation}
The block diagonal fluctuations $\phi^{a,b}$ correspond to strings
beginning and ending on the same stack, while the $\phi^{\mp}$
fluctuations correspond to strings stretching from one stack to the
other, giving the charges shown in Table~\ref{table:charges}.  If
$\vev_{a}=\vev_{b}$ has no solution (e.g., if $\vev_{a} -\vev_{b}$ is
constant) then all the modes arising from $\phi^{\mp}$ are necessarily
massive.  However, if the branes do intersect, then $\phi^{\mp}$ will
(partially) describe the massless bifundamental fields localized at
the intersection.  Because the string tension is proportional to its
length, for intersecting $\uD 7$-branes the massless modes of
$\phi^{\mp}$ should be localized around the points of
intersection. Therefore, to capture the dynamics of these fields it
suffices to approximate $\vev_{a,b}$ by linear functions (see
Fig.~\ref{fig:intersection})
\begin{equation}
  \label{eq:angle}
  \vev=\begin{pmatrix} M_{3}^{\left(a\right)}\lambda^{-1}\mathbb{I}_{N_{a}}
    z^{2} & \\
    & M_{3}^{\left(b\right)}\lambda^{-1}\mathbb{I}_{N_{b}}z^{2}
  \end{pmatrix}.
\end{equation}
so that the intersection curve is given by $z_2 = z_3 = 0$ and the
intersection is described by an $\SU{2}$ rotation on the
$z^{2}$-$z^{3}$ plane, in agreement with the results
of~\cite{Berkooz:1996km}. In the following we will take
$M_{3}^{\left(a\right)}>M_{3}^{\left(b\right)}$ though there are no
significant changes if we flip the inequality.

\begin{table}[t]
  \begin{center}
  \begin{tabular}{c|c|c}
    field & $\U{N_{a}}$ & $\U{N_{b}}$ \\
    \hline
    $\phi^{a}$ & $\mathbf{adj}$ & $\mathbf{1}$ \\
    $\phi^{b}$ & $\mathbf{1}$ & $\mathbf{adj}$ \\
    $\phi^{-}$ & $\square$ & $\ovl{\square}$ \\
    $\phi^{+}$ & $\ovl{\square}$ & $\square$ \\
  \end{tabular}
  \caption{\label{table:charges}Charges for the fields
    in~\eqref{eq:flucs}.}
  \end{center}
\end{table}

As mentioned above, the non-Abelian $D$-flatness conditions are
derived by essentially neglecting the dependence of bulk fields on
$\Phi$.  That is, in general a bulk field $\Psi$ should be interpreted
as a functional of the adjoint-valued transverse
fluctuations~\cite{Myers:1999ps}
\begin{equation}
  \label{eq:non_abelian_taylor_expansion}
  \Psi\bigl[\Phi\bigr] =\sum_{n=0}^{\infty} \lambda^{n}\Phi^{i_{1}}\cdots
  \Phi^{i_{n}}\Psi_{i_{1}\ldots i_{n}}.
\end{equation}
While higher powers of $\Phi$ contain higher powers of $\alpha'$,  
$\langle \Phi \rangle = \vev$ contains a factor of $\lambda^{-1}$ 
via ~\eqref{eq:angle} and so, schematically, at the level of the background 
we have
\begin{equation}
  \Psi\sim \sum_{n=0}^{\infty}
  \left(M_{3}^{\left(a,b\right)}\right)^{n}\Psi_{n}
  \bigl(z^{2},\bar{z}^{\ovl{2}}\bigr).
\end{equation}
Therefore, neglecting higher terms in the expansion
(\ref{eq:non_abelian_taylor_expansion}) is reliable in the limit where
$M_{3}^{\left(a,b\right)}$  (that is, the intersection angles), are small. 
Nevertheless, as shown in Appendix~\ref{app:corrections}, taking into account 
the full non-Abelian pull-back \eqref{eq:non_abelian_taylor_expansion} 
 does not modify the form of the non-Abelian $D$-term equation, and all the 
 corrections can be absorbed in a redefinition of the warping $\alpha$.

Finally, in order to have a 4d chiral spectrum, the intersection curve
must support a non-vanishing magnetic flux~\cite{Manton:1981es,
  *Chapline:1982wy, *RandjbarDaemi:1982hi,
  *Wetterich:1982ed,*Frampton:1984px,*Frampton:1984pk,*Witten:1983ux,*Witten:1984dg,
  *Pilch:1984ur, *Pilch:1985pm}.  As $[\vev, \bar{\vev}]=0$ the
$F$-term~\eqref{eq:f_flatness_1} and the truncated
$D$-term~\eqref{eq:d_flatness} conditions require that $F_{2}$ is
self-dual, just as in the Abelian case.  Let us for simplicity choose
a magnetic flux that does not further break down the gauge group, such
as
\begin{equation}
  \label{eq:magnetic_flux}
 \langle  F_{2}\rangle =\sum_{m=1}^{2}\frac{\pi\ui}{\im\tau_{m}}
  \begin{pmatrix}
    M^{\left(a\right)}_{m}\mathbb{I}_{N_{a}} & \\
    & M^{\left(b\right)}_{m}\mathbb{I}_{N_{b}}
  \end{pmatrix}\ud z^{m}\wedge\ud\bar{z}^{\bar{m}}.
\end{equation}
Imposing self-duality then amounts to satisfying the condition
\begin{equation}
  \label{eq:self-dual}
  \frac{M_{1}^{\left(a,b\right)}}{\cV_{1}}+
  \frac{M_{2}^{\left(a,b\right)}}{\cV_{2}}=0, \quad \quad
  \cV_{m}=\left(2\pi R_{m}\right)^{2}\im\tau_{m}.
\end{equation}
%

\section{\label{sec:nonchiral}Unmagnetized intersections}

Given the $\uD 7$-brane supersymmetry conditions derived in the
previous section, the equations of motion for the open string zero modes can be 
obtained by expanding these BPS conditions to first order in fluctuations. 
In the following, we will apply this observation to analyze the zero modes at the
intersection of two unmagnetized stacks of $\uD 7$-branes. Notice that such 
intersection, unless placed at a singularity, will yield a non-chiral 4d spectrum
upon dimensional reduction. Nevertheless, this simple case already demonstrates 
the non-trivial effect that warping has on open strings at $\uD 7$-brane intersections,
and will serve as a useful warmup for the more general magnetized case.

\subsection{\label{sec:nonchiral_eom}Equations of motion}

Open strings localized at the intersection correspond to bifundamental  fluctuations around 
the vev~\eqref{eq:angle} and $\langle F_{2}\rangle =0$. Since this open string background
is supersymmetric, the zero mode fluctuations should also satisfy the BPS 
conditions~\eqref{eq:f_flatness_1}, \eqref{eq:f_flatness_2}, and~\eqref{eq:d_flatness}.  
Let us write these fluctuations as
\begin{equation}
  \label{eq:param_flucs}
  A_{m}=\sqrt{2\pi}R_{m}a_{m},\qquad
  \Phi=\vev+\frac{2}{\sqrt{2\pi}R_{3}\lambda}\phi,
\end{equation}
where $\vev$ is given in~\eqref{eq:angle}.  We are interested only
in the bifundamental fluctuations so we take
\begin{equation}
  a_{m}=\begin{pmatrix}
    0 & \bigl(\phi^{+}_{m}\bigr)^{\dagger} \\
    \bigl(\phi^{-}_{m}\bigr)^{\dagger} & 0
  \end{pmatrix},\qquad
  \phi=\begin{pmatrix}
    0 & \phi_{3}^{-} \\ \phi^{+}_{3} & 0
  \end{pmatrix},
\end{equation}
so that we then have
\begin{equation}
  \bar{a}_{\ovl{m}}=\begin{pmatrix}
    0 & \phi^{-}_{m} \\ \phi^{+}_{m} & 0
  \end{pmatrix},\qquad
  \bar{\phi}=\begin{pmatrix}
    0 & \left(\phi^{+}_{3}\right)^{\dagger} \\
    \left(\phi^{-}_{3}\right)^{\dagger} & 0
  \end{pmatrix}.
\end{equation}
The labeling of the fluctuations and prefactors are introduced for latter convenience, as 
upon dimensional reduction $\bar{a}_{\bar{m}}$ and $\phi$ correspond to the bosonic d.o.f. 
of the left-handed 4d chiral multiplets, and ${a}_{m}$ and $\bar{\phi}$ to its CPT 
conjugates.\footnote{Indeed, note also that upon T-duality on the transverse coordinate $z^3$,
mapping intersecting $\uD 7$-branes to magnetized $\uD 9$-branes, $(\bar{A}_{\bar{1}},\bar{A}_{\bar{2}},\Phi)$ is mapped to $(\bar{A}_{\bar{1}},\bar{A}_{\bar{2}},\bar{A}_{\bar{3}})$, from where left-handed chiral fields arise.}

Plugging \eqref{eq:param_flucs} back into the BPS conditions of the
previous section and expanding them up to linear order in
fluctuations, we obtain the equations of motion for
$\bar{a}_{\bar{m}}$ and $\phi$. In particular, the F-term condition
$F^{\left(0,2\right)} = 0$ reads
\begin{equation}
  F_{\bar{1}\bar{2}}=2\pi R_{1}R_{2}
  \begin{pmatrix}
    0 & \hat{\partial}_{1}^{\ast}\phi^{-}_{2}-
    \hat{\partial}_{2}^{\ast}\phi^{-}_{1} \\
    \hat{\partial}_{1}^{\ast}\phi^{+}_{2}
    -\hat{\partial}_{2}^{\ast}\phi^{+}_{1} & 0
  \end{pmatrix}=0,
\end{equation}
where we have defined
\begin{equation}
  \label{eq:hatted_derivatives}
  \hat{\partial}_{m}=\frac{1}{\sqrt{2\pi}R_{m}}\partial_{m}.
\end{equation}
The F-term condition $D_{\bar{m}}\Phi=0$ gives
\begin{align}
  D_{\bar{m}}\Phi
  =&\frac{2}{\sqrt{2\pi}R_{3}\lambda}\partial_{\bar{m}}\phi
  -\ui\sqrt{2\pi}R_{m}\bigl[\bar{a}_{\ovl{m}},\vev\bigr] \notag \\
  =&\frac{2 R_{m}}{R_{3}\lambda}
  \begin{pmatrix}
    0 & \hat{\partial}_{m}^{\ast}
    \phi^{-}_{3}
    +\frac{\ui}{2}\sqrt{2\pi}R_{3}{2}I_{3}^{\left(ab\right)}z^{2}
    \phi_{m}^{-} \\
    \hat{\partial}_{m}^{\ast}
    \phi^{+}_{3}
    -\frac{\ui}{2}\sqrt{2\pi}R_{3}I_{3}^{\left(ab\right)}z^{2}
    \phi^{+}_{m} & 0
  \end{pmatrix}=0,
\end{align}
where we have defined
\begin{equation}
  \label{eq:define_I3}
  I_{3}^{\left(ab\right)}=M_{3}^{\left(a\right)}-M_{3}^{\left(b\right)}.
\end{equation}
Finally, for the $D$-flatness condition~\eqref{eq:d_flatness}, we use the fact that
\begin{equation}
  \mathfrak{J}\wedge F_{2}=
  \ue^{-2\alpha}\frac{\ui\alpha'}{2}
  \biggl\{\bigl(2\pi R_{1}\bigr)^{2}F_{2\bar{2}}
  + \bigl(2\pi R_{2}\bigr)^{2}F_{1\bar{1}}\biggr\}\ud^{4}z
\end{equation}
so that in terms of fluctuations around $\langle F_2\rangle
=0$,~\eqref{eq:d_flatness} becomes
\begin{align}
\label{eq:unmagnetized_d_term}
  0=\frac{\ui\alpha'}{2}\biggl\{&
  \sum_{m=1}^{2}\begin{pmatrix}
    0 &
    \hat{\partial}_{m}
    \phi_{m}^{-}-
    \hat{\partial}_{m}^{\ast}
    \bigl(\phi_{m}^{+}\bigr)^{\dagger} \\
    \hat{\partial}_{m}
    \phi_{m}^{+}-
    \hat{\partial}_{m}^{\ast}\bigl(\phi_{m}^{-}\bigr)^{\dagger}
    & 0
  \end{pmatrix}\notag \\
  &-\ue^{-4\alpha}\frac{\ui}{2}\sqrt{2\pi}R_{3}I_{3}^{\left(ab\right)}
  \begin{pmatrix}
    0 & \bar{z}^{\bar{2}}\phi_{3}^{-}-
    z^{2}\bigl(\phi^{+}_{3}\bigr)^{\dagger} \\
    -\bar{z}^{\bar{2}}\phi_{3}^{+}+
    z^{2}\bigl(\phi^{-}_{3}\bigr)^{\dagger} & 0
  \end{pmatrix}
  \biggr\}.
\end{align}
To sum up, the conditions for $F$ and $D$-flatness on the $\uD
7$-brane bosonic fluctuations are
\begin{subequations}
\label{eq:unmagnetized_bps_eom_2}
\begin{align}
  \label{eq:unmagnetized_bps_eom_2a}
  0=&\bigl(\hat{D}_{1}^{\pm}\bigr)^{\dagger}\phi^{\mp}_{2}-
  \bigl(\hat{D}_{2}^{\pm}\bigr)^{\dagger}\phi^{\mp}_{1},\\
  \label{eq:unmagnetized_bps_eom_2b}
  0=&\bigl(\hat{D}_{2}^{\pm}\bigr)^{\dagger}\phi_{3}^{\mp}-
  \bigl(\hat{D}^{\pm}_{3}\bigr)^{\dagger}\phi_{2}^{\mp},\\
  \label{eq:unmagnetized_bps_eom_2c}
  0=&\bigl(\hat{D}_{3}^{\pm}\bigr)^{\dagger}\phi^{\mp}_{1}
  -\bigl(\hat{D}_{1}^{\pm}\bigr)^{\dagger}\phi^{\mp}_{3},\\
  \label{eq:unmagnetized_bps_eom_2d}
  0=&\hat{D}_{1}^{\mp}\phi^{\mp}_{1}
  +\hat{D}_{2}^{\mp}\phi^{\mp}_{2}
  +\ue^{-4\alpha}\hat{D}_{3}^{\mp}\phi^{\mp}_{3},
\end{align}
\end{subequations}
where we have defined the operators\footnote{In the magnetized case of
 Section~\ref{sec:chiral}, $\hat{D}_{1,2}^{\mp}$ will be modified to
take into account the $\uD 7$ worldvolume flux.}
\begin{equation}
  \label{eq:unmagnetized_gauge_covariant_derivatives}
  \hat{D}_{1,2}^{\mp}=\hat{\partial}_{1,2},\qquad
  \hat{D}_{3}^{\mp}=\mp\frac{\ui}{2}
  \sqrt{2\pi}R_{3}I_{3}^{\left(ab\right)}\bar{z}^{\bar{2}}.
\end{equation}
This notation is motivated by the T-dual picture of magnetized $\uD
9$-branes, in which the intersection angle between $\uD 7$-branes
becomes a magnetic flux (see \cite{afim} for more details). In the
$\uD 9$-brane picture, $\hat{D}_{m}^{\mp}$ are nothing but the set of
normalized covariant derivatives that appear in the (unwarped) Laplace
and Dirac operators after assuming that the wavefunctions do not
depend on the $(z^3,\bar{z}^{\bar{3}})$ coordinates and so
$\partial_{3}=0$. As we show in Appendix~\ref{app:bosonic_eom}, the
eom resulting from the DBI and CS actions are satisfied whenever
eqs.\eqref{eq:unmagnetized_bps_eom_2} are satisfied.

In the absence of warping,~\eqref{eq:unmagnetized_bps_eom_2} are
straightforward to solve.  Indeed, the $F$-term
equations~\eqref{eq:unmagnetized_bps_eom_2a}-\eqref{eq:unmagnetized_bps_eom_2c}
are solved by taking the ansatz 
\begin{equation}
\label{eq:f_term_solving_ansatz}
  \phi_{m}^{\mp}=\bigl(\hat{D}_{m}^{\pm}\bigr)^{\dagger}f^{\mp}
\end{equation}
for some arbitrary functions $f^{\pm}$.  Then for $\ue^{-4\alpha}=1$,
the $D$-term equation~\eqref{eq:unmagnetized_bps_eom_2d} becomes
\begin{equation}
  \label{eq:quadratic_eom_unwarped}
  \biggl\{
  \hat{D}_{1}^{\mp}\bigl(\hat{D}_{1}^{\pm}\bigr)^{\dagger}+
  \hat{D}_{2}^{\mp}\bigl(\hat{D}_{2}^{\pm}\bigr)^{\dagger}+
  \hat{D}_{3}^{\mp}\bigl(\hat{D}_{3}^{\pm}\bigr)^{\dagger}
  \biggr\}f^{\mp}=0.
\end{equation}
As \eqref{eq:quadratic_eom_unwarped} only depends on the
intersection coordinates $(z^1, \bar{z}^{\bar{1}})$ through
derivatives, one expects the zero modes to be independent of them. In
particular, if we take the ansatz
\begin{equation}
  f^{\mp}=\frac{1}{z^{2}}g^{\mp}\bigl(\abs{z^{2}}^{2}\bigr),
\end{equation}
we find the solution to~\eqref{eq:quadratic_eom_unwarped} to be~\cite{Katz:1996xe,
  Hashimoto:2003xz, Nagaoka:2003zn}
\begin{equation}
  \label{eq:unwarped_sol_g}
  g^{\mp}=\ue^{-\kappa\absb{z^{2}}^{2}},\quad\kappa=\frac{1}{2}2\pi
  R_{2}R_{3}I_{3}^{\left(ab\right)},
\end{equation}
giving
\begin{equation}
  \label{eq:unwarped_solution}
  \phi_{1}^{\mp}=0,\quad
  \phi_{2}^{\mp}=-\frac{\kappa}{\sqrt{2\pi}R_{2}}
  \sigma^{\mp}\left(x^{\mu}\right)
  \ue^{-\kappa\left\vert z^{2}\right\vert^{2}},\quad
  \phi^{\mp}_{3}=\mp\frac{\ui\kappa}{\sqrt{2\pi}R_{2}}
  \sigma^{\mp}\left(x^{\mu}\right)\ue^{-\kappa\left\vert z^{2}\right\vert^{2}},
\end{equation}
where we have introduced the function $\sigma^{\mp}$ that depends on
the external coordinates $x^{\mu}$ and carries (suppressed)
bifundamental gauge indices.  Note that as a consequence of the
ansatz~\eqref{eq:f_term_solving_ansatz} the same function
$\sigma^{\mp}$ appears in both $\phi_{2}^{\mp}$ and $\phi_{3}^{\mp}$.
We then conclude that at the intersection there are only two
independent complex scalar fields, one transforming under a
bifundamental representation of $\U{N_{a}}\times\U{N_{b}}$ and the
other transforming under the conjugate representation.  The other
linearly independent solution to~\eqref{eq:quadratic_eom_unwarped} is
$g^{\mp}=\exp\left(\kappa\absb{z^{2}}^{2}\right)$ which is not peaked
at the intersection and so is discarded when we consider normalizable
modes. Finally, note that the space transverse to the matter curve is
in general compact, and so one may wonder wether the wavefunctions
ought to satisfy some periodicity conditions; however, since we expect
the wavefunctions to be highly peaked around the intersection (as the
above Gaussian solutions show) such constraints can be safely
neglected in our analysis.

Let us now consider the case of non-trivial warping. As one would
expect from holomorphicity of the superpotential, the $F$-term
equations remain unmodified, so one may again consider the
ansatz~\eqref{eq:f_term_solving_ansatz}. Plugging it into the warped
$D$-term equation gives
\begin{equation}
  \label{eq:quadratic_eom_warped}
  \biggl\{
  \hat{D}_{1}^{\mp}\bigl(\hat{D}_{1}^{\pm}\bigr)^{\dagger}+
  \hat{D}_{2}^{\mp}\bigl(\hat{D}_{2}^{\pm}\bigr)^{\dagger}+
  \ue^{-4\alpha}\hat{D}_{3}^{\mp}\bigl(\hat{D}_{3}^{\pm}\bigr)^{\dagger}
  \biggr\}f^{\mp}=0
\end{equation}
whose only warping dependence arises from the factor $e^{-4\alpha}$. 
As we will now see, the same kind of equation arises when one
considers fermionic wavefunctions in a warped background.

\paragraph{Fermionic equations of motion}

A useful check of the equations of
motion~\eqref{eq:unmagnetized_bps_eom_2} is to consider the equations
for the fermionic degrees of freedom.  In the Abelian case, the
fermionic action for a single $\uD 7$-brane on $\mathcal{S}_{4}$
is~\cite{Marolf:2003ye,*Marolf:2003vf, Martucci:2005rb}
\begin{equation}
\label{eq:warped_Dirac_action}
  S_{\uD 7}^{\mathrm{f}}=\frac{1}{2g_{8}^{2}}\int_{\mathcal{W}}\ud^{8}x\,
  \sqrt{\tilde{g}}\,
  \bar{\theta}\biggl\{\ue^{-a}\slashed{\partial}_{\mathbb{R}^{1,3}}
  + \ue^{a}\slashed{\partial}_{\cS_{4}}
  +\ue^{a}\frac{1}{2}\slashed{\partial}_{\cS_{4}}
  a\bigl(1+2\Gamma_{\mathrm{extra}}\bigr)\biggr\}\theta,
\end{equation}
where $\theta$ is a 10d Majorana-Weyl spinor\footnote{Our conventions
  are spelled out in Appendix~\ref{app:conventions}.  Note that the
  spinors differ by a multiplicative factor of $\lambda$ compared
  to~\cite{Martucci:2005rb}.} and the 8d Yang-Mills coupling is
related to the $\uD 7$-brane tension by $g_{8}^{-2}=\tau_{\uD
  7}\lambda^{2}$.  The warp factor has explicitly been factored out
from the $\Gamma$-matrices so that
\begin{equation}
\label{slashp}
  \slashed{\partial}_{\mathbb{R}^{1,3}}=
  \tilde{\Gamma}^{\mu}\partial_{\mu},\qquad
  \slashed{\partial}_{\cS_{4}}=
  \tilde{\Gamma}^{b}\partial_{b},
\end{equation}
where $\tilde{\Gamma}^{M}$ are unwarped $\Gamma$-matrices, $\mu$ runs
over the external dimensions and $b$ over
$\mathcal{S}_{4}$.\footnote{If the 4-cycle ${\cal S}_4$ has a non-flat
  metric then, globally, we need to replace $\partial_b \rightarrow
  \nabla_b$, with $\nabla_b$ the pull-back of the ambient space
  covariant derivative, see~\cite{Marchesano:2008rg}. However, when
  analyzing wavefunctions in a local coordinate system such
  that~\eqref{eq:warped_kahler_form_restricted} holds, one may locally
  work in flat coordinates as in~\eqref{slashp}.}  Note that in
writing~\eqref{eq:warped_Dirac_action}, we have assumed that the
dilaton is constant.  The effect of the $5$-form flux is encoded in
$\Gamma_{\mathrm{extra}}$, the chirality matrix on
$\mathcal{S}_{4}$~\cite{Marchesano:2008rg}. As elaborated upon in
Appendix~\ref{app:conventions}, the internal spinors can be written as
$\eta_{\epsilon_{1}\epsilon_{2}\epsilon_{3}}$ where $\epsilon_{m}=\pm$
and if $\epsilon_{m}=+\left(-\right)$ then
$\eta_{\epsilon_{1}\epsilon_{2}\epsilon_{3}}$ is annihilated by
$\Gamma^{m}$ ($\Gamma^{\bar{m}}$).  Then
\begin{equation}
  \Gamma_{\mathrm{extra}}\eta_{\epsilon_{1}\epsilon_{2}\epsilon_{3}}
  =\epsilon_{1}\epsilon_{2}\eta_{\epsilon_{1}\epsilon_{2}\epsilon_{3}}.
\end{equation}

As follows from our previous discussion, in order to describe the
non-trivial intersection we need a non-Abelian generalization
of~\eqref{eq:warped_Dirac_action}.  For general backgrounds, the
fermionic analogue of the Myers action~\cite{Myers:1999ps} is not
known. However, to leading order in $\alpha'$, the non-Abelian version
of~\eqref{eq:warped_Dirac_action} can be obtained by promoting
derivatives to gauge-covariant derivatives and including the Yukawa
coupling that appears in the Super Yang-Mills action,
\begin{equation}
  \label{eq:wynants}
  S_{\uD 7}^{\mathrm{f}}=\frac{1}{2g_{8}^{2}}
  \int_{\mathcal{W}}\ud^{8}x\,\sqrt{\tilde{g}}\,\tr\biggl\{
  \bar{\theta}\biggl[\ue^{-a}\slashed{D}_{\mathbb{R}^{1,3}}
  +\ue^{a}\slashed{D}_{\cS_{4}}
  +\ue^{a}\frac{1}{2}\slashed{\partial}_{\cS_{4}}a
  \bigl(1+2\Gamma_{\mathrm{extra}}\bigr)\biggr]\theta
  -\ui\bar{\theta}\ue^{-a}\Gamma_{i}
  \bigl[\Phi^{i},\theta\bigr]\biggr\},
\end{equation}
where $i,j$ run over the coordinates that are transverse to the
brane.  One can explicitly check~\cite{Wynants} that, to leading order
in $\alpha'$, this is the supersymmetrization of the bosonic action.
As was done in considering the BPS equations, in
writing~\eqref{eq:wynants}, we have neglected the non-Abelian nature
of the bulk field $a$ and have evaluated it at $\Phi=0$.

It is
useful to define
\begin{equation}
  \hat{\cD}_{m}^{\mp}=\hat{D}^{\mp}_{m}+\frac{1}{2}
  \hat{\partial}_{m}a\bigl(1+2\Gamma_{\mathrm{extra}}\bigr),
\end{equation}
where $\hat{D}^{\mp}_{m}$ is defined in the unmagnetized case
in~\eqref{eq:unmagnetized_gauge_covariant_derivatives} and we have
used the fact that, since this is defined on $\cS_{4}$, $\partial_{3}$
acting on anything vanishes so that $\hat{\cD}_{3}^{\mp}=\hat{D}_{3}^{\mp}$.  
Separating terms based on internal chirality, the equation of motion resulting
from~\eqref{eq:wynants} to linear order in fluctuations gives
\begin{subequations}
\label{eq:fermion_eom}
\begin{align}
  0=&\ui
  \slashed{\partial}_{\mathbb{R}^{3,1}}
  \bigl(\psi^{\pm}_{0}\bigr)^{\dagger}
  \begin{pmatrix} 0 \\ \sigma_{2}\xi^{\dagger}\end{pmatrix}
  -\sqrt{\frac{2}{\pi\alpha'}}\ue^{2a}
  \begin{pmatrix} \xi \\ 0\end{pmatrix}
  \biggl(\hat{\cD}^{\mp}_{1}\psi^{\mp}_{1}
  +\hat{\cD}_{2}^{\mp}\psi^{\mp}_{2}
  +\ue^{-2a}\hat{\cD}_{3}^{\mp}\psi^{\mp}_{3}\biggr),\\
  0=&\ui
  \slashed{\partial}_{\mathbb{R}^{3,1}}
  \bigl(\psi^{\pm}_{1}\bigr)^{\dagger}
  \begin{pmatrix} 0 \\ \sigma_{2}\xi^{\ast}\end{pmatrix}
  +\sqrt{\frac{2}{\pi\alpha'}}\ue^{2a}
  \begin{pmatrix} \xi \\ 0\end{pmatrix}
  \biggl(\hat{\cD}^{\mp}_{1}\psi^{\mp}_{0}
  +\bigl(\hat{\cD}^{\pm}_{2}\bigr)^{\dagger}\psi^{\mp}_{3}
  -\ue^{-2a}\bigl(\hat{\cD}^{\pm}_{3}\bigr)^{\dagger}\psi^{\mp}_{2}
  \biggr),\\
  0=&\ui
  \slashed{\partial}_{\mathbb{R}^{3,1}}
  \bigl(\psi^{\pm}_{2}\bigr)^{\dagger}
  \begin{pmatrix} 0 \\ \sigma_{2}\xi^{\ast}\end{pmatrix}
  -\sqrt{\frac{2}{\pi\alpha'}}\ue^{2a}
  \begin{pmatrix} \xi \\ 0\end{pmatrix}
  \biggl(\bigl(\hat{\cD}^{\pm}_{1}\bigr)^{\dagger}\psi^{\mp}_{3}
  -\hat{\cD}^{\mp}_{2}\psi^{\mp}_{0}
  -\ue^{-2a}\bigl(\hat{\cD}^{\pm}_{3}\bigr)^{\dagger}\psi^{\mp}_{1}
  \biggr),\\
  0=&\ui
  \slashed{\partial}_{\mathbb{R}^{3,1}}
  \bigl(\psi^{\pm}_{3}\bigr)^{\dagger}
  \begin{pmatrix} 0 \\ \sigma_{2}\xi^{\ast}\end{pmatrix}
  +\sqrt{\frac{2}{\pi\alpha'}}\ue^{2a}
  \begin{pmatrix} \xi \\ 0\end{pmatrix}
  \biggl(\bigl(\hat{\cD}^{\pm}_{1}\bigr)^{\dagger}\psi^{\mp}_{2}
  -\bigl(\hat{\cD}^{\pm}_{2}\bigr)^{\dagger}\psi^{\mp}_{1}
  +\ue^{-2a}\hat{\cD}_{3}^{\mp}\psi^{\mp}_{0}\biggr),
\end{align}
\end{subequations}
where $\hat{\cD}_{m}^{\mp}\psi^{\mp}_{n}$ should be understood to mean
\begin{equation}
  \hat{\cD}_{m}^{\mp}\psi^{\mp}_{n}
  =\begin{cases}
    \bigl(\hat{D}_{m}^{\mp}+\frac{3}{2}\hat{\partial}_{m}a\bigr)
    \psi^{\mp}_{n}&
    n=0,3\\
    \bigl(\hat{D}_{m}^{\mp}-\frac{1}{2}\hat{\partial}_{m}a\bigr)
    \psi^{\mp}_{n}&n=1,2
  \end{cases}.
\end{equation}
Here $\psi_{0}$ is the 4d gaugino, $\psi_{3}$ is the modulino, the
superpartner of the complexified transverse scalar $\phi$ and
$\psi_{1}$ and $\psi_{2}$ are the Wilsonini, the superpartners of the
complexified Wilson lines $\phi_{1}$ and $\phi_{2}$; the
former pair have positive $\mathcal{S}_{4}$-chirality while the latter
have negative chirality.  In writing~\eqref{eq:fermion_eom}, we have
made use of the fact that the Clifford algebra following
from~\eqref{eq:warped_kaehler_form} implies that the $\Gamma$-matrices
have explicit factors of the ${\cal S}_4$ metric and the warp factor.

To compare the above result with the eom for bosonic wavefunctions,
let us relate them as in the Abelian case by~\cite{Marchesano:2008rg}
\begin{equation}
  \label{eq:MMS1ansatz}
  \psi_{0}^{\mp}=\ue^{-3a/2}\phi_{0}^{\mp},\quad
  \psi_{1}^{\mp}=\ue^{a/2}\phi_{1}^{\mp},\quad
  \psi_{2}^{\mp}=\ue^{a/2}\phi_{2}^{\mp},\quad
  \psi_{3}^{\mp}=\ue^{-3a/2}\phi_{3}^{\mp},
\end{equation}
The zero mode equations then become
\begin{subequations}
\label{eq:bosonic_zero_mode_eqs}
\begin{align}
  0=&\hat{D}^{\mp}_{1}\phi_{1}^{\mp}+
  \hat{D}^{\mp}_{2}\phi_{2}^{\mp}+
  \ue^{-4a}\hat{D}^{\mp}_{3}\phi_{3}^{\mp},\\
  0=&\hat{D}^{\mp}_{1}\phi^{\mp}_{0}
  +\bigl(\hat{D}^{\pm}_{2}\bigr)^{\dagger}\phi_{3}^{\mp}
  -\bigl(\hat{D}^{\pm}_{3}\bigr)^{\dagger}\phi_{2}^{\mp},\\
  0=&\bigl(\hat{D}^{\pm}_{1}\bigr)^{\dagger}\phi_{3}^{\mp}
  -\hat{D}^{\mp}_{2}\phi^{\mp}_{0}
  -\bigl(\hat{D}^{\pm}_{3}\bigr)^{\dagger}\phi_{1}^{\mp},\\
  0=&\bigl(\hat{D}^{\pm}_{1}\bigr)^{\dagger}\phi_{2}^{\mp}-
  \bigl(\hat{D}^{\pm}_{2}\bigr)^{\dagger}\phi_{1}^{\mp}
  +\ue^{-4a}
  \hat{D}_{3}^{\mp}\phi^{\mp}_{0},
\end{align}
\end{subequations}
exactly reproducing \eqref{eq:unmagnetized_bps_eom_2} in the vanishing
dilaton case up to the degree of freedom given by the gaugino-like
component $\psi^{\mp}_{0}$. Its bosonic partner $A^{\mp}_{\mu}$ was
not present in our previous discussion of the BPS $\uD 7$-brane
conditions by simple 4d Poincar\'e invariance. Since the gauge group
is given by $U(N_a) \times U(N_b)$, we always expect to be able to
consistently set $A^{\mp}_{\mu}=0$ at the massless level and so, if
our background is supersymmetric, the same should be true for
$\psi^{\mp}_{0}$. As we will see, this is the case for all the
wavefunctions obtained below, none of them containing any
$\psi^{\mp}_{0}$ piece.

As mentioned above,~\eqref{eq:bosonic_zero_mode_eqs} were derived assuming
a constant dilaton background.  The complication in
moving to the more general case is that the appearance of the axio-dilaton
in the fermionic action of~\cite{Marolf:2003ye,Marolf:2003vf,
  Martucci:2005rb} modify the equations of motion for the fermionic
wavefunctions in a non-trivial way (see,
e.g.~\cite{Marchesano:2008rg}).  However,
as~\eqref{eq:bosonic_zero_mode_eqs} precisely
reproduce~\eqref{eq:unmagnetized_bps_eom_2} in the case of constant
dilaton, we expect~\eqref{eq:bosonic_zero_mode_eqs} to hold in the
case of varying dilaton as well after the replacement $a\to
\alpha=a+\frac{1}{4}\log\left(\im{\tau}\right)$.  Furthermore, as
mentioned above $\psi_{0}^{\mp}$ ought to vanish for the warped zero
modes in which case~\eqref{eq:bosonic_zero_mode_eqs} applies.

While either \eqref{eq:bosonic_zero_mode_eqs} or
\eqref{eq:unmagnetized_bps_eom_2} can then be taken to the form
\eqref{eq:quadratic_eom_warped}, the latter does not seem to admit an
exact solution for general warp factor. One should then use an
approximation scheme in order to express the warped wavefunctions. The
scheme that we develop below is based on the spectrum of massive modes
at the intersection, which we now turn to analyze.

\subsection{\label{subsec:nonchiralMM}Unwarped massive spectrum}

For generic warp factors, the
equations~\eqref{eq:bosonic_zero_mode_eqs} do not seem to admit a
simple analytic solution. However, given a complete set of functions
that satisfy the same boundary conditions as the warped zero mode, we
can always expand the latter in terms of this set. In this sense,
solving the equations of motion amounts to solving for the
coefficients of this expansion.

In our case one may realize this expansion as follows. Let us first write the warped zero mode as a vector
\begin{equation}
  \label{eq:zero_mode}
  \phi_{m}^{\mp}=\sigma^{\mp}\left(x^{\mu}\right)
  \chi_{m}^{\mp}\left(x^{a}\right),\qquad
  \mathbf{X}^{\mp}=\bigl(\chi^{\mp}_{0},\chi_{1}^{\mp},
  \chi_{2}^{\mp},\chi_{3}^{\mp}\bigr)^{\mathrm{T}},
\end{equation}
where $\sigma^{\mp}$ again carry suppressed gauge indices while
$\chi_{m}^{\mp}$ do not. Then~\eqref{eq:bosonic_zero_mode_eqs} takes
the form $\sigma^{\mp}\hat{\mathbf{D}}^{\mp}\mathbf{X}^{\mp}=0$, where
\begin{equation}
  \label{eq:vector_diff_op}
  \hat{\mathbf{D}}^{\mp}=
  \begin{pmatrix}
    0 & \ue^{4\alpha}\hat{D}^{\mp}_{1} &
    \ue^{4\alpha}\hat{D}^{\mp}_{2} & \hat{D}^{\mp}_{3} \\
    -\hat{D}^{\mp}_{1} & 0 &
    \bigl(\hat{D}^{\pm}_{3}\bigr)^{\ast} &
    -\bigl(\hat{D}^{\pm}_{2}\bigr)^{\ast} \\
    -\hat{D}^{\mp}_{2} &
    -\bigl(\hat{D}^{\pm}_{3}\bigr)^{\ast} & 0 &
    \bigl(\hat{D}^{\pm}_{1}\bigr)^{\ast} \\
    -\hat{D}_{3}^{\mp} &
    \ue^{4\alpha}\bigl(\hat{D}^{\pm}_{2}\bigr)^{\ast} &
    -\ue^{4\alpha}\bigl(\hat{D}^{\pm}_{1}\bigr)^{\ast} & 0
  \end{pmatrix}.
\end{equation}
Denoting the complete set of functions as
$\bigl\{\bs{\Phi}_{\lambda}^{\mp}\bigr\}$, we take the expansion
\begin{equation}
  \label{eq:MMexpansion}
  \mathbf{X}^{\mp}=\sum_{\lambda}
  c^{\mp}_{\lambda}\boldsymbol{\Phi}_{\lambda}^{\mp},
\end{equation}
where $c^{\mp}_{\lambda}$ are the coefficients for which we wish to solve.

In general, a complete set of wavefunctions with the same boundary
conditions is given by the full tower of massive modes within the same
open string sector, which in our case are the tower of strings stretched
between the two intersecting $\uD 7$-branes. We may thus take as a
set those wavefunctions that correspond to the unwarped massive modes
at the intersection, and then expand the warped zero mode in this
basis.  Such a spectrum of unwarped massive modes can be deduced
from~\eqref{eq:fermion_eom}, which in the absence of warping gives the
following equation of motion
\begin{equation}
  \label{eq:MM_eom1}
  \ui\hat{\mathbf{D}}_{0}^{\mp}\boldsymbol{\Phi}^{\mp}_{\lambda}=
  \sqrt{\frac{\pi\alpha'}{2}}m_{\lambda}
  \boldsymbol{\Phi}^{\pm\ast}_{\lambda},
\end{equation}
where $\hat{\mathbf{D}}^{\mp}_{0}$ is the unwarped version
of~\eqref{eq:vector_diff_op}.  Acting on~\eqref{eq:MM_eom1} with
its conjugate gives
\begin{equation}
  \label{eq:MM_eom2}
  \bigl(\hat{\mathbf{D}}_{0}^{\pm}\bigr)^{\ast}
  \hat{\mathbf{D}}_{0}^{\mp}\boldsymbol{\Phi}_{\lambda}^{\mp}
  =\frac{\pi\alpha'}{2}
  \abs{m_{\lambda}}^{2}\boldsymbol{\Phi}_{\lambda}^{\mp},
\end{equation}
and so we obtain an eigenvalue equation for the unwarped massive modes
at the intersection. Moreover, in our local description the
bifundamental fields at the intersection can be treated as living on
$\Sigma\times\mathbb{C}$ where $\Sigma$ is the matter curve, and
$\mathbb{C}$ has coordinate $z^{2}$.  We then impose that the massive
modes are well-defined on $\Sigma$ and vanish as
$\abs{z^{2}}^{2}\to\infty$.  With these boundary conditions, we can
further impose that the massive modes are orthonormal with respect to
the inner product
\begin{equation}
  \label{eq:vector_inner_product}
  \bigl<\boldsymbol{\Phi},\boldsymbol{\Psi}\bigr>=
  \im\tau_{2}
  \int_{\Sigma\times\mathbb{C}}\ud^{4}z\,
  \sqrt{\tilde{g}}\,\boldsymbol{\Phi}^{\ast}\cdot
  \boldsymbol{\Psi},
\end{equation}
with $\cdot$ the ordinary dot product for vectors. The prefactor
is introduced for later convenience.

In order to find the general solution to~\eqref{eq:MM_eom2} our
strategy will be to map the eigenvalue problem to that of the quantum
simple harmonic oscillator (QSHO) and then make use of basic
techniques of quantum mechanics to find the spectrum. This implies
using the non-trivial commutation relations between the covariant
derivatives, which in the unmagnetized case amount to
\begin{equation}
  \bigl[\bigl(\hat{D}^{\pm}_{2}\bigr)^{\ast},\hat{D}^{\mp}_{3}\bigr]=\mp
  \frac{\ui R_{3}I_{3}^{\left(ab\right)}}{2R_{2}}=: \mp \ui \hat{M}_{3}.
\end{equation}
Using this, we find
\begin{equation}
  \label{eq:unmagnetized_Laplacian}
  \bigl(\mathbf{D}_{0}^{\pm}\bigr)^{\ast}\mathbf{D}_{0}^{\mp}
  =-\triangle^{\mp}
  \pm\mathbf{B},
\end{equation}
where
\begin{equation}
  \triangle^{\mp}=\sum_{m=1}^{3}
  \bigl(\hat{D}^{\pm}_{m}\bigr)^{\ast}\hat{D}^{\mp}_{m},\quad
  \mathbf{B}=\begin{pmatrix}
    \phantom{a}0\phantom{a} & 0 & 0 & 0 \\
    0 & 0 & 0 & 0 \\
    0 & 0 & 0 & \ui\hat{M}_{3} \\
    0 & 0 & -\ui\hat{M}_{3} & 0
  \end{pmatrix}.
\end{equation}
Since the two pieces in~\eqref{eq:unmagnetized_Laplacian} commute,
they can be simultaneously diagonalized.  $\mathbf{B}$ clearly has a
non-trivial nullspace spanned by $\left(1, 0, 0,
  0\right)^{\mathrm{T}}$ and $\left(0, 1, 0, 0\right)^{\mathrm{T}}$.
In addition, there are two non-trivial eigenvalues, $-\hat{M}_{3}$ and
$+\hat{M}_{3}$ with respective eigenvectors
\begin{equation}
  \frac{1}{\sqrt{2}}
  \left(0, 0, 1, \ui\right)^{\mathrm{T}},\quad
  \frac{1}{\sqrt{2}}
  \left(0, 0, \ui, 1\right)^{\mathrm{T}}.
\end{equation}
The diagonalization of $\mathbf{B}$ is then effected by the
rotation
\begin{equation}
  \mathbf{J}
  =\begin{pmatrix}
    1 & 0 & 0 & 0 \\
    0 & 1 & 0 & 0 \\
    0 & 0 & 1/\sqrt{2} & \ui/\sqrt{2} \\
    0 & 0 & \ui/\sqrt{2} & 1/\sqrt{2}
  \end{pmatrix},
\end{equation}
so that we have\footnote{$\mathbf{\Phi}^{\mp}$ transforms as
  $\mathbf{\Phi}^{\mp}\to\mathbf{J}^{-1}\mathbf{\Phi}^{\mp}$, so to
  have
  $\bigl(\hat{\mathbf{D}}^{\pm}_{m}\bigr)^{\ast}\mathbf{\Phi}^{\mp}$
  transform as $\bs{\Phi}^{\pm\ast}$, we must have
  $\hat{\mathbf{D}}^{\mp}_{m}\to \mathbf{J}\hat{\mathbf{D}}^{\mp}_{m}
  \mathbf{J}$.  Note than when $\mathbf{J}$ is not symmetric, which
  occurs when, for example, warping is included, the transformation is
  $\hat{\mathbf{D}}^{\mp}_{m}\to\mathbf{J}^{\mathrm{T}}\hat{\mathbf{D}}^{\mp}_{m}\mathbf{J}$.}
\begin{equation}
\label{rotatedL}
  \mathbf{J}^{-1}\bigl(\hat{\mathbf{D}}_{0}^{\pm}\bigr)^{\ast}
  \hat{\mathbf{D}}_{0}^{\mp}\mathbf{J}
  =-\triangle^{\mp}\pm
  \mathrm{diag}\bigl(0,0,-\hat{M}_{3},\hat{M}_{3}\bigr).  
\end{equation}

To make use of QSHO techniques, we begin with a ground state.  This is
given by the unwarped zero mode~\eqref{eq:unwarped_solution}, though
it is useful to confirm this in this language.  In the rotated basis,
the unwarped zero mode satisfies
$\hat{\mathbf{D}}'^{\mp}_{0}\boldsymbol{\Phi}'^{\mp}_{0}=0$ where
$\bs{\Phi}'^{\mp}_{0} = \mathbf{J}^{-1} \bs{\Phi}^{\mp}_{0}$ is the 
unwarped zero mode in the rotated basis and
\begin{equation}
  \label{eq:rotated_vec_diff_op}
  \hat{\mathbf{D}}'^{\mp}_{0}=
  \mathbf{J}\hat{\mathbf{D}}^{\mp}_{0}\mathbf{J}
  =\begin{pmatrix}
    0 & \hat{D}'^{\mp}_{1} &
    \hat{D}'^{\mp}_{2} &
    \hat{D}'^{\mp}_{3} \\
    -\hat{D}'^{\mp}_{1} & 0 &
    \bigl(\hat{D}'^{\pm}_{3}\bigr)^{\ast} &
    -\bigl(\hat{D}'^{\pm}_{2}\bigr)^{\ast} \\
    -\hat{D}'^{\mp}_{2} &
    -\bigl(\hat{D}'^{\pm}_{3}\bigr)^{\ast} & 0 &
    \bigl(\hat{D}'^{\pm}_{1}\bigr)^{\ast} \\
    -\hat{D}'^{\mp}_{3} &
    \bigl(\hat{D}'^{\pm}_{2}\bigr)^{\ast} &
    -\bigl(\hat{D}'^{\pm}_{1}\bigr)^{\ast} & 0
  \end{pmatrix},
\end{equation}
with
\begin{subequations}
\label{eq:rotated_covariant_derivatives}
\begin{align}
  \hat{D}'^{\mp}_{1}=&\hat{D}^{\mp}_{1}=\hat{\partial}_{1},\\
  \hat{D}'^{\mp}_{2}=&\frac{1}{\sqrt{2}}\bigl(\hat{D}^{\mp}_{2}
  +\ui \hat{D}^{\mp}_{3}\bigr)=\frac{1}{\sqrt{2}\sqrt{2\pi}R_{2}}
  \bigl(\partial_{2}\pm\kappa\bar{z}^{\bar{2}}\bigr),\\
  \hat{D}'^{\mp}_{3}=&\frac{1}{\sqrt{2}}\bigl(\hat{D}^{\mp}_{3}
  +\ui \hat{D}^{\mp}_{2}\bigr)
  =\frac{\ui}{\sqrt{2}\sqrt{2\pi}R_{2}}\bigl(
  \partial_{2}\mp\kappa\bar{z}^{\bar{2}}\bigr).
\end{align}
\end{subequations}
with $\kappa$ defined as in~\eqref{eq:unwarped_sol_g}.
To look for a zero mode, we can now try the various eigenvectors of
$\mathbf{B}$.  We first consider something in the null space of
$\mathbf{B}$ and try to solve
\begin{equation}
  \hat{\mathbf{D}}'^{\mp}_{0}
  \begin{pmatrix}
    \vp^{\mp} \\ 0 \\ 0 \\ 0
  \end{pmatrix}
  =\begin{pmatrix}
    0 \\ -\hat{D}'^{\mp}_{1} \\ -\hat{D}'^{\mp}_{2} \\
    -\hat{D}'^{\mp}_{3}
  \end{pmatrix}\vp^{\mp}=0.
\end{equation}
For this to be a zero mode, we then need
$\hat{D}'^{\mp}_{3}\vp^{\mp}=\hat{D}'^{\mp}_{2}\vp^{\mp}=0$, which
together have only the trivial solution $\vp^{\mp}=0$.  A similar
statement applies for $\left(0, \vp^{\mp}, 0, 0\right)^{\mathrm{T}}$.
On the other hand, the $-\hat{M}_{3}$ eigenvector $\left(0, 0,
  \vp^{\mp}, 0\right)^{\mathrm{T}}$ will be a zero mode of
$\hat{\mathbf{D}}'^{\mp}_{0}$ if $\vp^{\mp}$ is in the kernel of
$\bigl(\hat{D}'^{\pm}_{1}\bigr)^{\ast}$,
$\bigl(\hat{D}'^{\pm}_{3}\bigr)^{\ast}$, and $\hat{D}'^{\mp}_{2}$.
This implies that $\vp^{\mp}$ is independent of $\bar{z}^{\bar{1}}$
(and hence, by periodicity, independent of $z^1$ as well) and satisfies
\begin{equation}
  \bigl(\partial_{2}\pm\kappa\bar{z}^{\bar{2}}\bigr)\vp^{\mp}=
  \bigl(\partial_{2}^{\ast}\pm \kappa z^{2}\bigr)\vp^{\mp}=0.
\end{equation}
These in turn imply
\begin{equation}
  \vp^{\mp}\sim\ue^{\mp\kappa\left\vert z^{2}\right\vert^{2}}.
\end{equation}
Requiring that the wavefunction goes to zero as $\abs{z^{2}}\to\infty$
implies that the $+$-sector has no non-trivial solutions.  However, in
the $-$-sector, there is a non-trivial zero mode given by
\begin{equation}
\label{zm-}
  \boldsymbol{\Phi}'^{-}_{0}=\bigl(
    0, 0,\vp_{0},0\bigr)^{\mathrm{T}},\quad
  \vp_{0}\sim \ue^{-\kappa\left\vert z^{2}\right\vert^{2}}.
\end{equation}
Similarly, there is a non-trivial zero mode in the $+$-sector in the
$+\hat{M}_{3}$ eigenspace of $\mathbf{B}$ given by
\begin{equation}
  \label{zm+}
    \boldsymbol{\Phi}'^{+}_{0}=
    \bigl(0, 0, 0, \vp_{0}\bigr)^{\mathrm{T}}.
\end{equation}
Requiring that these modes are normalized according
to~\eqref{eq:vector_inner_product} gives
\begin{equation}
  \vp_{0}=\sqrt{\frac{2\kappa}{\pi\cV_{1}\cV_{2}}}
  \ue^{-\kappa\left\vert z^{2}\right\vert^{2}}.
\end{equation}
Finally, after rotating back, these zero modes agree with~\eqref{eq:unwarped_solution}.

In order to find the higher modes of
$\bigl(\hat{\mathbf{D}}^{\pm}\bigr)^{\ast}\hat{\mathbf{D}}^{\mp}$, we
need to find the spectrum of $\triangle^{\mp}$.  To do this, we
re-express the problem of finding this spectrum of modes in the
language of a QSHO.  The rotated derivatives $\hat{D}'^{\mp}_{m}$
satisfy the commutation relations
\begin{equation}
  \label{eq:unmagnetized_QSHO_algebra}
  \bigl[\bigl(\hat{D}'^{\pm}_{2}\bigr)^{\ast},
  \hat{D}'^{\mp}_{2}\bigr]=\pm\hat{M}_{3},\qquad
  \bigl[\bigl(\hat{D}'^{\pm}_{3}\bigr)^{\ast},
  \hat{D}'^{\mp}_{3}\bigr]=\mp\hat{M}_{3},
\end{equation}
with other commutators vanishing, while $\triangle^{\mp}$ can be
expressed as
\begin{equation}
  \label{eq:rotated_Laplacians}
  \triangle^{\mp}=\sum_{m=1}^{3}\triangle'^{\mp}_{m},\qquad
  \triangle'^{\mp}_{m}=
  \frac{1}{2}\biggl\{
  \bigl(\hat{D}'^{\pm}_{m}\bigr)^{\ast},\hat{D}'^{\mp}_{m}\biggr\}.
\end{equation}
We then have the commutation relations
\begin{subequations}
\begin{align}
  \bigl[\triangle'^{\mp}_{2},\hat{D}'^{\mp}_{2}\bigr]=&
  \pm\hat{M}_{3}\hat{D}'^{\mp}_{2}, &
  \bigl[\triangle'^{\mp}_{2},\bigl(\hat{D}'^{\pm}_{2}\bigr)^{\ast}\bigr]=&
  \mp\hat{M}_{3}\bigl(\hat{D}'^{\pm}_{2}\bigr)^{\ast},\\
  \bigl[\triangle'^{\mp}_{3},\hat{D}'^{\mp}_{3}\bigr]=&
  \mp\hat{M}_{3}\hat{D}'^{\mp}_{3}, &
  \bigl[\triangle'^{\mp}_{3},\bigl(\hat{D}'^{\pm}_{3}\bigr)^{\ast}\bigr]=&
  \pm\hat{M}_{3}\bigl(\hat{D}'^{\pm}_{3}\bigr)^{\ast}.
\end{align}
\end{subequations}
Which give four independent QHSO algebras, using the rotated covariant derivatives 
as ladder operators and $\triangle'^{\mp}_{m=2,3}$ as Hamiltonians.  The ground state
wavefunction $\vp_{0}$ satisfies
\begin{equation}
  \label{eq:unmagnetized_gs_eigenvalues}
  \triangle'^{\mp}_{1}\vp_{0}=0,\quad
  \triangle'^{\mp}_{2}\vp_{0}=\triangle'^{\mp}_{3}\vp_{0}
  =-\frac{1}{2}\hat{M}_{3}\vp_{0}.
\end{equation}
from which $\triangle^{\mp}\varphi_{0}=-M_{3}\varphi_{0}$ and so \eqref{zm-}, 
\eqref{zm+} are zero modes of \eqref{rotatedL}. We can then build up the higher modes 
with raising operators acting on $\varphi_0$, just as is done for the QSHO.

Considering the $-$-sector, since the ground state $\vp_{0}$ is annihilated
by  $\hat{D}'^{-}_{2}$ and $\bigl(\hat{D}'^{+}_{3}\bigr)^{\ast}$, 
we have the lowering operators
\begin{equation}
  \ui \hat{D}'^{-}_{2},\quad \ui\bigl(\hat{D}'^{+}_{3}\bigr)^{\ast},
\end{equation}
whose adjoints with respect to the inner product
\begin{equation}
  \label{eq:component_inner_product}
  \bigl\langle\phi,\psi\bigr\rangle=
  \im\tau_{2}
  \int_{\Sigma\times\mathbb{C}}\ud^{4}z\,
  \sqrt{\tilde{g}}\,\phi^{\ast}\psi,
\end{equation}
are, respectively,
\begin{equation}
  \ui \bigl(\hat{D}'^{+}_{2}\bigr)^{\ast},\quad
  \ui \hat{D}'^{-}_{3},
\end{equation}
and will act as raising operators.  The algebra generated by
$\ui \hat{D}'^{-}_{2}$ and $\ui\bigl(\hat{D}'^{+}_{2}\bigr)^{\ast}$ is
independent from the algebra generated by
$\ui\bigl(\hat{D}'^{+}_{3}\bigr)^{\ast}$ and $\ui \hat{D}'^{-}_{3}$.
The higher eigenfunctions of $\triangle^{\mp}$ result from acting on
the zero modes with the raising operators
\begin{equation}
  \vp^{-}_{mnlp}=N^{-}_{lp}
  \bigl[\ui\bigl(\hat{D}'^{+}_{2}\bigr)^{\ast}\bigr]^{l}
  \bigl(\ui \hat{D}'^{-}_{3}\bigr)^{p}
  \vp_{mn00}^{-},
\end{equation}
where
\begin{equation}
  \vp_{mn00}^{-}=h_{mn}\bigl(z^{1},\bar{z}^{\bar{1}}\bigr)
  \vp_{0},
\end{equation}
and where $h_{mn}$ are Fourier modes that are discussed below.  The
proportionality constant is chosen so that $\vp_{mn00}^{-}$ is
normalized with respect to~\eqref{eq:component_inner_product}.  Using
the QSHO algebras, it is easy to verify that these modes satisfy
\begin{equation}
  \triangle\vp_{00lp}^{-}=-\hat{M}_{3}\bigl(l+p+1\bigr)\vp_{00lp}^{-}.
\end{equation}
From this it follows that after the eigenfunctions $\vp_{mn00}^{-}$
are normalized, $\vp_{mnlp}^{-}$ are normalized by taking
\begin{equation}
  N^{-}_{lp}=\frac{1}{\sqrt{\hat{M_{3}}^{l+p}l!p!}}.
\end{equation}

The massive eigenmodes will additionally have a non-trivial dependence on
$\Sigma$. For instance, in the case $\Sigma = \mathbb{T}^2$, because 
$\langle F_2 \rangle =0$  all fields need to be periodic in $\mathbb{T}^2$, and so 
the higher modes involve the Fourier modes
\begin{equation}
  \label{eq:fouriermodes}
  h_{mn}=
  \ue^{2\pi\ui\,\mathrm{Im}\left[\left(m-\bar{\tau}_{1}n\right)z^{1}\right]/
    \mathrm{Im}\,\tau_{1}}.
\end{equation}
The normalized eigenfunctions of $\triangle^{-}$ are then
\begin{equation}
  \label{eq:unmagnetized_component_MMs}
  \vp_{mnlp}^{-}=\sqrt{\frac{2\kappa}
    {\hat{M}_{3}^{p+l}\cV_{1}\cV_{2}\,p!\,l!}}
  h_{mn}\,
  \bigl[\ui\bigl(\hat{D}'^{+}_{2}\bigr)^{\ast}\bigr]^{l}
  \bigl(\ui \hat{D}'^{-}_{3}\bigr)^{p}
  \ue^{-\kappa\left\vert z^{2}\right\vert^{2}},
\end{equation}
with
\begin{equation}
  \triangle^{-}\vp_{mnlp}^{-}
  =-\biggl(\frac{2\pi^{3}\left\vert m-\tau_{1}n\right\vert^{2}}
  {\cV_{1}\im\tau_{1}}
  +\hat{M}_{3}\bigl(l+p+1\bigr)\biggr)\vp_{mnlp}^{-}.
\end{equation}
From this and~\eqref{eq:unmagnetized_Laplacian}, we find the following
spectrum for the $-$-sector
\begin{subequations}
\label{eq:unmagnetized_minus_sector_MM_spectrum}
\begin{align}
  \bigl\vert m^{-}_{0;mnlp}\bigr\vert^{2}=&\frac{2}{\pi\alpha'}
  \biggl(\hat{m}_{mn}^{2}
  +\hat{M}_{3}\bigl(l+p+1\bigr)\biggr)
  &
  \boldsymbol{\Phi}'^{-}_{0;mnlp}
  =&\bigl(\vp_{mnlp}^{-},0,0,0\bigr)^{\mathrm{T}},\\
  \bigl\vert m^{-}_{1;mnlp}\bigr\vert^{2}=&\frac{2}{\pi\alpha'}
  \biggl(\hat{m}_{mn}^{2}
  +\hat{M}_{3}\bigl(l+p+1\bigr)\biggr)
  &
  \boldsymbol{\Phi}'^{-}_{1;mnlp}
  =&\bigl(
    0, \vp_{mnlp}^{-}, 0,0\bigr)^{\mathrm{T}},\\
  \bigl\vert m^{-}_{2;mnlp}\bigr\vert^{2}=&\frac{2}{\pi\alpha'}
  \biggl(\hat{m}_{mn}^{2}
  +\hat{M}_{3}\bigl(l+p\bigr)\biggr)
  &
  \boldsymbol{\Phi}'^{-}_{2;mnlp}
  =&\bigl(0, 0,\vp_{mnlp}^{-},0\bigr)^{\mathrm{T}},\\
  \bigl\vert m^{-}_{3;mnlp}\bigr\vert^{2}=&\frac{2}{\pi\alpha'}
  \biggl(\hat{m}_{mn}^{2}
  +\hat{M}_{3}\bigl(l+p+2\bigr)\biggr)
  &
  \boldsymbol{\Phi}'^{-}_{3;mnlp}
  =&\bigl(0, 0,0,\vp_{mnlp}^{-}\bigr)^{\mathrm{T}},
\end{align}
\end{subequations}
where
\begin{equation}
  \label{eq:Fourier_mass}
  \hat{m}_{mn}^{2}=\frac{2\pi^{3}\absb{m-\tau_{1}n}^{2}}
  {\cV_{1}\,\im\tau_{1}}.
\end{equation}
Using the QSHO algebra, one can see that this is an orthonormal basis
with respect to~\eqref{eq:vector_inner_product}, that can be expressed
in terms of Hermite functions (see Appendix~\ref{app:hermite} for more
details).

For the $+$-sector, the lowering operators are
\begin{equation}
  \ui\bigl(\hat{D}'^{-}_{2}\bigr)^{\ast},\quad
  \ui \hat{D}'^{+}_{3},
\end{equation}
while the raising operators are
\begin{equation}
  \ui \hat{D}'^{+}_{2},\quad
  \ui\bigl(\hat{D}'^{-}_{3}\bigr)^{\ast}.
\end{equation}
An analogous calculation for the $+$-sector yields
\begin{subequations}
\label{eq:unmagnetized_plus_sector_MM_spectrum}
\begin{align}
  \bigl\vert m^{+}_{0;mnlp}\bigr\vert^{2}=&\frac{2}{\pi\alpha'}
  \biggl(\hat{m}_{mn}^{2}
  +\hat{M}_{3}\bigl(l+p+1\bigr)\biggr)
  &
  \boldsymbol{\Phi}'^{+}_{0;mnlp}
  =&\bigl(\vp_{mnlp}^{+},0,0,0\bigr)^{\mathrm{T}},\\
  \bigl\vert m^{+}_{1;mnlp}\bigr\vert^{2}=&\frac{2}{\pi\alpha'}
  \biggl(\hat{m}_{mn}^{2}
  +\hat{M}_{3}\bigl(l+p+1\bigr)\biggr)
  &
  \boldsymbol{\Phi}'^{+}_{1;mnlp}
  =&\bigl(
    0, \vp_{mnlp}^{+}, 0,0\bigr)^{\mathrm{T}},\\
  \bigl\vert m^{+}_{2;mnlp}\bigr\vert^{2}=&\frac{2}{\pi\alpha'}
  \biggl(\hat{m}_{mn}^{2}
  +\hat{M}_{3}\bigl(l+p+2\bigr)\biggr)
  &
  \boldsymbol{\Phi}'^{+}_{2;mnlp}
  =&\bigl(0, 0,\vp_{mnlp}^{+},0\bigr)^{\mathrm{T}},\\
  \bigl\vert m^{+}_{3;mnlp}\bigr\vert^{2}=&\frac{2}{\pi\alpha'}
  \biggl(\hat{m}_{mn}^{2}
  +\hat{M}_{3}\bigl(l+p\bigr)\biggr)
  &
  \boldsymbol{\Phi}'^{+}_{3;mnlp}
  =&\bigl(0, 0,0,\vp_{mnlp}^{+}\bigr)^{\mathrm{T}},
\end{align}
\end{subequations}
where $\vp^{+}_{mnlp} = (\vp^{-}_{mnlp})^*$, is as
in~\eqref{eq:unmagnetized_component_MMs} after replacing
$(\hat{D}'^{+}_{2})^{\ast} \rightarrow \hat{D}'^{+}_{2}$ and
$\hat{D}'^{-}_{3} \rightarrow (\hat{D}'^{-}_{3})^{\ast}$.

\subsection{\label{sec:unmagnetizedMM}Mode expansion of the warped zero mode}

For general warping, it is not always possible to solve for the
coefficients $c^{\mp}_{\lambda}$ appearing
in~\eqref{eq:MMexpansion}.  However, in cases of weak warping, we can
treat the deviation of the warp factor from constant as a perturbation
to the unwarped system.  That is, after a rescaling of coordinates, we
can write the warp factor as
\begin{equation}
  \label{eq:weak_warping}
  \ue^{-4\alpha}=1+\epsilon\beta,
\end{equation}
where $\beta$ is an $\mathcal{O}\left(1\right)$ function. If the
warping is weak in the sense that $\partial_{m}\alpha\ll 1$, then
$\epsilon\ll 1$ and so we can use $\epsilon$ as an expansion
coefficient. Indeed, in terms of $\epsilon$, the operator~\eqref{eq:vector_diff_op} can be
written
\begin{equation}
  \hat{\mathbf{D}}^{\mp}=
  \sum_{n}\epsilon^{n}\hat{\mathbf{D}}_{\left(n\right)}^{\mp},\qquad
  \hat{\mathbf{D}}_{\left(n\neq 0\right)}^{\mp}=
  \bigl(-\beta\bigr)^{n}\mathbf{K}^{\mp},
\end{equation}
where as before $\hat{\mathbf{D}}^{\mp}_{\left(0\right)} \equiv \mathbf{D}^{\mp}_{0}$ is
given by setting $\alpha=0$ in~\eqref{eq:vector_diff_op}, and $\mathbf{K}^{\mp}$ is
given by
\begin{equation}
  \label{eq:K}
  \mathbf{K}^{\mp}=
  \begin{pmatrix}
    0 & \hat{D}^{\mp}_{1} & \hat{D}^{\mp}_{2} & 0 \\
    0 & 0 & 0 & 0 \\
    0 & 0 & 0 & 0 \\
    0 & \bigl(\hat{D}^{\pm}_{2}\bigr)^{\ast} &
    -\bigl(\hat{D}^{\pm}_{1}\bigr)^{\ast} & 0
  \end{pmatrix}.
\end{equation}
Similarly, the zero mode can be written as
\begin{equation}
  \mathbf{X}^{\mp}=\sum_{n}\epsilon^{n}\mathbf{X}_{\left(n\right)}^{\mp}.
\end{equation}
As $\epsilon\to 0$, the warped zero mode should approach the unwarped
zero mode so we take the zeroth order term in the expansion
$\mathbf{X}_{\left(0\right)}^{\mp}$ to be the unwarped zero mode
$\bs{\Phi}_{0}^{\mp}$.  For $n\neq 0$, we expand in terms of the
unwarped massive modes,
\begin{equation}
  \mathbf{X}_{\left(n\right)}^{\mp}=\sum_{\lambda}c_{\lambda}^{\left(n\right)\mp}
  \bs{\Phi}_{\lambda}^{\mp}.
\end{equation}
The $\mathcal{O}\left(\epsilon^{n}\right)$ contribution to the warped
zero mode equation $\hat{\mathbf{D}}^{\mp}\mathbf{X}^{\mp}=0$ is
\begin{equation}
  0=\sum_{m=0}^{n}\hat{\mathbf{D}}_{\left(m\right)}^{\mp}
  \mathbf{X}_{\left(n-m\right)}^{\mp}.
\end{equation}
For $n=0$, this is satisfied with the choice
$\mathbf{X}_{\left(0\right)}^{\mp}=\bs{\Phi}_{0}^{\mp}$.  For $n>0$,
we can re-express this contribution as
\begin{equation}
  \hat{\mathbf{D}}_{0}^{\mp}\mathbf{X}_{\left(n\right)}^{\mp}
  =-\sum_{m=1}^{n}\hat{\mathbf{D}}_{\left(m\right)}^{\mp}
  \mathbf{X}_{\left(n-m\right)}^{\mp}.
\end{equation}
Acting on both sides with
$\bigl(\hat{\mathbf{D}}_{0}^{\pm}\bigr)^{\ast}$ and using the
orthonormality of~\eqref{eq:unmagnetized_minus_sector_MM_spectrum}
and~\eqref{eq:unmagnetized_plus_sector_MM_spectrum}, we find
\begin{equation}
  \label{eq:general_MM_coefficient}
  c_{\lambda}^{\left(n\right)\mp}
  =-\frac{2}{\pi\alpha'\absb{m_{\lambda}}^{2}}
  \sum_{m=1}^{n}\bigl(-1\bigr)^{m}
  \bigl\langle\bs{\Phi}_{\lambda}^{\mp},
  \bigl(\hat{\mathbf{D}}^{\pm}_{0}\bigr)^{\ast}
  \beta^{m}
  \mathbf{K}^{\mp}
  \mathbf{X}_{\left(n-m\right)}^{\mp}\bigr\rangle,
\end{equation}
the same expression also holding in the rotated basis.

Note that as is familiar from perturbation theory in quantum
mechanics, the coefficients $c_{0}^{\left(n\right)\mp}$ are not
determined by this procedure.  We will fix it by demanding that, to
all orders in $\epsilon$,
$\left\langle\mathbf{X}^{\mp},\bs{\Phi}_{\left(0\right)}^{\mp}\right\rangle=1$,
fixing $c_{0}^{\left(n\right)\mp}=0$ for all $n>0$.

We can write a particularly simple expression for the first order
correction.  From~\eqref{eq:general_MM_coefficient},
\begin{equation}
  \label{eq:MM_expansion_term_1}
  c_{\lambda}^{\left(1\right)}=\frac{2}{\pi\alpha'\absb{m_{\lambda}}^{2}}
  \bigl\langle
  \bs{\Phi}_{\lambda}^{\mp},\bigl(\hat{\mathbf{D}}^{\pm}_{0}
  \bigr)^{\ast}
  \beta\mathbf{K}^{\mp}\bs{\Phi}^{\mp}_{0}\bigr\rangle.
\end{equation}
Consider the $-$-sector where the zero mode in the unrotated basis is
written
\begin{equation}
  \boldsymbol{\Phi}^{-}_{0}=\frac{1}{\sqrt{2}}
  \begin{pmatrix}
  0 \\ 0 \\ 1 \\ \ui
    \end{pmatrix}
    \vp_{0}.
\end{equation}
From this,
\begin{equation}
  \beta\mathbf{K}^{-}\bs{\Phi}_{0}^{-}
  =\frac{1}{\sqrt{2}}\beta
   \begin{pmatrix}
  \hat{D}^{-}_{2} \\ 0 \\  0 \\  -\bigl(\hat{D}^{+}_{1}\bigr)^{\ast}
   \end{pmatrix}
  \vp_{0}.
\end{equation}
Since $\vp_{0}$ is independent of $z^{1}$, the lowest component
vanishes.  In the rotated basis we then get
\begin{equation}
  \label{eq:minus_sector_thing}
  \mathbf{J}^{-1}
  \bigl(\hat{\mathbf{D}}^{+}_{0}\bigr)^{\ast}\beta
  \mathbf{K}^{-}\boldsymbol{\Phi}_{0}^{-}
  =\frac{\ui}{2}
  \begin{pmatrix}
    0 \\
    \bigl(\hat{D}'^{+}_{1}\bigr)^{\ast} \\
    \bigl(\hat{D}'^{+}_{2}\bigr)^{\ast} \\
    \bigl(\hat{D}'^{+}_{3}\bigr)^{\ast}
  \end{pmatrix}
  \beta \hat{D}'^{-}_{3}\vp_{0},
\end{equation}
where we have used the fact that that $\hat{D}'^{-}_{2}$ annihilates
$\vp_{0}$.  In the $+$-sector, the zero mode is
\begin{equation}
  \bs{\Phi}_{0}^{+}=\frac{1}{\sqrt{2}}
 \begin{pmatrix}
  0 \\ 0 \\ \ui \\1
    \end{pmatrix}
    \vp_{0},
\end{equation}
and an analogous calculation gives
\begin{equation}
  \label{eq:plus_sector_thing}
  \mathbf{J}^{-1}
  \bigl(\hat{\mathbf{D}}^{-}_{0}\bigr)^{\ast}\beta
  \mathbf{K}^{+}\bs{\Phi}_{0}^{+}
  =-\frac{\ui}{2}
  \begin{pmatrix}
    0 \\
    \bigl(\hat{D}'^{-}_{1}\bigr)^{\ast} \\
    \bigl(\hat{D}'^{-}_{2}\bigr)^{\ast} \\
    \bigl(\hat{D}'^{-}_{3}\bigr)^{\ast}
  \end{pmatrix}
  \beta \hat{D}'^{+}_{2}\vp_{0}.
\end{equation}
In both cases, the vector boson component (i.e. the top
entry) is not excited by the warping.  Similarly, if the warp factor
is independent of $z^{1}$ (and by periodicity independent of $\bar{z}^{1}$ 
as well) then the the second entry, corresponding to the Wilson 
line along the matter curve, is not excited.

\subsection{\label{sec:unmagnetized_examples}Examples}

In this subsection we illustrate the massive-mode expansion by
considering specific simple examples.  A complementary analysis is
also given in Appendix~\ref{app:examples} where we consider exact
solutions for a few cases including those where there is no
weak-warping limit.

\paragraph{Constant warp factor}

Let us first consider the case of constant warping $\beta=1$.  For
simplicity of presentation, we will focus on the $-$-sector.  The
first order corrections to the wavefunctions come
from~\eqref{eq:MM_expansion_term_1}.  In the rotated basis,
\begin{equation}
  \bigl(\hat{\mathbf{D}}'^{+}_{0}\bigr)^{\ast}
  \beta\mathbf{K}'^{-}
  \bs{\Phi}'^{-}_{0}
  =\frac{\ui}{2}
  \begin{pmatrix}
    0 \\
    \bigl(\hat{D}'^{+}_{1}\bigr)^{\ast} \\
    \bigl(\hat{D}'^{+}_{2}\bigr)^{\ast} \\
    \bigl(\hat{D}'^{+}_{3}\bigr)^{\ast}
  \end{pmatrix}
  \hat{D}'^{-}_{3}\vp_{0}.
\end{equation}
Since $\vp_{0}$ is independent of $\bar{z}^{\ovl{1}}$, we have
$\bigl(\hat{D}'^{+}_{1}\bigr)^{\ast}\hat{D}'^{-}_{3}\vp_{0}=0$.
Making use of~\eqref{eq:unmagnetized_component_MMs}, we have
$\bigl(\hat{D}'^{+}_{2}\bigr)^{\ast}\hat{D}'^{-}_{3}\vp_{0}=-\hat{M}_{3}\vp_{0011}^{-}$.
Finally,~\eqref{eq:unmagnetized_gs_eigenvalues} and the QHSO
algebra~\eqref{eq:unmagnetized_QSHO_algebra} give
$\bigl(\hat{D}'^{+}_{3}\bigr)^{\ast}\hat{D}'^{-}_{3}\vp_{0}=-\hat{M}_{3}\vp_{0}$.
Thus, in terms of the
modes~\eqref{eq:unmagnetized_minus_sector_MM_spectrum}, we have
\begin{equation}
  \bigl(\hat{\mathbf{D}}'^{+}_{0}\bigr)^{\ast}
  \beta\mathbf{K}'^{-}
  \bs{\Phi}'^{-}_{0}
  =-\frac{\ui\hat{M}_{3}}{2}\bs{\Phi}'^{-}_{2;0011}
  -\frac{\ui\hat{M}_{3}}{2}\bs{\Phi}'^{-}_{3;0000}.
\end{equation}
Each of these modes has a mass
\begin{equation}
  \abs{m_{\lambda}}^{2}=\frac{2}{\pi\alpha'}2\hat{M}_{3},
\end{equation}
so in the rotated basis the first order correction to the zero mode is
\begin{equation}
  \label{eq:constant_warping_unmagnetized_first_order}
  \mathbf{X}'^{-}_{\left(1\right)}=
  -\frac{\ui}{4}\boldsymbol{\Phi}'^{-}_{2;0011}
  -\frac{\ui}{4}\boldsymbol{\Phi}'^{-}_{3;0000}
  =-\frac{\ui}{4}\bigl(
    0, 0,\vp_{0011}^{-},\vp_{0}\bigr)^{\mathrm{T}}.
\end{equation}

Now, using~\eqref{eq:unmagnetized_component_MMs}, we get
\begin{equation}
  \label{eq:unwarped_0011}
  \vp_{0011}^{-}=\ui\bigl(1-2\kappa\abs{z^{2}}^{2}\bigr)\vp_{0}.
\end{equation}
Then the $-$-sector warped zero mode in the unrotated basis up
through $\mathcal{O}\left(\epsilon\right)$ is
\begin{subequations}
\label{eq:constant_warping_unmagnetized_MM_answer}
\begin{align}
  \chi_{2}^{-}=
  \biggl\{1
  +\frac{\epsilon}{2}\bigl(1-\kappa\bigl\vert z^{2}\bigr\vert^{2}\bigr)
  \biggr\}
  \frac{\vp_{0}}{\sqrt{2}},\qquad
  \chi_{3}^{-}=
  \biggl\{1-\frac{\epsilon}{2}\kappa\bigl\vert z^{2}\bigr\vert^{2}
  \biggr\}
  \frac{\ui\vp_{0}}{\sqrt{2}}.
\end{align}
\end{subequations}

Since the warping is constant, it can be absorbed into a redefinition
of the coordinates and so there is a simple analytic solution.  The
solution to~\eqref{eq:quadratic_eom_warped} for constant warping is
\begin{equation}
  f^{-}=\frac{\mathcal{N}}{z^{2}}
  \ue^{-\kappa_{\mathrm{w}}\left\vert z^{2}\right\vert^{2}},
\end{equation}
where $\kappa_{\mathrm{w}}=\kappa\ue^{-2\alpha}$.  Then
using~\eqref{eq:f_term_solving_ansatz}, we get
\begin{equation}
  \label{eq:constant_warping_unnormed_exact}
  \chi_{2}^{-}=-\frac{\mathcal{N}}{\sqrt{2\pi}R_{2}}\kappa_{\mathrm{w}}
  \ue^{-\kappa_{\mathrm{w}}\left\vert z^{2}\right\vert^{2}},\qquad
  \chi_{3}^{-}
  =-\frac{\ui\mathcal{N}}{\sqrt{2\pi}R_{2}}\kappa
  \ue^{-\kappa_{\mathrm{w}}\left\vert z^{2}\right\vert^{2}}.
\end{equation}
In order to compare this exact answer to the answer resulting from the
massive-mode expansion, the two solutions need to normalized in the
same way.\footnote{Here we use the unwarped norm for the purpose of
  comparison.  Of course, when calculating more physical data like
  K\"ahler metrics, the warp factor will generally appear in the
  measure of the integral.}
From~\eqref{eq:constant_warping_unmagnetized_first_order}, we have
that~\eqref{eq:constant_warping_unmagnetized_MM_answer} is normalized
to unity up to terms quadratic in $\epsilon$.  For the exact
solution~\eqref{eq:constant_warping_unnormed_exact}, we find
\begin{equation}
  \left\Vert \mathbf{X}^{-}\right\Vert^{2}
  =\frac{\mathcal{N}^{2}\kappa\cV_{1}\cV_{2}}{2\pi^{2} R_{2}^{2}}
  \cosh 2\alpha
  =\frac{\mathcal{N}^{2}\kappa\cV_{1}\cV_{2}}{2\pi^{2} R_{2}^{2}}
  \bigl(1+\mathcal{O}\left(\epsilon^{2}\right)\bigr),
\end{equation}
so that we take
\begin{equation}
  \mathcal{N}=-\frac{\sqrt{2\pi}R_{1}}{\sqrt{\kappa}}
  \sqrt{\frac{2\kappa}{\pi\cV_{1}\cV_{2}}}.
\end{equation}
Then using the fact that
\begin{equation}
  \ue^{-\kappa_{\mathrm{w}}\absb{z^{2}}^{2}}
  =\ue^{-\kappa\absb{z^{2}}^{2}}
  \left(1-\frac{\epsilon}{2}\kappa\abs{z^{2}}^{2}
    +\mathcal{O}\bigl(\epsilon^{2}\bigr)\right),
\end{equation}
we find that to leading order in
$\epsilon$,~\eqref{eq:constant_warping_unnormed_exact} agrees
with~\eqref{eq:constant_warping_unmagnetized_MM_answer}.

\paragraph{Constant warping along the matter curve}

A less trivial case is when the warp factor is non-constant but does
not depend on the position along the matter curve $\Sigma$,
\begin{equation}
  \ue^{-4\alpha}=1+\epsilon\beta\bigl(z^{2},\bar{z}^{\ovl{2}}\bigr),
\end{equation}
where we have neglected the dependence of $\alpha$ on $z^{3}$ 
(see Appendix~\ref{app:corrections} for a discussion on this approximation). 
Since we are treating the space
transverse to the matter curve as non-compact, we can neglect
requirements of periodicity of the warp factor.  We will also suppose
for simplicity that the background is arranged such that $\beta$
depends on $z^{2}$ and $\bar{z}^{\ovl{2}}$ only through the modulus
$\absb{z^{2}}^{2}$.  The requirement that the warping is weak then
implies that $\beta$ must admit a Taylor expansion in
$\abs{z^{2}}^{2}$.  Let us first consider a warp factor of the form
\begin{equation}
  \label{eq:quadratic_warp_factor}
  \beta=L^{-2}\bigl\vert z^{2}\bigr\vert^{2}.
\end{equation}
and later generalize our computation to a general polynomial on  $\abs{z^{2}}^2$.

Considering again the $-$-sector, for the
$\mathcal{O}\left(\epsilon\right)$ corrections we have
\begin{equation}
  \bigl(\hat{\mathbf{D}}'^{+}_{0}\bigr)^{\ast}
  \beta
  \mathbf{K}'^{-}
  \boldsymbol{\Phi}'^{-}_{0}
  =\frac{\ui}{2L^{2}}
  \begin{pmatrix}
    0 \\
    \bigl(\hat{D}'^{+}_{1}\bigr)^{\ast} \\
    \bigl(\hat{D}'^{+}_{2}\bigr)^{\ast} \\
    \bigl(\hat{D}'^{+}_{3}\bigr)^{\ast}
  \end{pmatrix}
  \bigl\vert z^{2}\bigr\vert^{2}
  \hat{D}'^{-}_{3}\vp_{0}.
\end{equation}
To calculate the coefficients of the massive-mode expansion, we could
use~\eqref{eq:MM_expansion_term_1} and explicitly calculate the
overlap integral, using the fact that the massive modes are
related to the standard Hermite functions as discussed in
Appendix~\ref{app:hermite}.  Alternatively, we can express the warp
factor in terms of the raising and lowering operators acting on the
$-$-sector.  From~\eqref{eq:rotated_covariant_derivatives}, we have
\begin{equation}
  z^{2}=-\frac{\sqrt{2}\sqrt{2\pi}R_{2}}{2\kappa}
  \biggl\{\bigl(\hat{D}'^{+}_{2}\bigr)^{\ast}
  -\ui\bigl(\hat{D}'^{+}_{3}\bigr)^{\ast}\biggr\},\qquad
  \bar{z}^{\ovl{2}}=\frac{\sqrt{2}\sqrt{2\pi}R_{2}}{2\kappa}
  \biggl\{\hat{D}'^{-}_{2}
  +\ui \hat{D}'^{-}_{3}\biggr\}.
\end{equation}
As a consistency check, one can easily confirm that these operators
commute.  In terms of these, the warp
factor~\eqref{eq:quadratic_warp_factor} can be written as
\begin{equation}
  \beta=L^{-2}z^{2}\bar{z}^{\ovl{2}}
  =-\frac{2\left(2\pi R_{2}^{2}\right)}{4\kappa^{2}L^{2}}
  \biggl\{\triangle'^{-}_{2}+\triangle'^{-}_{3}
  +\ui\bigl(\bigl(\hat{D}'^{+}_{2}\bigr)^{\ast}\hat{D}'^{-}_{3}
  -\bigl(\hat{D}'^{+}_{3}\bigr)^{\ast}\hat{D}'^{-}_{2}\bigr)\biggr\}.
\end{equation}
Then using the fact that in the unmagnetized case $\kappa=2\pi
R_{2}^{2}\hat{M}_{3}$,
\begin{equation}
  \beta\vp_{mnlp}^{-}=
  \frac{\ui}{2\kappa L^{2}}
  \biggl(\sqrt{\left(l+1\right)\left(p+1\right)}
  \vp_{00,l+1,p+1}^{-}
  -\ui\left(l+p+1\right)\vp_{00lp}^{-}
  -\sqrt{lp}\vp_{00,l-1,p-1}^{-}\biggr).
\end{equation}
Following a procedure similar to the constant warping case, we find
that the first order correction to the warped zero mode is
\begin{equation}
  \label{eq:nonchiral_quadratic_solution}
  \mathbf{X}'^{-}_{\left(1\right)}
  =\frac{1}{8\kappa L^{2}}
  \biggl(\bs{\Phi}'^{-}_{2;0022}
  +\bs{\Phi}'^{-}_{3;0011}-2\ui\bs{\Phi}'^{-}_{2;0011}
  -2\ui\bs{\Phi}'^{-}_{3;0000}\biggr).
\end{equation}
Using~\eqref{eq:unwarped_0011} and
\begin{equation}
  \vp_{0022}^{-}=-\bigl(1-4\kappa\abs{z^{2}}^{2}
  +2\kappa^{2}\abs{z^{2}}^{4}\bigr)\vp_{0},
\end{equation}
we get in the unrotated basis
\begin{subequations}
\begin{align}
  \chi_{2}^{-}=&
  \biggl\{1+\frac{\epsilon}{4\kappa L^{2}}
  \bigl(1+\kappa\abs{z^{2}}^{2}-\kappa^{2}\abs{z^{2}}^{2}\bigr)\biggr\}
  \vp_{0},\\
  \chi_{3}^{-}=&\biggl\{1-\frac{\epsilon}{4\kappa L^{2}}
  \kappa\abs{z^{2}}^{2}\bigl(1+\kappa\abs{z^{2}}^{2}\bigr)\biggr\}
  \frac{\ui\vp_{0}}{\sqrt{2}}.
\end{align}
\end{subequations}

Remarkably, this case also possesses an exact solution given in terms of an Airy
function.  Taking~\eqref{eq:f_term_solving_ansatz}, the solution to
the $D$-term equation~\eqref{eq:quadratic_eom_warped} is
\begin{equation}
  \label{eq:quadratic_exact}
  f^{-}=\frac{\mathcal{N}}{z^{2}}
  \mathrm{Ai}\bigl(a+b\abs{z^{2}}^{2}\bigr),\quad
  a=\frac{\kappa^{2/3} L^{1/3}}{\epsilon^{2/3}},\quad
  b=\frac{\kappa^{2/3}\epsilon^{1/3}}{L^{2/3}}.
\end{equation}
After normalizing and performing an expansion in $\epsilon$, the exact
solution agrees with the result from the perturbative analysis.

Let us now generalize the last two examples by considering
\begin{equation}
  \beta=L^{-2r}\abs{z^{2}}^{2r},
\end{equation}
where $r$ is a positive integer.\footnote{The case of negative powers
  is not easily addressable in this formalism.  Since $z^{2}$ and
  $\bar{z}^{\bar{2}}$ are expressed in terms of creation and
  annihilation operators, one would like to do the same for the
  inverse operators $1/z^{2}$ and $1/\bar{z}^{\bar{2}}$.  However,
  since the annihilation operator is not invertible, this cannot be
  done in a straightforward way.  Some exact solutions with inverse
  powers are given in Appendix~\ref{app:examples}.} One can show that
when at least one of $l$ or $p$ is zero, then
\begin{equation}
  \label{eq:general_power_warp_factor}
  \beta\vp_{nmlp}^{-}=\biggl(\frac{1}{2\kappa L^{2}}\biggr)^{r}
  \sum_{s=0}^{r}\ui^{s}\binom{r}{s}
  \sqrt{\frac{\left(l+s\right)!}{l!}
    \frac{\left(p+s\right)!}{p!}}
  \frac{\left(l+p+r\right)!}{\left(l+p+s\right)!}
  \vp_{nm,l+s,p+s}^{-},
\end{equation}
with an analogous expression holding for the $+$-sector.  The general
form is more involved when both $l$ and $p$ are non-vanishing.
However, for the first-order corrections resulting
from~\eqref{eq:minus_sector_thing} and~\eqref{eq:plus_sector_thing},
at least one of these oscillators is unexcited and the other is
excited to only the first level and the expression becomes relatively
simple
\begin{equation}
  \label{eq:general_power_thing}
  \hat{\mathbf{D}}'^{+}_{0}\beta\mathbf{K}'^{-}\bs{\Phi}'_{0}
  =-\frac{\ui}{2}\frac{\left(r+1\right)!\hat{M}_{3}}
  {\left(2\kappa L^{2}\right)^{r}}
  \sum_{s=0}^{r}\ui^{s}
  \binom{r}{s}
  \biggl\{\bs{\Phi}'^{-}_{2;00,s+1,s+1}+\bs{\Phi}'^{-}_{3;00ss}\biggr\}.
\end{equation}
Making use of~\eqref{eq:MM_expansion_term_1} gives the first order
correction to the wavefunction
\begin{equation}
  \label{eq:unmagnetized_monomial_result}
  \mathbf{X}'^{-}_{\left(1\right)}=\frac{\ui}{4}\frac{\left(r+1\right)!}
  {\left(2\kappa L^{2}\right)^{r}}
  \sum_{s=0}^{r}\ui^{s}\binom{r}{s}
  \frac{1}{s+1}
  \biggl\{\bs{\Phi}'^{-}_{2;00,s+1,s+1}+\bs{\Phi}'^{-}_{3;00ss}\biggr\},
\end{equation}
and so the higher the power $r$, the more massive KK modes are excited.

Finally, let us consider a general expansion of the form
\begin{equation}
  \label{eq:polynomial_wf}
  \beta=\sum_{r}\beta_{r}L^{-2r}\abs{z^{2}}^{2r},
\end{equation}
for which application of~\eqref{eq:unmagnetized_monomial_result} gives
\begin{equation}
  \label{eq:polynomial_MM_solution}
  \mathbf{X}'^{-}_{\left(1\right)}
  =\frac{\ui}{4}\sum_{s}C_{s}
  \biggl\{\bs{\Phi}'^{-}_{2;00,s+1,s+1}+\bs{\Phi}'^{-}_{3;00ss}\biggr\},
  \quad
  C_{s}=\frac{\ui^{s}}{s+1}\sum_{r=s}^{\infty}
  \binom{r}{s}
  \frac{\left(r+1\right)!\beta_{r}}{\left(2\kappa L^{2}\right)^{r}}.
\end{equation}
For example, for an exponential warp factor
\begin{equation}
  \beta=\ue^{-\absb{z^{2}}^{2}L^{-2}},\quad
  \beta_{r}=\frac{\left(-1\right)^{r}}{r!}.
\end{equation}
Then,
\begin{equation}
  C_{s}=\frac{\ui^{s}}{\left(s+1\right)!}
  \sum_{r=s}^{\infty}
  \frac{\left(-1\right)^{r}}{\left(2\kappa L^{2}\right)^{r}}
  \frac{\left(r+1\right)!}{\left(r-s\right)!}
  =\frac{\left(-\ui\right)^{s}\left(2\kappa
      L^{2}\right)^{2}} {\left(1+2\kappa L^{2}\right)^{2+s}}.
\end{equation}
A related example is
\begin{equation}
  \beta=\cos\bigl(L^{-2}\abs{z^{2}}^{2}\bigr).
\end{equation}
Then,
\begin{align}
  \label{eq:cosine_example_coefficients}
  C_{s}=&\frac{\ui^{s}}{\left(s+1\right)!}
  \sum_{k=\left\ulcorner s/2\right\urcorner}
  \frac{\left(-1\right)^{k}}{\left(2\kappa L^{2}\right)^{2k}}
  \frac{\left(2k+1\right)!}{\left(2k-s\right)!}\notag \\
  =&\frac{\ui^{s}\left(-1\right)^{\left\ulcorner s/2\right\urcorner}
    \left(2\kappa L^{2}\right)^{2}} {\left(1+\left(2\kappa
        L^{2}\right)^{2}\right)^{1+s/2}}
  \begin{cases}
    \cos\bigl[\bigl(2+s\bigr) \mathrm{arccot}\bigl(2\kappa
    L^{2}\bigr)\bigr] &
    \text{$s$ even} \\
    \sin\bigl[\bigl(2+s\bigr) \mathrm{arccot}\bigl(2\kappa
    L^{2}\bigr)\bigr] & \text{$s$ odd}
  \end{cases},
\end{align}
where $\left\ulcorner\,\right\urcorner$ denotes the ceiling function.
Note that the solution~\eqref{eq:constant_warping_unmagnetized_first_order} 
for the constant warp factor example suggests that the $s=0$ modes can be
eliminated by a rescaling of the coordinates.  Indeed
from~\eqref{eq:cosine_example_coefficients}, it is easy to see that by
redefining $R_{2}$ such that $2\kappa L^{2}=1$, $C_{s=0}$ can be made
to vanish.

\paragraph{Variable warping along the matter curve}

Finally, we consider the case where the warp factor varies
non-trivially along the matter curve, which we again consider to be 
a two-torus $\Sigma = \mathbb{T}^2$.  As is the case for the
wavefunctions, the warp factor must be well-defined over the compact
space and thus must admit an expansion in the Fourier
modes~\eqref{eq:fouriermodes}
\begin{equation}
  \label{eq:fourier_transformed_warp_factor}
  \beta=\sum_{mn}\tilde{\beta}_{mn}\bigl(z^{2},\bar{z}^{\bar{2}}\bigr)
  h_{mn}.
\end{equation}
For simplicity of presentation, we consider the case where the
$\tilde{\beta}_{mn}$ are constants; allowing the $\tilde{\beta}_{mn}$
to depend on $z^{2}$ and $\bar{z}^{\ovl{2}}$ would not introduce any
significant complication beyond that encountered in the last example.
Again, since we are interested in the warp factor only on the
worldvolume, the $z^{3}$ and $\bar{z}^{\bar{3}}$ dependence can be
suppressed.

For the first order corrections to the warped zero mode, we need to
calculate
\begin{equation}
  \frac{\ui}{2}
  \sum_{mn}\tilde{\beta}_{mn}
  \begin{pmatrix}
    0 \\
    \bigl(\hat{D}'^{+}_{1}\bigr)^{\ast} \\
    \bigl(\hat{D}'^{+}_{2}\bigr)^{\ast} \\
    \bigl(\hat{D}'^{+}_{3}\bigr)^{\ast}
  \end{pmatrix}
  h_{mn}
  \hat{D}'^{-}_{3}\vp_{0}.
\end{equation}
Carrying over from the constant warping case, we have
\begin{equation}
  \bigl(\hat{D}'^{+}_{2}\bigr)^{\ast}h_{mn}
  \hat{D}'^{-}_{3}\vp_{0}
  =-\hat{M}_{3}\vp_{mn11}^{-},\qquad
  \bigl(\hat{D}'^{+}_{3}\bigr)^{\ast}h_{mn}
  \hat{D}'^{-}_{3}\vp_{0}
  =-\hat{M}_{3}\vp_{mn00}^{-}.
\end{equation}
Defining
\begin{equation}
  \label{eq:define_t_mn}
  t_{mn}=-\frac{\pi\left(m-\tau_{1}n\right)}
  {\sqrt{2\pi}R_{1}\im \tau_{1}}
\end{equation}
we also have
\begin{equation}
  \bigl(\hat{D}'^{+}_{1}\bigr)^{\ast}
  h_{mn}\hat{D}'^{-}_{3}\vp_{0}
  =-\ui\, t_{mn}\hat{M}_{3}^{1/2}
  \vp_{mn01}^{-}.
\end{equation}
The first order correction is then
\begin{equation}
  \mathbf{X}'^{-}_{\left(1\right)}
  =-\frac{\ui}{2}
  \sum_{mn}
  \frac{\tilde{\beta}_{mn}\hat{M}_{3}^{1/2}}
  {\hat{m}_{mn}^{2}+2\hat{M}_{3}}\biggl\{\hat{M}_{3}^{1/2}
  \bs{\Phi}_{2;mn11}^{-}
  +\hat{M}_{3}^{1/2}
  \bs{\Phi}_{3;mn00}^{-}
  +\ui t_{mn}\bs{\Phi}_{1;mn01}^{-}\biggr\}.
\end{equation}
Again, the higher the Fourier mode $\beta_{mn}$, the
higher open string massive modes that is involved in the warped zero mode.
Note also that since the warp factor is no longer independent of $z^{1}$
and $\bar{z}^{\ovl{1}}$, the warped zero mode now includes the Wilson
line along the matter curve.

\section{\label{sec:chiral}Magnetized intersections}

We now turn our attention towards the case of non-trivial magnetic
flux.  Unlike the unmagnetized case where the $-$-sector zero mode was
accompanied by a $+$-sector zero mode, the presence of a magnetic flux 
$\langle F_2 \rangle$ selects one of the two sectors, inducing 4d chirality.
 This non-trivial flux, however, also causes the
equations of motion to become more involved.

Analogously to the previous section, we consider fluctuations about the self-dual, 
quantized flux background~\eqref{eq:magnetic_flux} and the non-trivial
intersection~\eqref{eq:angle}.  To produce the magnetic
flux~\eqref{eq:magnetic_flux}, we take the background connection to be
\begin{equation}
  \label{eq:connection}
  \bigl\langle A\bigr\rangle=\cA=
  \sum_{m=1}^{2}\frac{\pi}{2\ui \im\tau_{m}}
  \begin{pmatrix}
    M_{m}^{\left(a\right)}\mathbb{I}_{N_{a}} & \\
    & M_{m}^{\left(b\right)}\mathbb{I}_{N_{b}}
  \end{pmatrix}
  \bigl(\bar{z}^{\bar{m}}\ud z^{m}-z^{m}\ud\bar{z}^{\bar{m}}\bigr).
\end{equation}

As in the unmagnetized case, the equations of motion for the zero
modes fluctuations~\eqref{eq:param_flucs} can be inferred from the
$F$-flatness and $D$-flatness conditions discussed in
Sec~\ref{sec:setup} or by considering the fermionic
action~\eqref{eq:wynants}.  From the latter, we again
find~\eqref{eq:fermion_eom} but now the covariant derivatives include
contributions from the magnetic flux,
\begin{equation}
  \label{eq:magnetized_covariant_derivs}
  \hat{D}^{\mp}_{m=1,2}=
  \hat{\partial}_{1,2}\mp\frac{\sqrt{2\pi}I_{m}^{\left(ab\right)}}
  {4R_{m}\im\tau_{m}}\bar{z}^{\ovl{m}},\quad
  \hat{D}^{\mp}_{3}=\mp\frac{\ui}{2}\sqrt{2\pi}R_{3}I_{3}^{\left(ab\right)},
\end{equation}
where
\begin{equation}
  I_{m}^{\left(ab\right)}=M_{m}^{\left(a\right)}-M_{m}^{\left(b\right)}.
\end{equation}
Similarly, taking the ansatz~\eqref{eq:MMS1ansatz} again
gives~\eqref{eq:bosonic_zero_mode_eqs} with the same modification of
the covariant derivatives.  As in the unmagnetized case, the warped
zero mode equation cannot in general be solved in a simple analytic
way and we will consider an expansion in terms of the unwarped
massive modes.

\subsection{Unwarped chiral spectrum}

The unwarped chiral spectrum again follows
from~\eqref{eq:MM_eom1}.  A crucial difference between the
unmagnetized and the magnetized case is the richer algebra satisfied
by the covariant derivatives in the latter.  We define
\begin{equation}
  \bigl[\bigl(\hat{D}^{\pm}_{n}\bigr)^{\ast},\hat{D}_{m}^{\mp}\bigr]
  =\mp 2\ui\hat{\cF}_{\bar{n}m},\quad
  \bigl[\hat{D}^{\mp}_{n},\hat{D}^{\mp}_{m}\bigr]=\mp 2\ui\hat{\cF}_{nm},\quad
  \bigl[\bigl(\hat{D}^{\pm}_{n}\bigr)^{\ast},
  \bigl(\hat{D}^{\pm}_{m}\bigr)^{\ast}\bigr]
  =\mp 2\ui\hat{\cF}_{\bar{n}\bar{m}}.
\end{equation}
Then the non-vanishing components of $\hat{\cF}$ are
\begin{equation}
  \label{eq:D9_flux}
  \hat{\cF}_{1\bar{1}}=-\hat{\cF}_{2\bar{2}}=
  \frac{\ui I^{\left(ab\right)}_{1}}{4R_{1}^{2}\im\tau_{1}},\qquad
  \hat{\cF}_{2\bar{3}}=\frac{R_{3}I^{\left(ab\right)}_{3}}{4R_{2}},
\end{equation}
where we have made use of the self-duality condition on $F_{2}$.
As a result of this richer algebra, when
writing~\eqref{eq:MM_eom2}, we again
have~\eqref{eq:unmagnetized_Laplacian} but now with
\begin{equation}
  \triangle^{\mp}=\frac{1}{2}\sum_{m=1}^{3}
  \biggl\{\bigl(\hat{D}^{\pm}_{m}\bigr)^{\ast},
  \hat{D}^{\mp}_{m}\biggr\},\quad
  \mathbf{B}=
  -2\ui\begin{pmatrix}
    \sigma_{+++} & \hat{\cF}_{\bar{3}{\bar{2}}} &
    \hat{\cF}_{\bar{1}\bar{3}} & \hat{\cF}_{\bar{2}\bar{1}} \\
    \hat{\cF}_{23} & \sigma_{+--} &
    \hat{\cF}_{2\bar{1}} & \hat{\cF}_{3\bar{1}} \\
    \hat{\cF}_{31} & \hat{\cF}_{1\bar{2}} & \sigma_{-+-} & \hat{\cF}_{3\bar{2}} \\
    \hat{\cF}_{12} & \hat{\cF}_{1\bar{3}} & \hat{\cF}_{2\bar{3}} & \sigma_{--+}
  \end{pmatrix},
\end{equation}
where
\begin{equation}
  \sigma_{\epsilon_{1}\epsilon_{2}\epsilon_{3}}
  =\frac{1}{2}\bigl(
  \epsilon_{1}\hat{\cF}_{1\bar{1}}+
  \epsilon_{2}\hat{\cF}_{2\bar{2}}+
  \epsilon_{3}\hat{\cF}_{2\bar{3}}\bigr).
\end{equation}
In our case,
\begin{equation}
  \mathbf{B}=
  \begin{pmatrix}
    0 & 0 & 0 & 0 \\
    0 & \hat{M}_{1} & 0 & 0 \\
    0 & 0 & -\hat{M}_{1} & \ui\hat{M}_{3} \\
    0 & 0 & -\ui\hat{M}_{3} & 0
  \end{pmatrix},
\quad
  \hat{M}_{1}=\frac{I_{1}^{\left(ab\right)}}{2R_{1}^{2}\im\tau_{1}},\quad
  \hat{M}_{3}=\frac{R_{3}I^{\left(ab\right)}_{3}}{2R_{2}}.
\end{equation}
Just as in the unmagnetized case, to find the massive-mode expansion,
we will look for simultaneous eigenvectors of $\mathbf{B}$ and
$\triangle^{\mp}$.  A non-vanishing magnetic flux increases the rank
of $\mathbf{B}$ so that the nullspace is now only dimension 1, spanned
by $\left(1, 0, 0, 0\right)^{\mathrm{T}}$ while the non-trivial
spectrum includes an eigenvalue $\hat{M}_{1}$ with eigenvector
$\left(0, 1, 0, 0\right)^{\mathrm{T}}$.  The magnetic flux also breaks
the degeneracy of the remaining spectrum.  Defining
\begin{equation}
  \rho^{\pm}=\frac{\hat{M}_{1}}{2}\pm
  \sqrt{\left(\frac{\hat{M}_{1}}{2}\right)^{2}+\hat{M}_{3}^{2}},
\end{equation}
the other non-trivial eigenvalues are $-\rho^{+}$ and $-\rho^{-}$ with
respective eigenvectors
\begin{equation}
  \bigl(0, 0, c, -\ui s\bigr)^{\mathrm{T}},\quad
  \bigl(0,0,-\ui s,c\bigr)^{\mathrm{T}},
\end{equation}
where we have introduced the angle $\delta$ defined by the relations
\begin{equation}
  c:=\cos\delta=
  \frac{\hat{M}_{3}}{\sqrt{\left(\rho^{-}\right)^{2}+\hat{M}_{3}^{2}}},
  \qquad
  s:=\sin\delta=
  \frac{\rho^{-}}{\sqrt{\left(\rho^{-}\right)^{2}+\hat{M}_{3}^{2}}}.
\end{equation}
In the unmagnetized case, $\delta=-\pi/4$.  We note here the useful
relation $\rho^{+}\rho^{-}=-\hat{M}_{3}^{2}$.  With the magnetic flux,
$\mathbf{B}$ is now diagonalized by
\begin{equation}
  \mathbf{J}=\begin{pmatrix}
    \phantom{a}1\phantom{a} & 0 & 0 & 0 \\
    0 & 1 & 0 & 0 \\
    0 & 0 & c & -\ui s \\
    0 & 0 & -\ui s & c
  \end{pmatrix},
\end{equation}
giving
\begin{equation}
  \label{eq:rotated_vector_Laplacian}
  \mathbf{J}^{-1}
  \bigl(\hat{\mathbf{D}}^{\pm}_{0}\bigr)^{\ast}
  \hat{\mathbf{D}}^{\mp}_{0}\mathbf{J}
  =-\triangle^{\mp}
  \pm\mathrm{diag}\bigl(0, \hat{M}_{1},
  -\rho^{+},-\rho^{-}\bigr).
\end{equation}

The zero modes in the rotated basis are again in the kernel
of~\eqref{eq:rotated_vec_diff_op} except due to the magnetic flux, the
rotated covariant derivatives are now
\begin{align}
  \label{eq:magnetized_rotated_covariant_derivatives}
  \hat{D}'^{\mp}_{1}=&\hat{D}^{\mp}_{1},\\
  \hat{D}'^{\mp}_{2}=&c\hat{D}^{\mp}_{2}-\ui s\hat{D}^{\mp}_{3}
  =\frac{c}{\sqrt{2\pi}R_{2}}\bigl(\partial_{2}\pm\kappa\bar{z}^{\bar{2}}\bigr),\\
  \hat{D}'^{\mp}_{3}=&c\hat{D}^{\mp}_{2}-\ui s\hat{D}^{\mp}_{2}
  =\frac{-\ui s}{\sqrt{2\pi}R_{2}}
  \bigl(\partial_{2}\mp\kappa\bar{z}^{\bar{2}}\bigr),
\end{align}
where the width defined in~\eqref{eq:unwarped_sol_g} is modified by
the magnetic flux
\begin{equation}
  \label{eq:magnetized_unwarped_width}
  \kappa=2\pi R_{2}^{2}\sqrt{\left(\frac{\hat{M}_{1}}{2}\right)^{2}
    +\hat{M}_{3}^{2}}.
\end{equation}
We now look for a zero mode among the eigenvectors of $\mathbf{B}$.
As in the unmagnetized case, a mode with 0 or $\hat{M}_{1}$
$\mathbf{B}$-eigenvalue has no non-trivial zero modes.  On the other
hand, a $-\rho^{+}$ eigenvector  $\left(0,0,\vp^{\mp},0\right)^{\mathrm{T}}$
will be a zero mode if
\begin{equation}
  \hat{D}'^{\mp}_{2}\vp^{\mp}=\bigl(\hat{D}'^{\pm}_{3}\bigr)^{\ast}\vp^{\mp}
  =\bigl(\hat{D}'^{\pm}_{1}\bigr)^{\ast}\vp^{\mp}=0.
\end{equation}
This is satisfied for
\begin{equation}
  \vp^{\mp}=\ue^{\mp \pi R_{1}^{2}\hat{M}_{1}\absb{z^{1}}^{2}}
  \zeta\bigl(z^{1}\bigr)
  \ue^{\mp \kappa\absb{z^{2}}^{2}},
\end{equation}
where $\zeta$ is a holomorphic function of $z^{1}$ that will be
determined momentarily.  As in the unmagnetized case, demanding that
the wavefunction is normalizable locally selects the solution that vanishes as
$\abs{z^{2}}\to\infty$, and so only the solution for the $-$-sector survives.

In the unmagnetized case, the wavefunctions are required to be
periodic along $\Sigma = \mathbb{T}^2$.  However, the non-trivial
magnetic flux results from a potential that is not periodic along
$\Sigma$, but is instead periodic only up to a gauge
transformation
\begin{subequations}
\begin{align}
  \cA\bigl(z^{1}+1\bigr)=&
  \cA\bigl(z^{1}\bigr)+\frac{\pi}{\im\tau_{1}}
  \begin{pmatrix} M_{1}^{\left(a\right)} & \\
    & M_{1}^{\left(b\right)}
  \end{pmatrix}\im\ud z^{1},\\
  \cA\bigl(z^{1}+\tau_{1}\bigr)=&
  \cA\bigl(z^{1}\bigr)+\frac{\pi}{\im\tau_{1}}
  \begin{pmatrix} M_{1}^{\left(a\right)} & \\
    & M_{1}^{\left(b\right)}
  \end{pmatrix}\im\bigl(\bar{\tau}_{1}\ud z^{1}\bigr).
\end{align}
\end{subequations}
For a field $\omega^{\mp}$ in the $\mp$-sector, this implies the
quasi-periodicity conditions
\begin{subequations}
\label{eq:quasi-periodicity}
\begin{align}
  \omega^{\mp}\bigl(z^{1}+1\bigr)=&
  \ue^{\pm\frac{\ui\pi}{\mathrm{Im}\,\tau_{1}}
    I^{\left(ab\right)}_{1}\mathrm{Im}\,z^{1}}
  \omega^{\mp}\bigl(z^{1}\bigr),\\
  \omega^{\mp}\bigl(z^{1}+\tau_{1}\bigr)=&
  \ue^{\pm\frac{\ui\pi}{\mathrm{Im}\,\tau_{1}}
    I^{\left(ab\right)}_{1}\mathrm{Im}\left(\bar{\tau}_{1} z^{1}\right)}
  \omega^{\mp}\bigl(z^{1}\bigr).
\end{align}
\end{subequations}
With this condition, there are
$\absb{I^{\left(ab\right)}_{1}}$ independent solutions given
in terms of $\vartheta$-functions with
characteristics~\cite{Cremades:2004wa}
\begin{equation}
  \label{eq:minus_sector_zero_mode}
  \vp^{j,-}_{0}=\mathcal{N}^{j,-}
  \ue^{-\kappa\left\vert z^{2}\right\vert}
  \ue^{\pi\ui I^{\left(ab\right)}_{1}z^{1}\mathrm{Im}\, z^{1}/
    \mathrm{Im}\, \tau_{1}}
  \vartheta\begin{bmatrix}
    j/I^{\left(ab\right)}_{1} \\
    0
  \end{bmatrix}
  \bigl(I^{\left(ab\right)}_{1}z^{1},
  I^{\left(ab\right)}_{1}\tau_{1}\bigr),
\end{equation}
where $j=0,\ldots,\absb{I^{\left(ab\right)}_{1}}-1$ and
\begin{equation}
  \label{eq:theta_function_defined}
  \vartheta\begin{bmatrix}a \\ b\end{bmatrix}
  \bigl(\nu,\tau\bigr)=
  \sum_{l\in\mathbb{Z}}
  \ue^{\pi\ui\left(a+l\right)^{2}\tau}
  \ue^{2\pi\ui\left(a+l\right)\left(\nu+b\right)}.
\end{equation}
From~\eqref{eq:theta_function_defined} and using the fact that
$\tau_{1}$ has a positive definite imaginary part and that we have
already established that this is only valid for the $-$-sector, we see
that this converges only if $I^{\left(ab\right)}_{1}>0$.  That
is, when $I^{\left(ab\right)}_{1}>0$, we have the (rotated) zero modes
\begin{equation}
\label{eq:magnetized_unwarped_minus_zero_mode}
  \mathbf{\Phi}'^{j,-}_{0}=\bigl(
    0 , 0 , \vp^{j,-}_{0}, 0\bigr)^{\mathrm{T}},
\end{equation}
the functions $\vp^{j,-}_{0}$ satisfying
\begin{equation}
  \bigl(\hat{D}'^{+}_{1}\bigr)^{\ast}\vp^{j,-}_{0}=
  \hat{D}'^{-}_{2}\vp^{j,-}_{0}=
  \bigl(\hat{D}'^{+}_{3}\bigr)^{\ast}\vp^{j,-}_{0}=0.
\end{equation}
      
Similar arguments show that if $I^{\left(ab\right)}_{1}<0$,
there are zero modes in the $+$-sector of the form
\begin{equation}
\label{eq:magnetized_unwarped_plus_zero_mode}
  \boldsymbol{\Phi}'^{+,j}_{0}=\bigl(
    0, 0, 0,\vp_{0}^{j,+}\bigr)^{\mathrm{T}},
\end{equation}
where
\begin{equation}
  \vp^{j,+}_{0}=\mathcal{N}^{j,+}
  \ue^{-\kappa\left\vert z^{2}\right\vert}
  \ue^{-\pi\ui I^{\left(ab\right)}_{1}z^{1}\mathrm{Im}\, z^{1}/
    \mathrm{Im}\, \tau_{1}}
  \vartheta\begin{bmatrix}
    -j/I^{\left(ab\right)}_{1} \\
    0
  \end{bmatrix}
  \bigl(- I^{\left(ab\right)}_{1}z^{1},
  -I^{\left(ab\right)}_{1}\tau_{1}\bigr),
\end{equation}
and that they satisfy
\begin{equation}
  \bigl(\hat{D}'^{-}_{1}\bigr)^{\ast}\vp^{j,+}_{0}=
  \bigl(\hat{D}'^{+}_{2}\bigr)^{\ast}\vp^{j,+}_{0}=
  \hat{D}'^{-}_{3}\vp^{j,+}_{0}=0.
\end{equation}
Note that if $I^{\left(ab\right)}_{1}\neq 0$, then the zero modes consist
of only $+$-sector modes or $-$-sector modes and so the spectrum is
chiral.  Finally, the normalization constants are~\cite{Cremades:2004wa}
\begin{equation}
  \label{eq:normalization}
  \mathcal{N}^{j,\mp}=
  \left(\frac{2\kappa\sqrt{\pm 2I^{\left(ab\right)}_{1}
      \mathrm{Im}\,\tau_{1}}}{\pi\cV_{1}\cV_{2}}\right)^{1/2}.
\end{equation}

To find the massive modes of this configuration, we again map the problem to a QSHO problem.
Again, due to the non-vanishing flux, we have a richer algebra
\begin{equation}
  \bigl[\bigl(\hat{D}'^{\pm}_{1}\bigr)^{\ast},
  \hat{D}^{\mp}_{1}\bigr]=\mp\hat{M}_{1},\quad
  \bigl[\bigl(\hat{D}'^{\pm}_{2}\bigr)^{\ast},
  \hat{D}^{\mp}_{2}\bigr]=\pm\rho^{+},\quad
  \bigl[\bigl(\hat{D}'^{\pm}_{3}\bigr)^{\ast},
  \hat{D}^{\mp}_{3}\bigr]=\pm\rho^{-}.
\end{equation}
Writing $\triangle^{\mp}$ as in~\eqref{eq:rotated_Laplacians}, we also have
\begin{subequations}
\begin{align}
  \bigl[\triangle'^{\mp}_{1},\hat{D}'^{\mp}_{1}\bigr]
  =&\mp\hat{M}_{1}\hat{D}'^{\mp}_{1},
  &
  \bigl[\triangle'^{\mp}_{1},\bigl(\hat{D}'^{\pm}_{1}\bigr)^{\ast}\bigr]
  =&\pm\hat{M}_{1}\bigl(\hat{D}'^{\pm}_{1}\bigr)^{\ast}, \\
  \bigl[\triangle'^{\mp}_{2},\hat{D}'^{\mp}_{2}\bigr]
  =&\pm\rho^{+}\hat{D}'^{\mp}_{2},
  &
  \bigl[\triangle'^{\mp}_{2},\bigl(\hat{D}'^{\pm}_{2}\bigr)^{\ast}\bigr]
  =&\mp\rho^{+}\bigl(\hat{D}'^{\pm}_{2}\bigr)^{\ast}, \\
  \bigl[\triangle'^{\mp}_{3},\hat{D}'^{\mp}_{3}\bigr]
  =&\pm\rho^{-}\hat{D}'^{\mp}_{3},
  &
  \bigl[\triangle'^{\mp}_{3},\bigl(\hat{D}'^{\pm}_{3}\bigr)^{\ast}\bigr]
  =&\mp\rho^{-}\bigl(\hat{D}'^{\pm}_{3}\bigr)^{\ast}.
\end{align}
\end{subequations}
The zero modes then satisfy
\begin{equation}
  \triangle'^{\mp}_{1}\vp_{0}^{j,\mp}=\mp\frac{1}{2}\hat{M}_{1}\vp_{0}^{j,\mp},
  \qquad
  \triangle'^{\mp}_{2}\vp_{0}^{j,\mp}=-\frac{1}{2}\rho^{+}\vp_{0}^{j,\mp},
  \qquad
  \triangle'^{\mp}_{3}\vp_{0}^{j,\mp}=-\frac{1}{2}\rho^{-}\vp_{0}^{j,\mp},
\end{equation}
giving
\begin{equation}
  \triangle^{\mp}\vp_{0}^{j,\mp}=\mp \rho^{\pm}\vp_{0}^{j,\mp}.
\end{equation}

In the unmagnetized case, the problem of finding the massive modes was
reduced to the problem of a 2D quantum simple harmonic oscillator; the
massive excitations along the matter curve did not have this
algebra available.  However, when the index is non-vanishing, all of
the higher modes can be found using algebraic techniques.  Indeed, focusing in
the $-$-sector we now have three lowering operators
\begin{equation}
  \ui\bigl(\hat{D}'^{+}_{1}\bigr)^{\ast},\qquad
  \ui \hat{D}'^{-}_{2},\qquad
  \ui\bigl(\hat{D}'^{+}_{3}\bigr)^{\ast},
\end{equation}
and three raising operators
\begin{equation}
  \ui \hat{D}'^{-}_{1},\qquad
  \ui\bigl(\hat{D}'^{+}_{2}\bigr)^{\ast},\qquad
  \ui \hat{D}'^{-}_{3}.
\end{equation}
Using these, we build the normalized eigenstates of $\triangle^{-}$.
\begin{equation}
  \label{eq:magnetized_minus_sector_component_MMs}
  \vp^{j,-}_{nlp}
  =\sqrt{\frac{1}{n!l!p!\hat{M}_{1}^{n}\left(\rho^{+}\right)^{l}
      \left(-\rho^{-}\right)^p}}
  \bigl(\ui \hat{D}'^{-}_{1}\bigr)^{n}
  \bigl[\ui\bigl(\hat{D}'^{+}_{2}\bigr)^{\ast}\bigr]^{l}
  \bigl(\ui \hat{D}'^{-}_{3}\bigr)^{p}\vp_{0}^{j,-}.
\end{equation}

Similarly, in the $+$-sector we have the lowering operators
\begin{equation}
  \ui\bigl(\hat{D}'^{-}_{1}\bigr)^{\ast},\qquad
  \ui\bigl(\hat{D}'^{-}_{2}\bigr)^{\ast},\qquad
  \ui \hat{D}'^{+}_{3},
\end{equation}
and the raising operators
\begin{equation}
  \ui \hat{D}'^{+}_{1},\qquad
  \ui \hat{D}'^{+}_{2},\qquad
  \ui\bigl(\hat{D}'^{-}_{3}\bigr)^{\ast}.
\end{equation}
Then the eigenstates of $\triangle^{+}$ are,
\begin{equation}
\label{eq:magnetized_plus_sector_component_MMs}
  \vp^{j,+}_{nlp}
  =\sqrt{\frac{1}{n!l!p!\bigl(-\hat{M}_{1}\bigr)^{n}\left(\rho^{+}\right)^{l}
      \left(-\rho^{-}\right)^p}}
  \bigl(\ui \hat{D}'^{+}_{1}\bigr)^{n}
  \bigl(\ui \hat{D}'^{+}_{2}\bigr)^{l}
  \bigl[\ui\bigl(\hat{D}'^{-}_{3}\bigr)^{\ast}\bigr]^{p}
  \vp_{0}^{j,+}.
\end{equation}
The eigenvalues are
\begin{equation}
  \triangle'^{\mp}_{1}\vp_{nlp}^{j,\mp}
  =\mp\bigl(\frac{1}{2}+n\bigr)\hat{M}_{1},\quad
  \triangle'^{\mp}_{2}\vp_{nlp}^{j,\mp}
  =-\bigl(\frac{1}{2}+l\bigr)\rho^{+},\quad
  \triangle'^{\mp}_{3}\vp_{nlp}^{j,\mp}
  =\bigl(\frac{1}{2}+p\bigr)\rho^{-},
\end{equation}
so
\begin{equation}
  \triangle^{\mp}\vp_{nlp}^{j,\mp}
  =-\bigl(\pm n\hat{M}_{1}+l\rho^{+}-p\rho^{-}
  \pm \rho^{\pm}\bigr)\vp^{j,\mp}_{nlp}.
\end{equation}

The higher modes also get a mass from the magnetic fluxes in
$\mathbf{B}$.  The $-$-sector spectrum, which is valid for
$\hat{M}_{1}>0$, is
\begin{subequations}
\label{eq:magnetized_minus_sector_MM_spectrum}
\begin{align}
  \bigl\vert m^{-}_{0;nlp}\bigr\vert^{2}=&
  \frac{2}{\pi\alpha'}
  \biggl(n\hat{M}_{1}+\left(l+1\right)\rho^{+}
  -p\,\rho^{-}\biggr)
  &
  \bs{\Phi}'^{j,-}_{0;nlp}=
  \bigl(
    \vp^{j,-}_{nlp},0, 0, 0\bigr)^{\uT},\\
  \bigl\vert m^{-}_{1;nlp}\bigr\vert^{2}=&
  \frac{2}{\pi\alpha'}
  \biggl(\bigl(n+1\bigr)\hat{M}_{1}+\left(l+1\right)\rho^{+}
  -p\,\rho^{-}\biggr)
  &
  \bs{\Phi}'^{j,-}_{1;nlp}=
  \bigl(0,\vp^{j,-}_{nlp},0, 0\bigr)^{\uT},\\
  \bigl\vert m^{-}_{2;nlp}\bigr\vert^{2}=&
  \frac{2}{\pi\alpha'}
  \biggl(n\hat{M}_{1}+l\rho^{+}
  -p\,\rho^{-}\biggr)
  &
  \bs{\Phi}'^{j,-}_{2;nlp}=
  \bigl(
    0, 0,\vp^{j,-}_{nlp},0 \bigr)^{\uT},\\
  \bigl\vert m^{-}_{3;nlp}\bigr\vert^{2}=&
  \frac{2}{\pi\alpha'}
  \biggl(n\hat{M}_{1}+\left(l+1\right)\rho^{+}
  -\left(p+1\right)\,\rho^{-}\biggr)
  &
  \bs{\Phi}'^{j,-}_{3;nlp}=
  \bigl(0, 0, 0, \vp^{j,-}_{nlp}\bigr)^{\uT}.
\end{align}
\end{subequations}

In the $+$ sector, which is valid when the $\hat{M}_{1}<0$, we have
\begin{subequations}
\label{eq:magnetized_plus_sector_MM_spectrum}
\begin{align}
  \bigl\vert m^{+}_{0;nlp}\bigr\vert^{2}=&
  \frac{2}{\pi\alpha'}
  \biggl(-n\hat{M}_{1}+l\rho^{+}
  -\bigl(p+1\bigr)\,\rho^{-}\biggr)
  &
  \bs{\Phi}^{j,+}_{0;nlp}=
  \bigl(\vp^{j,+}_{nlp},0,0,0\bigr)^{\uT},\\
  \bigl\vert m^{+}_{1;nlp}\bigr\vert^{2}=&
  \frac{2}{\pi\alpha'}
  \biggl(-\bigl(n+1\bigr)\hat{M}_{1}+l\rho^{+}
  -\bigl(p+1\bigr)\,\rho^{-}\biggr)
  &
  \bs{\Phi}^{j,+}_{1;nlp}=
  \bigl(
    0 ,\vp^{j,+}_{nlp},0,0\bigr)^{\uT},\\
  \bigl\vert m^{+}_{2;nlp}\bigr\vert^{2}=&
  \frac{2}{\pi\alpha'}
  \biggl(n\hat{M}_{1}+\bigl(l+1\bigr)\rho^{+}
  -\bigl(p+1\bigr)\rho^{-}\biggr)
  &
  \bs{\Phi}^{j,+}_{2;nlp}=
  \bigl(0,0,\vp^{j,+}_{nlp},0\bigr)^{\uT},\\
  \bigl\vert m^{+}_{3;nlp}\bigr\vert^{2}=&
  \frac{2}{\pi\alpha'}
  \biggl(n\hat{M}_{1}+l\rho^{+}
  -p\,\rho^{-}\biggr)
  &
  \bs{\Phi}^{j,+}_{3;nlp}=
  \bigl(
    0, 0 ,0 , \vp^{j,+}_{nlp}\bigr)^{\uT}.
\end{align}
\end{subequations}

We note that in each sector, there are
$\absb{I^{\left(ab\right)}_{1}}$ towers of massive modes, labeled by $j$,
that are independent in the sense that the raising and lowering
operators do not move from one tower to another.  Using that the zero
modes are orthogonal, it is straightforward to show that the
massive modes are orthonormal in that
\begin{equation}
  \bigl\langle\bs{\Phi}^{j,\mp}_{\lambda},\bs{\Phi}^{j',\mp}_{\lambda'}
  \bigr\rangle=\delta^{jj'}\delta_{\lambda\lambda'}.
\end{equation}

Finally, because of the identity $\rho^{-}\rho^{+}=-\hat{M}_{3}^{2}$,
the $z^{2}$ dependence of the unmagnetized wavefunction
$\vp_{mnll}^{\mp}$ is identical to that of the magnetized wavefunction 
$\vp^{j,\mp}_{nll}$ up to the modification of the
width~\eqref{eq:magnetized_unwarped_width}.  The analogous statement
for $\vp_{nlp}^{j,\mp}$ and $\vp_{mnlp}^{j,\mp}$ for $l\neq p$ does
not hold.

The connection between these modes and the Hermite functions are briefly discussed in Appendix~\ref{app:hermite}.

\subsection{\label{subsec:warped_chiral_wfs}Warped chiral wavefunctions}

We consider again the warped zero modes satisfying
$\hat{\mathbf{D}}'^{\mp}\mathbf{X}'^{j,\mp}=0$.  As discussed in the
previous subsection, in the unwarped case the magnetic flux gives rise
to family replication so that $j$ runs from $0$ to
$\left(\absb{I^{\left(ab\right)}_{1}}-1\right)$.  Although this
multiplicity should not be effected by the warping, there is in
general no obvious relationship between the family index $j$ appearing
in the warped case and index appearing in the unwarped case.  However,
if we again consider the special case of weak
warping~\eqref{eq:weak_warping} then we can write a warped zero mode
as
\begin{equation}
  \mathbf{X}^{j,\mp}=\sum_{n}\epsilon^{n}\mathbf{X}_{\left(n\right)}^{j,\mp},
\end{equation}
and relate the warped and unwarped family indices by taking
$\mathbf{X}_{\left(0\right)}^{j,\mp}=\bs{\Phi}_{0}^{j,\mp}$.  In
general however, the perturbations to this zero mode will involve
modes from different families.  It is straightforward to confirm that
the unwarped massive modes satisfy the same boundary conditions as the
warped zero mode, that is, they vanish as $\absb{z^{2}}\to\infty$ and
satisfy the quasi-periodicity
conditions~\eqref{eq:quasi-periodicity}. We can then take the
expansion
\begin{equation}
  \mathbf{X}_{\left(n\right)}^{j,\mp}=
  \sum_{\lambda,k}c_{\lambda}^{\left(n\right)jk,\mp}
  \bs{\Phi}_{\lambda}^{k,\mp}.
\end{equation}
Following the steps that lead up to~\eqref{eq:general_MM_coefficient},
we find
\begin{equation}
  \label{eq:general_magnetic_MM_coefficient}
  c_{\lambda}^{\left(n\right)jk,\mp}
  =-\frac{2}{\pi\alpha'\absb{m_{\lam}}^{2}}
  \sum_{m=1}^{n}\bigl(-1\bigr)^{m}
  \bigl\langle\bs{\Phi}_{\lam}^{k,\mp},
  \bigl(\hat{\mathbf{D}}^{\pm}_{0}\bigr)^{\ast}
  \beta^{m}
  \mathbf{K}^{\mp}
  \mathbf{X}_{\left(n-m\right)}^{j\mp}\bigr\rangle,
\end{equation}
where $\mathbf{K}^{\mp}$ takes the same form as~\eqref{eq:K} using
now~\eqref{eq:magnetized_rotated_covariant_derivatives}.  As in the
unmagnetized case, we take $c^{\left(n\right)jj,\mp}_{0}=0$ for $n>0$.
Note that the method also does not determine the coefficients
$c^{\left(n\right)jk,\mp}_{0}$ for $k\neq j$ as is typical from
degenerate perturbation theory.  However, because the number of zero
modes is a topological number, it will not be modified by inclusion of
warping effects\footnote{More precisely, it is the net chirality of
  the zero modes that is related to the instanton number, and in
  principle in the presence of warping there could be additional
  vector-like zero modes.  However, at least in the weak-warping limit
  we view it as reasonable to assume that the number of massless modes
  does not become modified.} and we can find linear
combinations of the warped zero modes that additionally satisfy
$c^{\left(n\right)jk,\mp}_{0}=0$.  With this choice, we now revisit
the examples of section~\ref{sec:unmagnetized_examples} to explore how
warping will effect chiral matter wavefunctions.  The exact solutions
presented in Appendix~\ref{app:examples} apply in the magnetized case
as well.

\paragraph{Constant warping}

The case of constant warping again provides a toy example to
demonstrate the perturbative expansion on massive modes.  Focusing once more on the
$-$-sector, from~\eqref{eq:general_magnetic_MM_coefficient}, we have
\begin{equation}
  c^{\left(1\right)jk,-}_{\lambda}=
  \frac{2}{\pi\alpha'\abs{m_{\lambda}}^{2}}
  \bigl\langle
  \bs{\Phi}_{\lambda}^{k,-},
  \bigl(\hat{\mathbf{D}}_{0}^{+}\bigr)^{\ast}
  \beta\mathbf{K}^{-}
  \bs{\Phi}_{0}^{j,-}\bigr\rangle.
\end{equation}
The right-hand side involves
\begin{equation}
  \bigl(\hat{\mathbf{D}}_{0}^{+}\bigr)^{\ast}
  \beta\mathbf{K}^{-}
  \bs{\Phi}_{0}^{j,-}
  =-\ui sc\begin{pmatrix}
    0 \\ \bigl(\hat{D}'^{+}_{1}\bigr)^{\ast}\\
    \bigl(\hat{D}'^{+}_{2}\bigr)^{\ast}\\
    \bigl(\hat{D}'^{+}_{3}\bigr)^{\ast}
  \end{pmatrix}
  \beta \hat{D}'^{-}_{3}\vp_{0}^{j,-}
  =\ui sc\begin{pmatrix}
    0 \\ 0 \\ M_{3}\vp^{j,-}_{011} \\ -\rho^{-}\vp^{j,-}_{0}
  \end{pmatrix},
\end{equation}
where for the second equality we have taken $\beta=1$.
From~\eqref{eq:magnetized_minus_sector_MM_spectrum}, we find
\begin{equation}
  \mathbf{X}'^{j,-}_{\left(1\right)}
  =-\frac{\ui\hat{M}_{3}}{\hat{\kappa}^{2}}
  \biggl\{\hat{M}_{3}\bs{\Phi}'^{j,-}_{2;011}
  -\rho^{-}\bs{\Phi}'^{j,-}_{3;000}\biggr\},
\end{equation}
where
\begin{equation}
  \hat{\kappa}=\frac{2\kappa}{2\pi R_{2}^{2}}=\rho^{+}-\rho^{-}.
\end{equation}

As expected, the warping does not cause the vector boson or matter
curve Wilson line components to be mixed into the zero mode.
Furthermore, since the warping is constant, the warping does not cause
the families to intermix; the $j$th zero mode is perturbed only by the
addition of other members of the $j$th tower of massive modes.

In the unrotated basis, the solution can be written through
$\cO\left(\epsilon\right)$ as
\begin{align}
  \label{eq:constant_warping_magnetized_MM_soln}
  \chi_{2}^{j,-}=
  \biggl\{
  c+ \frac{\epsilon\hat{M}_{3}}{\hat{\kappa}^{2}}\biggl[
  \bigl(c\hat{M}_{3}+s\rho^{-}\bigr)
  -2c\hat{M}_{3}\kappa\abs{z^{2}}^{2}\biggr]\biggr\}\vp^{j,-}_{0},\ \ 
  \chi_{3}^{j,-}=
  -\ui\biggl\{
  s-\frac{2\epsilon s\hat{M}_{3}^{2}}{\hat{\kappa}^{2}}\kappa\abs{z^{2}}^{2}
  \biggr\}\vp_{0}^{j,-}.
\end{align}

As in the unmagnetized case, the constant warping can be absorbed into
a redefinition of $R_{3}$ and hence the equations of motion admit an
exact solution.  Taking $\chi_{0}^{\mp}=0$
and~\eqref{eq:f_term_solving_ansatz} where the covariant derivatives
are now~\eqref{eq:magnetized_covariant_derivs}, we again find from the
$D$-term equation~\eqref{eq:quadratic_eom_warped}.  The exact
$\absb{I^{\left(ab\right)}_{1}}$ solutions are
$f^{j,\mp}=\frac{1}{z^{2}}\vp_{\mathrm{w}}^{j,\mp}$ where
\begin{equation}
  \vp^{j,\mp}_{\mathrm{w}}=\mathcal{N}_{\mathrm{w}}^{j,\mp}
  \ue^{-\kappa_{\mathrm{w}}\left\vert z^{2}\right\vert}
  \ue^{\pm\pi I^{\left(ab\right)}_{1}z^{1}\mathrm{Im}\, z^{1}/
    \mathrm{Im}\, \tau_{1}}
  \vartheta\begin{bmatrix}
     \pm j/I^{\left(ab\right)}_{1} \\
    0
  \end{bmatrix}
  \bigl(\pm I^{\left(ab\right)}_{1}z^{1},
  \pm I^{\left(ab\right)}_{1}\tau_{1}\bigr),
\end{equation}
and the warped inverse width is
\begin{equation}
  \kappa_{\mathrm{w}}=2\pi R_{2}^{2}
  \sqrt{\left(\frac{\hat{M}_{1}}{2}\right)^{2}
    +\ue^{-4\alpha}\hat{M}_{3}^{2}}.
\end{equation}
Then,
\begin{align}
  \chi_{2}^{j,\mp}=-\sqrt{2\pi}R_{2}
  \biggl(\frac{\kappa_{\mathrm{w}}}{2\pi R_{2}^{2}}
  +\frac{1}{2}\hat{M}_{1}\biggr)\vp_{\mathrm{w}}^{j,\mp},\quad
  \chi_{3}^{j,\mp}=-\ui\sqrt{2\pi R_{2}}\hat{M}_{3}\vp_{\mathrm{w}}^{j,\mp}.
\end{align}
After normalizing and then expanding in a power series in $\epsilon$,
we find a result that agrees with the perturbative
analysis~\eqref{eq:constant_warping_magnetized_MM_soln}.

\paragraph{Constant warping along the matter curve}

We can consider again an example where the warp factor varies along
$X_{6}$ but is constant along the matter curve
\begin{equation}
  \beta=L^{-2}\abs{z^{2}}^{2}.
\end{equation}
The inclusion of $z^{3}$ dependence can be handled as in
Appendix~\ref{app:corrections}.  As in the magnetized case, this warp
factor can be expressed in terms of the ladder operators.  Indeed,
\begin{equation}
  z^{2}=-\frac{\sqrt{2\pi}R_{2}}{2\kappa}
  \biggl\{
  \frac{1}{c}\bigl(\hat{D}'^{+}_{2}\bigr)^{\ast}
  +\frac{\ui}{s}\bigl(\hat{D}'^{+}_{3}\bigr)^{\ast}\biggr\},\quad
  \bar{z}^{\bar{2}}=\frac{\sqrt{2\pi}R_{2}}{2\kappa}
  \biggl\{\frac{1}{c}\hat{D}'^{-}_{2}
  -\frac{\ui}{s}\hat{D}'^{-}_{3}\biggr\},
\end{equation}
giving
\begin{equation}
  \beta=-\frac{2\pi R_{2}^{2}}{4\kappa^{2}L^{2}}
  \biggl\{
  \frac{1}{c^{2}}\bigl(\hat{D}'^{+}_{2}\bigr)^{\ast}
  \bigl(\hat{D}'^{-}_{2}\bigr)+
  \frac{1}{s^{2}}\bigl(\hat{D}'^{+}_{3}\bigr)^{\ast}
  \hat{D}'^{-}_{3}
  +\frac{\ui}{sc}\biggl[
  \bigl(\hat{D}'^{+}_{3}\bigr)^{\ast}
  \hat{D}'^{-}_{2}-
  \bigl(\hat{D}'^{+}_{2}\bigr)^{\ast}
  \hat{D}'^{-}_{3}\biggr]\biggr\}.
\end{equation}
Then,
\begin{equation}
  \beta\vp^{j,-}_{nlp}
  =-\frac{\ui\left(2\pi R_{2}^{2}\right)\hat{M}_{3}}
  {4\kappa^{2}L^{2}sc}
  \biggl\{
  \sqrt{\left(l+1\right)\left(p+1\right)}
  \vp_{n,l+1,p+1}^{j,-}
  -\ui\left(l+p+1\right)\vp_{nlp}^{j,-}
  -\sqrt{lp}\vp_{n,l-1,p-1}^{j,-}
  \biggr\}.
\end{equation}
The resulting first order correction is
\begin{equation}
  \label{eq:chiral_quadratic_solution}
  \mathbf{X}'^{j,-}_{\left(1\right)}
  =\frac{\hat{M}_{3}}{\left(2\pi R_{2}^{2}\right)\hat{\kappa}^{3}L^{2}}
  \biggl\{\hat{M}_{3}\bs{\Phi}'^{j,-}_{2;022}
  -2\ui\hat{M}_{3}\bs{\Phi}'^{j,-}_{2;011}
  -\rho^{-}\bs{\Phi}'^{j,-}_{3;011}
  +2\ui\rho^{-}\bs{\Phi}'^{j,-}_{3;000}\biggr\}.
\end{equation}
In the unrotated basis and through $\mathcal{O}\left(\epsilon\right)$,
this gives
\begin{subequations}
\label{eq:quadratic_warping_magnetized_MM-soln}
\begin{align}
  \chi_{3}^{j,\mp}=&\biggl\{c
  + \frac{\epsilon\hat{M}_{3}}{2\pi R_{2}^{2}\hat{\kappa}^{3}L^{2}}
  \biggl[\bigl(c\hat{M}_{3}+s\rho^{-}\bigr)
  +2s\rho^{-}\kappa\abs{z^{2}}^{2}
  -2c\hat{M}_{3}\kappa^{2}\abs{z^{2}}^{4}\biggr]\biggr\}\vp_{0}^{j,\mp},\\
  \chi_{4}^{j,\mp}=&-\ui\biggl\{
  s -\frac{\epsilon\hat{M}_{3}}{2\pi R_{2}^{2}\hat{\kappa}^{3}L^{2}}
  \biggl[
  2c\rho^{-}\kappa\abs{z^{2}}^{2}+2s\hat{M}_{3}\kappa^{2}\abs{z^{2}}^{4}\biggr]
  \biggr\}\vp_{0}^{j,\mp}.
\end{align}
\end{subequations}
As in the previous case, the vector boson and matter-curve Wilson line
components are not excited when warping is introduced and the warping
does not mix different families.

This warp factor also admits a relatively simple exact solution
following from
\begin{equation}
  f^{j,\mp}
  =\frac{\mathcal{N}_{\mathrm{w}}^{j,\mp}}{z^{2}}
  \mathrm{Ai}\bigl(a
  +b\abs{z^{2}}^{2}\bigr)
  \ue^{\pm\pi\ui I^{\left(ab\right)}_{1}z^{1}\mathrm{Im}\, z^{1}/
    \mathrm{Im}\, \tau_{1}}
  \vartheta\begin{bmatrix}
     \pm j/I^{\left(ab\right)}_{1} \\
    0
  \end{bmatrix}
  \bigl(\pm I^{\left(ab\right)}_{1}z^{1},
  \pm I^{\left(ab\right)}_{1}\tau_{1}\bigr),
\end{equation}
where, as in the unmagnetized case,
$a=\kappa^{2/3}L^{1/3}\epsilon^{-2/3}$ and
$b=\kappa^{2/3}\epsilon^{1/3}L^{-2/3}$.  When correctly normalized, the
$\epsilon$ expansion agrees
with~\eqref{eq:quadratic_warping_magnetized_MM-soln}.

Considering general positive powers
\begin{equation}
  \beta=L^{-2r}\abs{z^{2}}^{2r},
\end{equation}
we have, as in the unmagnetized
case~\eqref{eq:general_power_warp_factor}
\begin{equation}
  \beta\vp^{j,-}_{nlp}=\biggl(\frac{1}{2\kappa L^{2}}\bigg)^{r}
  \sum_{s=0}^{r}\ui^{s}\binom{r}{s}
  \sqrt{\frac{\left(l+s\right)!}{l!}
    \frac{\left(p+s\right)!}{p!}}
  \frac{\left(l+p+r\right)!}{\left(l+p+s\right)!}
  \vp^{j,-}_{n,l+s,p+s},
\end{equation}
where again we use the magnetized width given
in~\eqref{eq:magnetized_unwarped_width}.  From this we find the first
order correction
\begin{equation}
  \mathbf{X}_{\left(1\right)}'^{j,-}
  =\frac{\ui\hat{M}_{3}}{\hat{\kappa}^{2}}
  \left(r+1\right)!
  \left(\frac{2\pi R_{2}^{2}}{\hat{\kappa}L^{2}}\right)^{r}
  \sum_{s=0}^{r}\ui^{s}\binom{r}{s}\frac{1}{s+1}
  \biggl\{\hat{M}_{3}\bs{\Phi}'^{j,-}_{2;0,s+1,s+1}-
  \rho^{-}\bs{\Phi}'^{j,-}_{3;0ss}\biggr\}.
\end{equation}
Finally, for a polynomial warp factor~\eqref{eq:polynomial_wf}, we similarly have
\begin{equation}
  \mathbf{X}_{\left(1\right)}'^{j,-}
  =\frac{\ui\hat{M}_{3}}{\hat{\kappa}^{2}}\sum_{s}C_{s}
  \biggl\{\hat{M}_{3}\bs{\Phi}'^{j,-}_{2;0,s+1,s+1}
  -\rho^{-}\bs{\Phi}'^{j,-}_{2;0ss}\biggr\},
\end{equation}
with the coefficients $C_{s}$ are again given
by~\eqref{eq:polynomial_MM_solution} after the definition of $\kappa$ given 
by~\eqref{eq:magnetized_unwarped_width}.

\paragraph{Variable warping along the matter curve}

Since the warp factor is neutral under the residual gauge group,
even in the presence of magnetic flux it is periodic on the matter
curve $\Sigma$.  We then reconsider a warp factor of the
form~\eqref{eq:fourier_transformed_warp_factor} with constant
$\tilde{\beta}_{nm}$.  The first order corrections to the $-$-sector
wavefunctions follow from
\begin{equation}
  \label{eq:z1_warped_magnetized_eom}
  -\ui sc \sum_{mn} \tilde{\beta}_{mn}
  \begin{pmatrix}
    0 \\
    \bigl(\hat{D}'^{+}_{1}\bigr)^{\ast} \\
    \bigl(\hat{D}'^{+}_{2}\bigr)^{\ast} \\
    \bigl(\hat{D}'^{+}_{3}\bigr)^{\ast}
  \end{pmatrix}
  h_{mn}
  \hat{D}'^{-}_{3}\vp_{0}^{j,-}.
\end{equation}
Unlike the previous warp factors, which depended only
on $z^{2}$ and $\bar{z}^{\ovl{2}}$, this warp factor cannot be
expressed in terms of only the raising and lowering operators acting
on the $-$-sector.  That is, in the $-$-sector, the only ladder
operators involving $z^{1}$ and $\bar{z}^{\bar{1}}$ are
$\ui\bigl(\hat{D}'^{+}_{1}\bigr)^{\ast}$ and $\ui \hat{D}'^{-}_{1}$.
Neither $z^{1}$ nor $\bar{z}^{\bar{1}}$ can be written in terms of
these operators without using their conjugates which do not act
naturally on $-$-sector.  This was also true in the unmagnetized case,
but there we had the simple fact that
\begin{equation}
  h_{m'n'}\vp^{\mp}_{mnlp}=\vp^{\mp}_{m+m',n+n',lp}.
\end{equation}
However, with magnetic flux, the fact that the warp factor is periodic
while the wavefunction is quasi-periodic makes this relationship a
little more involved.

Let us for instance consider the bottom component of~\eqref{eq:z1_warped_magnetized_eom}.
We have that
\begin{equation}
  -\ui sc\tilde{\beta}_{mn}
  \bigl(\hat{D}'^{+}_{3}\bigr)^{\ast}
  h_{mn}\hat{D}'^{-}_{3}\vp_{0}^{j,-}
  =-\ui sc\rho^{-}\tilde{\beta}_{mn}h_{mn}\vp_{0}^{j,-}.
\end{equation}
Since $h_{mn}$ is periodic, $h_{mn}\vp_{0}^{j,-}$ is quasi-periodic
and so admits an expansion in terms of the $\vp_{q00}^{k,-}$.
Unlike the cases where the warp factor did not vary over the matter
curve, we expect that the warping will mix different families and so
take an expansion
\begin{equation}
  \label{eq:fourier_theta}
  h_{mn}\vp_{0}^{j,-}=\sum_{q,k}B_{mnq}^{kj,-}\vp_{q00}^{k,-}.
\end{equation}
We show in Appendix~\ref{app:fourier_theta_overlap} that if $n=k-j\mod
I^{\left(ab\right)}_{1}$, then
\begin{equation}
  B_{mnq}^{kj,-}=
    \frac{\bigl(\ui t_{mn}\bigr)^{q}}{
      \hat{M}_{1}^{q}q!}
    \ue^{\mp\hat{m}^{2}_{mn}\cV_{1}/4\pi^{2}I^{\left(ab\right)}_{1}}
    \ue^{\mp\pi \ui m\left(k+j\right)/2I^{\left(ab\right)}_{1}},
\end{equation}
while otherwise $B_{mnq}^{kj,-}$ vanishes. Here $t_{mn}$ and
$\hat{m}_{mn}^{2}$ are defined in~\eqref{eq:define_t_mn}
and~\eqref{eq:Fourier_mass}.

Similar expansions apply for the other entries and we have the
$\mathcal{O}\left(\epsilon\right)$ correction
\begin{equation}
  \mathbf{X}'^{j,-}_{\left(1\right)}
  =-\ui sc\sum_{qmn,k}
  \frac{B^{kj,-}_{qmn}}{q\hat{M_{1}}+\hat{\kappa}}
  \biggl\{
  \sqrt{-l\hat{M}_{1}\rho^{-}}\bs{\Phi}'^{k,-}_{1;\left(q-1\right);10}
  +\hat{M}_{3}\bs{\Phi}'^{k,-}_{2;q11} -
  \rho^{-}\bs{\Phi}'^{k,-}_{3;q00}\biggr\}.
\end{equation}
Although the vector boson component remains unexcited, in contrast to
the previous cases, the warped zero mode now receives a contribution
from the Wilson line along the matter curve.  Additionally, the
warping causes the warped zero mode for one family to involve the
unwarped zero modes of other families.

\section{\label{sec:eft}Warped effective field theory}

A direct application of computing open string wavefunctions is to determine 4d
effective action describing the low-energy dynamics of the corresponding fields. 
This can be done by way of a standard dimensional reduction of the 
fermionic~\eqref{eq:wynants} and bosonic~\eqref{eq:bosonic_action} actions from 
the worldvolume $\cW$ to $\mathbb{R}^{1,3}$. The purpose of this section is to
perform such dimensional reduction in terms of the warped wavefunctions for 
matter fields at $\uD 7$-brane intersections. However, unlike in the 
adjoint case analyzed in~\cite{Marchesano:2008rg}, the wavefunctions for 
bifundamental matter depend on the detailed form of the warp factor and the 
intersection, and so they will differ from one intersecting $\uD 7$-brane model to another.
We will therefore limit ourselves to describe some general features on the warped effective 
field theory, leaving a more detailed case-by-case study for future work.

\subsection{Warped non-chiral matter metrics}

Let us first consider the unwarped case without any magnetic flux.  The
zero mode is a mixture of transverse fluctuations and one of the
Wilson lines.  The 4d kinetic terms for the open string fields follows
from the DBI action~\eqref{eq:DBI}.  In particular, after restoring
the axio-dilaton we have
\begin{align}
  \label{eq:kinetic_term}
  S^{\mathrm{DBI}}_{\uD 7}\ni S^{\mathrm{kin}}=
  & -\frac{1}{g_{8}^{2}}\int_{\cW}\ud^{8}x\,
  \sqrt{\tilde{g}}\bigl(\im\tau\bigr)^{-1} \tr\biggl\{
  \frac{1}{2}\eta^{\mu\nu}g^{ab}F_{\mu a}F_{\nu b}
  +\frac{1}{2}\eta^{\mu\nu}g_{ij}D_{\mu}\Phi^{i}D_{\nu}\Phi^{j}\biggr\}\notag \\
  =&-\frac{2}{\lambda g_{8}^{2}}\int_{\cW}\ud^{8}x\,\sqrt{\tilde{g}}
  \bigl(\im\tau\bigr)^{-1}
  \tr\biggl\{
  \partial_{\mu}a_{2}\partial^{\mu}\bar{a}_{\bar{2}}+
  \partial_{\mu}\phi\partial^{\mu}\bar{\phi}\biggr\},
\end{align}
where in the second line we have used the
parametrization~\eqref{eq:param_flucs} and have truncated to
quadratic order in fluctuations.

We can move to the 4d Einstein frame by the Weyl transformation
\begin{equation}
  \eta_{\mu\nu}\to\frac{\cV_{0}}{\cV}\eta_{\mu\nu},
\end{equation}
where $\cV$ is the volume of the internal space $X_{6}$ with fiducial
value $\cV_{0}$.  Such a Weyl transformation gives a canonical
Einstein-Hilbert action with 4d gravitational constant
$\kappa_{4}=\kappa_{10}\cV_{0}^{-1/2}$ where the 10d gravitational
constant is given by $2\kappa_{10}^{2}=8\pi^{3}\lambda^{4}g_{\us}^{2}$.
The 4d kinetic term for the bifundamental matter in this frame is then
\begin{equation}
  S^{\mathrm{kin}}_{4\ud} = -\int_{\mathbb{R}^{1,3}}\ud^{4}x\,
  \tr\biggl\{\mathcal{K}^{-}_{\sigma\bar{\sigma}}
  \partial_{\mu}\sigma^{-}\bigl(\partial^{\mu}\sigma^{-}\bigr)^{\dagger}
  + \mathcal{K}^{+}_{\sigma\bar{\sigma}}
  \partial_{\mu}\sigma^{+}\bigl(\partial^{\mu}\sigma^{+}\bigr)^{\dagger}
  \biggr\},
\end{equation}
where the K\"ahler metric is
\begin{align}
  \label{eq:ourmetric}
  \mathcal{K}^{\mp}_{\sigma\bar{\sigma}}
  =&\frac{2\lambda g_{\us}}{\kappa_{4}^{2}\cV}
  \int_{\mathcal{S}_{4}}\ud^{4}y\,\sqrt{\tilde{g}}
  \bigl(\im\tau\bigr)^{-1}
  \bigl(\mathbf{X}^{\mp}\bigr)^{\ast}\cdot\mathbf{X}^{\mp}\notag \\
  =&\frac{2\cN^{2}\lambda g_{\us}}{\kappa_{4}^{2}\cV}
  \int_{\mathcal{S}_{4}}\ud^{4}y\,\sqrt{\tilde{g}}
  \bigl(\im\tau\bigr)^{-1}
  \ue^{-2\kappa\absb{z^{2}}^{2}},
\end{align}
in which $\cN$ is a normalization constant the unwarped zero
mode~\eqref{eq:unwarped_solution}.  Note that although this is an
integral over $\mathcal{S}_{4}$, due to the exponential localization
the integral is sensitive only to field values near $z^{2}=0$.
Indeed, if we take the $\alpha'\to 0$ limit while keeping the physical
volumes $\alpha'\cV_{m}$ constant, then the norm-squared of the
internal wavefunction becomes a $\delta$-function,
\begin{equation}
  \label{eq:wf_to_delta}
  \frac{2\kappa\im\tau_{2}}{\pi\cV_{2}}
  \ue^{-2\kappa\absb{z^{2}}^{2}}\to 
  \frac{\delta^{2}\bigl(z^{2},\bar{z}^{\ovl{2}}\bigr)}
  {\sqrt{\tilde{g}_{\mathbb{T}_{2}^{2}}}},
\end{equation}
where the
$\delta$-function has been normalized according to
\begin{equation}
  \int\ud^{2}z^{2}\, 
  \delta^{2}\bigl(z^{2},\bar{z}^{\ovl{2}}\bigr)=1.
\end{equation}

The K\"ahler metric for bifundamental matter localized on intersecting
D-branes has previously been calculated via dimensional reduction and
worldsheet
methods~\cite{Ibanez:1998rf,Cvetic:2003ch,Lust:2004cx,*Lust:2004fi,
Lust:2004dn,Font:2004cx,Bertolini:2005qh}.
For the moment, instead of two stacks of intersecting branes specified
by~\eqref{eq:angle} consider a pair of $\uD 7$-branes filling $\Sigma$
and intersecting at angles $\pi\theta^{2,3}$ in the $z^{2,3}$ plane.
Then from~\cite{Lust:2004fi}, the metric for the bifundamental matter
is
\begin{equation}
  \label{eq:lustmetric}
  \breve{\cK}_{\sigma\bar{\sigma}}=\frac{\cV_{1}^{1/2}}{\kappa_{4}^{2}
  \cV^{1/2}\im\tau}
  \prod_{m=1}^{2}
  \bigl(\im\tau_{m}\bigr)^{-\theta^{m}}
  \sqrt{\frac{\Gamma\left(\theta^{m}\right)}
    {\Gamma\left(1-\theta^{m}\right)}},
\end{equation}
where $\Gamma$ is the usual $\Gamma$-function and we have suppressed
numerical coefficients.  To compare our result
to~\eqref{eq:lustmetric}, we first perform a coordinate redefinition
so that the geometry of~\eqref{eq:angle} is similar to the geometry
used to derive~\eqref{eq:lustmetric}.  To do so, first specialize to
the case where $\tau_{2}=\tau_{3}=\ui$, and then define a new complex
structure by
\begin{equation}
  u^{1}=z^{1},\quad
  u^{2}=y^{5}+\ui y^{6},\qquad
  u^{3}=y^{8}+\ui y^{9}.
\end{equation}
The two stacks of branes then intersect at the small angle $2\arctan
\frac{R_{3}I^{\left(ab\right)}_{3}}{2R_{2}}\sim \hat{M}_{3}$ in each of the
$u^{2}$ and $u^{3}$ planes.  Note that the product $\cV_{2}\cV_{3}$
takes the same value in either set of coordinates.
Then~\eqref{eq:lustmetric} gives
\begin{equation}
  \breve{\cK}_{\sigma\bar{\sigma}}\sim
  \frac{1}{\kappa_{4}^{2}\cV_{2}^{1/2}\cV_{3}^{1/2}
    \hat{M}_{3}\im\tau}.
\end{equation}
On the other hand,~\eqref{eq:ourmetric} gives
\begin{equation}
  \cK_{\sigma\bar{\sigma}}\sim\frac{1}{\kappa_{4}^{2}\cV_{2}\cV_{3}
    \hat{M}_{3}\im\tau}
\end{equation}
where we have used the fact that $\kappa\propto
\hat{M}_{3}\cV_{2}/\im\tau_{2}$. Evidently, in order
for~\eqref{eq:ourmetric} to agree with~\eqref{eq:lustmetric}, we must
perform a field redefinition
$\sigma\to\cV_{2}^{1/4}\cV_{3}^{1/4}\sigma$.  Since in our analysis we
have treated the closed string background as fixed, such a field
redefinition does not change the behavior of the wavefunctions.

In the warped case, the metrics appearing
in~\eqref{eq:kinetic_term} are replaced with the warped metrics and
the volume with the warped volume
\begin{equation}
  \cV_{\mathrm{w}}=\int_{X_{6}}\ud^{6}x\sqrt{\tilde{g}}\ue^{-4\alpha},
\end{equation}
giving
\begin{align}
  \cK^{\mp}_{\sigma\bar{\sigma}}
  =&\frac{2\lambda g_{\us}}{\kappa_{4}^{2}\cV_{\mathrm{w}}}
  \int_{\cS_{4}}\ud^{4}y\,
  \sqrt{\tilde{g}}\bigl(\im\tau\bigr)^{-1}
  \bigl(\mathbf{X}^{\mp}\bigr)^{\ast}\cdot\bigl(\ue^{\#\alpha}
  \mathbf{X}^{\mp}\bigr),
\end{align}
where we have defined
\begin{equation}
  \label{eq:warp_factor_matrix}
  \ue^{\#\alpha}=
  \mathrm{diag}\bigl(
  \ue^{-4\alpha},1,1,\ue^{-4\alpha}\bigr),
\end{equation}
and used the fact that the gauge boson component vanishes.  We can
then consider the weak warping approximation~\eqref{eq:weak_warping}.
Using the fact that the integral over $\mathcal{S}_{4}$ is proportional to
the inner product and using the normalization condition $\bigl\langle
\mathbf{X}^{\mp},\bs{\Phi}_{0}^{\mp}\bigr\rangle=\bigl\langle
\mathbf{X}_{\left(0\right)}^{\mp},\bs{\Phi}_{0}^{\mp}\bigr\rangle$, we get that
first-order correction to the warped K\"ahler metric is
\begin{align}
  \label{eq:warped_achiral_metric_one}
  \mathcal{K}^{\mp}_{\sigma\bar{\sigma}\left(1\right)}= \frac{2\lambda
    g_{\us}\epsilon}{\kappa_{4}^{2}\cV}
  \int_{\mathcal{S}_{4}}\ud^{4}y\,\sqrt{\tilde{g}}
  \bigl(\im\tau\bigr)^{-1}
  \bigl(\chi_{3\left(0\right)}^{\mp}\bigr)^{\ast}
  \beta\bigl(\chi_{3\left(0\right)}^{\mp}\bigr)
  -\frac{\delta\cV}{\cV}\mathcal{K}^{\mp}_{\sigma\bar{\sigma}\left(0\right)},
\end{align}
where $\chi_{m\left(0\right)}^{\mp}$ and
$\mathcal{K}^{\mp}_{\sigma\bar{\sigma}\left(0\right)}$ are the
unwarped wavefunction and K\"ahler metric~\eqref{eq:ourmetric} and
$\delta\cV:=\cV_{\mathrm{w}}-\cV\sim\epsilon$.  Note that if we now
take~\eqref{eq:wf_to_delta}, then through
$\mathcal{O}\left(\ep\right)$
\begin{equation}
  \cK^{\mp}_{\sigma\bar{\sigma}}
  \sim \frac{\left(\cV_{1}+\cV_{1,\uw}\right)}{\kappa_{4}^{2}\cV_{\uw}\hat{M}_{3}
    \im\tau},
\end{equation}
where
\begin{equation}
  \cV_{1,\uw}=\int_{\Sigma}\ud^{2}y\sqrt{\tilde{g}}\ue^{-4\alpha},
\end{equation}
is the warped volume of the matter curve.  The fact that it is the
average of the unwarped volume and warped volume of the matter curve
that appears is a result of the fact that the bifundamental zero modes
are mixtures of the deformation modulus and a Wilson line; in the zero
angle case, the kinetic term for the former involves the warped volume
of the 4-cycle while for the latter it is the unwarped volume that
appears~\cite{Marchesano:2008rg}.

Although~\eqref{eq:warped_achiral_metric_one} already takes into
account some non-trivial warping modifications, it uses only the
unwarped zero modes.  The $\mathcal{O}\left(\epsilon\right)$
corrections to the warped zero mode provide corrections to the
K\"ahler metric at $\mathcal{O}\left(\epsilon^{2}\right)$,
\begin{multline}
  \cK^{\mp}_{\sigma\bar{\sigma}\left(2\right)}
  =
  \\\frac{2\lambda g_{\us}\epsilon^{2}}{\kappa_{4}^{2}}
  \int_{\cS_{4}}\ud^{4}y\,\sqrt{\tilde{g}}
  \bigl(\im\tau\bigr)^{-1}
  \biggl\{\bigl(\chi_{3\left(1\right)}^{\mp}\bigr)^{\ast}
  \beta\chi_{3\left(0\right)}^{\mp}
  + \bigl(\chi_{3\left(0\right)}^{\mp}\bigr)^{\ast}\beta\chi_{3\left(1\right)}^{\mp}
  + \bigl(\mathbf{X}^{\mp}_{\left(1\right)}\bigr)^{\ast}\cdot
  \mathbf{X}^{\mp}_{\left(1\right)}\biggr\}
  -\frac{\delta\cV}{\cV}\cK^{\mp}_{\sigma\bar{\sigma}\left(1\right)}.
\end{multline}
  
\subsection{Warped chiral matter metrics}

Let us again first consider  the unwarped (chiral) case.  There are
$\absb{I^{\left(ab\right)}_{1}}$ zero modes and, when
$I^{\left(ab\right)}_{1}>0$, we write the $-$-sector zero mode as
\begin{equation}
  \mathbf{X}^{-}=\bigl(0, 0, c, -\ui s\bigr)^{\uT}
  \sum_{j}\sigma^{j,-}\bigl(x^{\mu}\bigr)\vp_{0}^{j,-}\bigl(x^{a}\bigr),
\end{equation}
where $\vp_{0}^{j,-}$ is given in~\eqref{eq:minus_sector_zero_mode}.
A similar expansion applies for the $+$-sector.  Since $\vp_{0}^{j,\mp}$
is orthogonal to $\vp_{0}^{j',\mp}$ for $j\neq j'$, we find in the 4d
Einstein frame
\begin{equation}
  \label{eq:chiral_unwarped_4d_kinetic_term}
  S_{4\uD}^{\mathrm{kin}}=-\int_{\mathbb{R}^{1,3}}\ud^{4}x\,\tr\biggl\{
  \sum_{j}
  \cK^{\mp}_{j\bar{j}}
  \partial_{\mu}\sigma^{j,\mp}\bigl(\partial^{\mu}\sigma^{j,\mp}\bigr)^{\dagger}
  \biggr\},
\end{equation}
where the K\"ahler metric is
\begin{equation}
  \cK^{\mp}_{j\bar{j}}
  =\frac{2\lambda g_{\us}}{\kappa_{4}^{2}\cV}
  \int_{\cS_{4}}\ud^{4}y\,
  \sqrt{\tilde{g}}\bigl(\im\tau\bigr)^{-1}
  \bigl(\vp_{0}^{j,\mp}\bigr)^{\ast}
  \vp_{0}^{j,\mp}.
\end{equation}
Although both the $+$- and $-$-sectors are included
in~\eqref{eq:chiral_unwarped_4d_kinetic_term}, as discussed in
section~\ref{sec:chiral} only one sector is present for
$I^{\left(ab\right)}_{1}\neq 0$.  From~\eqref{eq:normalization} we get
\begin{equation}
  \label{eq:unwarped_chiral_metric}
  \cK_{j\bar{k}}^{\mp}=\delta_{j\bar{k}}
  \frac{2\lambda g_{\us}\cN^{2}}{\kappa_{4}^{2}\cV_{3}}
  \frac{\pi}{2\kappa\,\im\tau_{2}\sqrt{\pm 2I^{\left(ab\right)}_{1}
    \im\tau_{1}}}.
\end{equation}
This result cannot be directly compared to the kinetic terms appearing
in~\cite{Ibanez:1998rf,Lust:2004cx,Lust:2004fi} since there the
assumption was that in the T-dual $\uD 9$-picture, where the
intersection is turned into magnetic flux, $F_{m\bar{n}}=0$ if $m\neq
n$, a relationship which is not satisfied by the
angle~\eqref{eq:angle} and flux~\eqref{eq:magnetic_flux}
(see~\eqref{eq:D9_flux}).  However, the analysis was reconsidered for
a more general magnetic flux and angles in~\cite{Bertolini:2005qh}.
The angular dependence is again recovered by~\eqref{eq:lustmetric} but
replacing the angles $\theta_{i}$ with the eigenvalues of
$\mathbf{B}$.  In terms of these
eigenvalues,~\eqref{eq:unwarped_chiral_metric} behaves as
\begin{equation}
  \cK_{j\bar{k}}^{\mp}\sim\frac{1}
  {\left(\rho^{+}-\rho^{-}\right)\sqrt{\hat{M}_{1}}}.
\end{equation}
On the other hand the analysis of~\cite{Bertolini:2005qh} suggests
\begin{equation}
  \label{eq:Bertolini_metric}
  \breve{\cK}_{j\bar{k}}^{\mp}\sim\frac{1}{\hat{M}_{3}{\sqrt{\hat{M}_{1}}}}.
\end{equation}
Although the angular dependence agrees in the case where $\hat{M}_{1}$
can be neglected in comparison to $\hat{M}_{3}$, the fields again need
to be redefined as was done in the unmagnetized case in order to agree 
with~\eqref{eq:Bertolini_metric}.

As discussed in section~\ref{subsec:warped_chiral_wfs}, when the warp
factor varies non-trivially along the matter curve, the warping will
generally mix together different families.  Although this mixing
occurred at the first order correction and so occurs at second order
in the kinetic terms, there is another mixing that occurs at first
order.  We have,
\begin{equation}
  \cK^{\mp}_{j\bar{k}}
  =\frac{2\lambda g_{\us}}{\kappa_{4}^{2}\cV_{\uw}}
  \int_{\cS_{4}}\ud^{4}y\sqrt{\tilde{g}}
  \bigl(\im\tau\bigr)^{-1}
  \bigl(\mathbf{X}^{k,\mp}\bigr)^{\ast}\cdot
  \bigl(\ue^{\#\alpha}\mathbf{X}^{j,\mp}\bigr),
\end{equation}
where $\mathbf{X}^{j,\mp}$ is now the warped zero mode.  The first
order correction to the K\"ahler metric is then
\begin{equation}
  \cK^{\mp}_{j\bar{k}\left(1\right)}
  =\frac{2\lam g_{\us}\ep}{\kap_{4}^{2}\cV}
  \int_{\cS_{4}}\ud^{4}y\,\sqrt{\tilde{g}}
  \bigl(\im\tau\bigr)^{-1}
  \bigl(\chi^{k,\mp}_{3\left(0\right)}\bigr)^{\ast}
  \beta
  \chi^{j,\mp}_{3\left(0\right)}
  -\frac{\delta \cV}{\cV}
  \cK^{\mp}_{j\bar{k}\left(0\right)}.
\end{equation}
The warp factor $\beta$ will generally admit a Fourier
transformation~\eqref{eq:fourier_transformed_warp_factor}.  As shown
in Appendix~\ref{app:fourier_theta_overlap}, the Fourier mode $h_{mn}$
will connect different families $\vp_{0}^{j,\mp}$ and
$\vp_{0}^{k,\mp}$ if $k-j=n\mod I^{\left(ab\right)}_{1}$.  Thus for
generic warping $\cK_{j\bar{k}}$ is not simultaneously diagonalizable
with its unwarped counterpart.  Therefore in the construction of
phenomenologically viable compactifications, one must carefully take
into account the effects of warping; a model that is free of dangerous
flavor-changing neutral currents in the unwarped case may not
automatically be so once warping is taken into account.

The problem of diagonalization becomes even more complex once we move
to higher order in perturbation theory.  The second order correction is
\begin{multline}
  \cK^{\mp}_{j\bar{k}\left(2\right)}=\\
  -\frac{\delta\cV}{\cV}
  \cK^{\mp}_{j\bar{k}\left(1\right)}
  +
  \frac{2\lambda g_{\us}\ep^{2}}{\kap_{4}^{2}\cV}
  \int_{\cS_{4}}\ud^{4}y\,\sqrt{\tilde{g}}
  \bigl(\im\tau\bigr)^{-1}
  \biggl\{\bigl(\chi_{3\left(0\right)}^{k,\mp}\bigr)^{\ast}\beta
  \chi_{3\left(1\right)}^{j,\mp}+
  \bigl(\chi_{3\left(1\right)}^{k,\mp}\bigr)^{\ast}\beta
  \chi_{3\left(0\right)}^{j,\mp}+
  \bigl(\mathbf{X}^{k,\mp}_{\left(1\right)}\bigr)^{\ast}\cdot
  \mathbf{X}_{\left(1\right)}^{j,\mp}\biggr\}.
\end{multline}

\subsection{\textit{D}-terms}

An interesting way of interpreting the warped chiral wavefunctions obtained 
in the previous sections is by considering how they affect the $D$-term at
the level of the 4d effective field theory. As discussed in 
subsection~\ref{sec:nonchiral_eom}, one obtains such $D$-term by plugging the 
chiral wavefunctions into~\eqref{eq:truncated_d_term}. For weak warping one 
may write~\eqref{eq:truncated_dterm_2} schematically as
\begin{equation}
D_\alpha \, =\, D_0 + \epsilon D_\beta \, =\, 
D_0  + \epsilon \lambda^2 \int_{{\cal S}_4} \beta [\Phi, \bar{\Phi}]
\label{contam}
\end{equation}
where we have used~\eqref{eq:weak_warping}. Here $D_0$ stands for the $D$-term 
with trivial warp factor, while $D_\beta$ is an extra contribution arising from the non-trivial
piece of the warping $\beta$.

From the viewpoint of the unwarped spectrum, $D_\beta$ is a perturbation of the 
$D$-term that spoils its usual structure. Indeed, if we plug in the full tower of massive,
unwarped modes
\begin{equation}
  \phi^{\mp}_{m}=\sum_{\lambda}\sigma_{\lambda}^{\mp}\bigl(x^{\mu}\bigr)
  \chi^{\mp}_{m}\bigl(y^{a}\bigr)
\end{equation}
into $D_0$, we obtain the following expression at quadratic order in
fluctuations
\begin{align}
  D=&-\frac{\lambda^{2}}{2\pi^{2}\im\tau_{2}}
  \sum_{j,j',\lambda,\lambda'}
  \biggl\{
  \bigl\langle\mathbf{X}_{\lambda}^{j,-},
  \mathbf{X}_{\lambda'}^{j',-}\bigr\rangle
  \begin{pmatrix}
    \sigma_{\lambda}^{j,-}\sigma_{\lambda'}^{j',-\dagger} & 0 \\
    0 & -\sigma_{\lambda'}^{j',-\dagger}\sigma^{j,-}_{\lambda}
  \end{pmatrix}\notag \\
  &\phantom{\frac{\lambda^{2}}{2\pi^{2}\im\tau_{2}}
    \sum_{j,j',\lambda,\lambda'} \biggl\{} -\bigl\langle
  \mathbf{X}^{j,+}_{\lambda},
  \mathbf{X}_{\lambda'}^{j',+}\bigr\rangle
  \begin{pmatrix}
    \sigma_{\lambda}^{j,+\dagger}\sigma_{\lambda'}^{j',+} & 0 \\
    0 & -\sigma_{\lambda'}^{j',+}\sigma^{j,+\dagger}_{\lambda}
  \end{pmatrix}
  \biggr\}\notag \\
  &=
  -\frac{\lambda^{2}}{2\pi^{2}\im\tau_{2}} \sum_{j,\lambda}
   \begin{pmatrix}
    \sigma_{\lambda}^{j,-}\sigma_{\lambda}^{j,-\dagger} -   \sigma_{\lambda}^{j,+\dagger}\sigma_{\lambda}^{j,+}  & 0 \\
    0 &\sigma_{\lambda}^{j,+}\sigma^{j,+\dagger}_{\lambda}- \sigma_{\lambda}^{j,-\dagger}\sigma^{j,-}_{\lambda}
  \end{pmatrix},
    \label{eq:quadratic_D_term_3}
  \end{align}
where in the second equality we have used the orthogonality property of the 
unwarped zero modes and canonically normalized our fields. 

For a non-vanishing $\epsilon$, the operator $D_\beta$ will spoil this
diagonal structure and, at this quadratic order in fluctuations,
induce a mixing between unwarped massive modes with different index
$\lambda$. In order to recover the diagonal
structure~\eqref{eq:quadratic_D_term_3} of the unwarped case, one
needs to consider a new set of modes, which are a particular linear
combination of the unwarped modes $\chi_m^\pm$.  Such new set of modes
are nothing but the linear combination of unwarped zero modes
described in sections \ref{subsec:nonchiralMM} and
\ref{subsec:warped_chiral_wfs}, which add up to build the warped zero
and massive modes in terms of the unwarped ones.

Indeed, notice that the warped $D$-term~\eqref{contam} at quadratic order in fluctuations 
can be written as
\begin{align}
  \label{eq:quadratic_D_term_1}
  D=-\frac{\lambda^{2}}{2\pi^{2}\im\tau_{2}}
  \sum_{j,j',\lambda,\lambda'}
  \biggl\{&
  \bigl\langle\mathbf{X}_{\lambda}^{j,-},\ue^{\#\alpha}
  \mathbf{X}_{\lambda'}^{j',-}\bigr\rangle
  \begin{pmatrix}
    \sigma_{\lambda}^{j,-}\sigma_{\lambda'}^{j',-\dagger} & 0 \\
    0 & -\sigma_{\lambda'}^{j',-\dagger}\sigma^{j,-}_{\lambda}
  \end{pmatrix}\notag \\
  -&\bigl\langle
  \mathbf{X}^{j,+}_{\lambda},\ue^{\#\alpha}
  \mathbf{X}_{\lambda'}^{j',+}\bigr\rangle
  \begin{pmatrix}
    \sigma_{\lambda}^{j,+\dagger}\sigma_{\lambda'}^{j',+} & 0 \\
    0 & -\sigma_{\lambda'}^{j',+}\sigma^{j,+\dagger}_{\lambda}
  \end{pmatrix}
  \biggr\},
\end{align}
where $\ue^{\#\alpha}$ is defined
in~\eqref{eq:warp_factor_matrix}. Since the equations of motion for
the warped massive modes are
\begin{equation}
  \label{eq:warped_massive_vector_eom}
  \ui\hat{\mathbf{D}}^{\mp}_{w}\bs{\Phi}_{\lambda}^{\mp}
  =\sqrt{\frac{\pi\alpha'}{2}}m_{\lambda}\ue^{\#\alpha}
  \bs{\Phi}_{\lambda}^{\pm\ast},
\end{equation}
where
\begin{equation}
  \mathbf{D}_{w}^{\mp}=\begin{pmatrix}
    0 &\hat{D}^{\mp}_{1} & \hat{D}^{\mp}_{2} & \ue^{-4\alpha}\hat{D}^{\mp}_{3} \\
    -\hat{D}^{\mp}_{1} & 0 & \bigl(\hat{D}^{\pm}_{3}\bigr)^{\ast} &
    -\bigl(\hat{D}^{\pm}_{2}\bigr)^{\ast} \\
    -\hat{D}^{\mp}_{2} &
    -\bigl(\hat{D}^{\pm}_{3}\bigr)^{\ast} & 0
    & \bigl(\hat{D}^{\pm}_{1}\bigr)^{\ast} \\
    -\ue^{-4\alpha}\hat{D}_{3}^{\mp}
    &
    \bigl(\hat{D}^{\pm}_{2}\bigr)^{\ast} &
    -\bigl(\hat{D}^{\pm}_{1}\bigr)^{\ast} & 0
  \end{pmatrix}
\end{equation}
and this can be
expressed as a Sturm-Liouville problem
\begin{equation}
  \bigl(\hat{\mathbf{D}}^{\pm}_{w}\bigr)^{\dagger}
  \bigl[\ue^{-\#\alpha}\hat{\mathbf{D}}^{\mp}_{w}\bs{\Phi}_{\lambda}^{\mp}\bigr]
  =\frac{\pi\alpha'}{2}\abs{m_{\lambda}}^{2}\ue^{\#\alpha}
  \bs{\Phi}_{\lambda}^{\mp},
\end{equation}
this guarantees that the modes are orthogonal for different values of
$\lambda$ for the scalar product in~\eqref{eq:quadratic_D_term_1}. 
We can then choose an orthogonal basis for the family index
$j$ and thus we get
\begin{align}
  \label{eq:quadratic_D_term_2}
  D=-\frac{\lambda^{2}}{2\pi^{2}\im\tau_{2}}
  \sum_{j,\lambda}
  \biggl\{&
  \bigl\langle\mathbf{X}_{\lambda}^{j,-},\ue^{\#\alpha}
  \mathbf{X}_{\lambda}^{j,-}\bigr\rangle
  \begin{pmatrix}
    \sigma_{\lambda}^{j,-}\sigma_{\lambda}^{j,-\dagger} & 0 \\
    0 & -\sigma_{\lambda}^{j,-\dagger}\sigma^{j,-}_{\lambda}
  \end{pmatrix}\notag \\
  -&\bigl\langle
  \mathbf{X}^{j,+}_{\lambda},\ue^{\#\alpha}
  \mathbf{X}_{\lambda}^{j,+}\bigr\rangle
  \begin{pmatrix}
    \sigma_{\lambda}^{j,+\dagger}\sigma_{\lambda}^{j,+} & 0 \\
    0 & -\sigma_{\lambda}^{j,+}\sigma^{j,+\dagger}_{\lambda}
  \end{pmatrix}
  \biggr\}.
\end{align}
Note that the inner product appearing in~\eqref{eq:quadratic_D_term_2}
is proportional to the K\"ahler metric appearing in the previous
subsection and so, after canonically normalizing, this gives the usual
$D$-term expression in $4$-dimensions.  

Hence, we find that the effect of (weak) warping can be considered as
a small perturbation of the $D$-term in the 4d effective field-theory.
That is, if $\phi_{m}^{\mp}$ were expanded in terms of the unwarped
massive modes, then~\eqref{eq:quadratic_D_term_1} would not simplify
to~\eqref{eq:quadratic_D_term_2} as the modes would no longer be
orthogonal.  In particular, the unwarped zero mode no longer has a
canonical $D$-term as
$\bigl\langle\tilde{\mathbf{X}}_{0}^{j,-},\ue^{\#\alpha}\tilde{\mathbf{X}}_{\lambda\neq
  0}^{j,-}\bigr\rangle\neq 0$, where $\tilde{\mathbf{X}}_{\lambda}$
here stand for the unwarped massive modes.  This warping-induced
mixing between the unwarped zero and massive modes provides a 4d
effective description of the expansion warped modes in terms of the
unwarped ones.

\section{\label{sec:conclusion}Conclusions and outlook}

In this paper we have analyzed the wavefunctions for the chiral
bifundamental degrees of freedom resulting from the open strings
stretching between intersecting $\uD 7$-branes in a warped
compactification.  While for arbitrary warp factors the equations of
motion do not seem to admit simple analytic solutions, we propose a
method for solving the wavefunctions systematically in the case of
weak warping.  Expanding the warped zero mode in terms of the massive
modes of the unwarped case, we can solve for the expansion
coefficients order by order in perturbation theory, a procedure which
we have illustrated with a few simple examples.  This analysis was
performed both with and without magnetic flux; the latter case
naturally gives rise to a chiral spectrum.\footnote{In the absence of
  worldvolume flux, chiral fermions can still be obtained if the
  7-branes are placed at a singularity.  See \cite{Marchesano:2008rg}
  for some examples of this kind.}  Such wavefunctions are necessary
for the derivation of a warped effective action for chiral fermions
via dimensional reduction.  Indeed, built on our results we take some
first steps in this direction, extending our earlier work
\cite{Marchesano:2008rg} for the adjoint matter fields.  Our results
can for instance be applied to the semi-realistic MSSM-like models of
\cite{Marchesano:2004yq, *Marchesano:2004xz}, where the MSSM sector
arises from intersecting $\uD 7$-branes, as well as other type IIB
models based on intersecting $\uD 7$-branes.

In~\cite{Marchesano:2008rg}, we were able to infer the warping
corrections to the full K\"ahler potential involving the adjoint open
string matter fields and the associated closed string degrees of
freedom, using the warped kinetic terms for the adjoint matter
\cite{Marchesano:2008rg} and the related terms for the closed string
modes \cite{Shiu:2008ry} and comparing with the unwarped results
in~\cite{Jockers:2004yj, *Jockers:2005pn}.  This approach reproduced
the warping modifications to the effective action found
in~\cite{Frey:2008xw,Martucci:2009sf}.  It would be interesting to
perform a similar analysis here, though even in the unwarped case, the
K\"ahler potential for chiral matter is only known through quadratic
order.

To illustrate our approach with explicit expressions, we 
have worked within a local framework and oftentimes considered
the $\uD 7$-branes intersection locus $\Sigma$ to be a two-torus.
A natural extension of our analysis would be to consider 
a more general matter curve, as well as a more global description of
the warped modes. In the Abelian case, the transverse scalar
$\Phi$ is globally described by a section in the normal bundle and so admits
an expansion~\cite{Jockers:2004yj}
\begin{equation}
  \Phi=\Phi^{A}\bigl(x^{\mu}\bigr)s_{A}\bigl(y^{a}\bigr)
  +\bar{\Phi}^{\bar{A}}\bigl(x^{\mu}\bigr)\bar{s}_{\bar{A}}\bigl(y^{a}\bigr),
\end{equation}
in which $\left\{s_{A}\right\}$ is a basis of the cohomology group
$H^{\left(2,0\right)}_{\bar{\partial}}\left({\cS}_{4}\right)$.  As in
our framework, in the non-Abelian case $\Phi$ is promoted to an
adjoint-valued field and the condition of $\bar{\partial}$-closure
replaced with $\bar{\partial}_{A}$-closure~\cite{Cecotti:2009zf}, and
a non-trivial intersection is also captured by giving a vev analogous
to~\eqref{eq:angle}.  All these similarities suggest that our explicit
expressions of warped zero modes in terms of unwarped modes should
hold for general intersecting $\uD 7$-branes in Calabi-Yau
compactifications, even when the matter curve is not a two-torus.  A
further generalization would be to consider not just the intersection
of $\uD 7$-branes, but the intersection of two general $7$-branes
described by more generic singularities of the F-theory fiber.
Although in~\cite{Beasley:2008dc}, an effective six-dimensional action
for such a matter curve was presented, the action did not include the
effects of warping.  While, as discussed in section~\ref{sec:setup},
such a general intersection cannot be simply described as Higgsing the
DBI action, we have argued that the corrections that we found for the
$D$-term equations of motion are still valid in the varying dilaton
case, and could in principle also hold for general F-theory setups.
Finally, as in many models the warping is sourced by bulk fluxes, it
would be useful to explore the influence of these fluxes on the open
string wavefunctions, following~\cite{Camara:2009xy,*Camara:2009zz}.

As an application of the techniques presented here, it would be useful
to revisit the problem of calculating the SUSY-breaking soft terms in
flux compactifications.  Although such soft terms were obtained from
worldsheet techniques in some previous works \cite{Lust:2004fi,
  Lust:2004dn}, in order to take into account non-trivial RR
backgrounds, one would have to resort to a dimensional reduction
analysis as in~\cite{Camara:2004jj,Benini:2009ff}.  Such an analysis,
which requires knowledge of the bi-fundamental wavefunctions, would
allow for an extension of the holographic description of gauge
mediation~\cite{Benini:2009ff,McGuirk:2009am} to include an explicit
realization of visible sector matter fields.  Finally, the
wavefunctions of the chiral bifundamental matter so obtained may also
find applications in other strong coupling extensions of the Standard
Model, such as technicolor like theories.

\acknowledgments

We would like to thank Francesco Benini, Alex Stuart and especially
Luca Martucci for useful discussions.  We also thank the Kavli
Institute for Theoretical Physics where some of this work was
completed during the ``Strings at the LHC and in the Early Universe"
workshop. PM and GS acknowledge the Institute for Advanced Study,
Princeton and the Institute for Advanced Study, Hong Kong University
of Science and Technology for hospitality while some preliminary
discussions were held, and FM would like to thank UW-Madison for
hospitality at the final stages of this work.  The work of PM and GS
was supported in part by a DOE grant DE-FG-02-95ER40896, a Cottrell
Scholar Award from Research Corporation, and a Vilas Associate Award.
PM was additionally supported by a String Vacuum Project Graduate
Fellowship, funded through NSF grant PHY-0917807.  FM is supported by
the MICINN Ram\'on y Cajal programme through the grant RYC-2009-05096,
and by the MICINN grant FPA2009-07908.

\appendix

\section{\label{app:conventions}Fermion conventions}

We make use of a Weyl basis for the $\gamma$-matrices on
$\mathbb{R}^{1,3}$
\begin{equation}
  \gamma^{\ul{\mu}}=\begin{pmatrix}
    0 & \bar{\sigma}^{\ul{\mu}} \\
    \sigma^{\ul{\mu}} & 0
  \end{pmatrix},
\end{equation}
where $\sigma^{\ul{\mu}}=\left(\mathbb{I}_{2},\bs{\sigma}\right)$
and $\bar{\sigma}^{\ul{\mu}}=\left(-\mathbb{I}_{2},\bs{\sigma}\right)$.
The 4d chirality operator is then
\begin{equation}
  \gamma_{\left(4\right)}=\ui \gamma^{\ul{0}}
  \gamma^{\ul{1}}\gamma^{\ul{2}}\gamma^{\ul{3}}
  =\begin{pmatrix}
    \mathbb{I}_{2} & 0 \\ 0 & -\mathbb{I}_{2}
  \end{pmatrix}.
\end{equation}
We also take the $\SO{6}$ $\gamma$-matrices
\begin{align}
  \tilde{\gamma}^{\ul{1}}=&\sigma_{1}\otimes\mathbb{I}_{2}\otimes\mathbb{I}_{2},&
  \tilde{\gamma}^{\ul{4}}=&\sigma_{2}\otimes\mathbb{I}_{2}\otimes\mathbb{I}_{2},
  \notag \\
  \tilde{\gamma}^{\ul{2}}=&\sigma_{3}\otimes\sigma_{1}\otimes\mathbb{I}_{2},&
  \tilde{\gamma}^{\ul{5}}=&\sigma_{3}\otimes\sigma_{2}\otimes\mathbb{I}_{2},&
  \\
  \tilde{\gamma}^{\ul{3}}=&\sig_{3}\otimes\sig_{3}\otimes\sig_{1},&
  \tilde{\gamma}^{\ul{6}}=&\sig_{3}\otimes\sig_{3}\otimes\sig_{2},\notag
\end{align}
and have the associated chirality operator
\begin{equation}
  \gamma_{\left(6\right)}=-\ui\tilde{\gamma}^{\ul{1}}
  \tilde{\gamma}^{\ul{2}}
  \tilde{\gamma}^{\ul{3}}
  \tilde{\gamma}^{\ul{4}}
  \tilde{\gamma}^{\ul{5}}
  \tilde{\gamma}^{\ul{6}}
  =\sig_{3}\otimes\sig_{3}\otimes\sig_{3}.
\end{equation}
In terms of these, we define the $\SO{1,9}$ $\Gamma$-matrices
\begin{equation}
  \Gamma^{\underline{\mu}}=\gamma^{\mu}\otimes\mathbb{I}_{8},\quad
  \Gamma^{\ul{m}}=\gamma_{\left(4\right)}\otimes\tilde{\gamma}^{m-3}.
\end{equation}
The associated chirality operator and Majorana matrix are then
\begin{align}
  \Gamma_{\left(10\right)}
  =&\Gamma^{\ul{0}}\cdots\Gamma^{\ul{10}}
  =\gamma_{\left(4\right)}\otimes\tilde{\gamma}_{\left(6\right)},\\
  \cB=&\Gamma^{\ul{2}}\Gamma^{\ul{7}}\Gamma^{\ul{8}}\Gamma^{\ul{9}}
  =\begin{pmatrix}
    0 & -\sig_{2} \\
    \sig_{2} & 0
  \end{pmatrix}
  \otimes\sigma_{2}\otimes\ui\sig_{1}\otimes\sig_{2}.
\end{align}

The Fermionic field $\theta$ appearing
in~\eqref{eq:warped_Dirac_action} and elsewhere is a 32-component 10d
spinor satisfying the Majorana and Weyl conditions
$\theta=\cB^{\ast}\theta^{\ast}$ and
$\theta=-\Gamma_{\left(10\right)}\theta$.  We thus consider spinors of
the form
\begin{subequations}
\begin{align}
  \theta_{0}=&\psi_{0}\begin{pmatrix}
    \xi_{+} \\ 0 \end{pmatrix}
  \otimes\eta_{---} -
  \ui\bigl(\psi_{0}\bigr)^{\ast}
  \begin{pmatrix}
    0 \\ \sigma_{2}\xi_{+}^{\ast}
  \end{pmatrix}
  \otimes\eta_{+++},\\
  \theta_{1}=&\psi_{1}\begin{pmatrix}
    \xi_{+} \\ 0 \end{pmatrix}
  \otimes\eta_{-++} +
  \ui\bigl(\psi_{1}\bigr)^{\ast}
  \begin{pmatrix}
    0 \\ \sigma_{2}\xi_{+}^{\ast}
  \end{pmatrix}
  \otimes\eta_{+--},\\
  \theta_{2}=&\psi_{2}\begin{pmatrix}
    \xi_{+} \\ 0 \end{pmatrix}
  \otimes\eta_{+-+} -
  \ui\bigl(\psi_{2}\bigr)^{\ast}
  \begin{pmatrix}
    0 \\ \sigma_{2}\xi_{+}^{\ast}
  \end{pmatrix}
  \otimes\eta_{-+-},\\
  \theta_{3}=&\psi_{3}\begin{pmatrix}
    \xi_{+} \\ 0 \end{pmatrix}
  \otimes\eta_{++-} +
  \ui\bigl(\psi_{3}\bigr)^{\ast}
  \begin{pmatrix}
    0 \\ \sigma_{2}\xi_{+}^{\ast}
  \end{pmatrix}
  \otimes\eta_{++-}.
\end{align}
\end{subequations}
Here,
\begin{equation}
  \eta_{\epsilon_{1}\epsilon_{2}\epsilon_{3}}
  =\eta_{\epsilon_{1}}\otimes\eta_{\epsilon_{2}}
  \otimes\eta_{\epsilon_{3}},\qquad
  \eta_{+}=\begin{pmatrix} 1 \\ 0 \end{pmatrix},\qquad
  \eta_{-}=\begin{pmatrix} 0 \\ 1\end{pmatrix}.
\end{equation}
Note that $\sig_{3}\eta_{\epsilon}=\epsilon\eta_{\epsilon}$.  Defining 
the complex coordinates as
\begin{equation}
  z^{m}=y^{m}+\ui y^{m+3},
\end{equation}
we have
\begin{align}
  \tilde{\gamma}^{\ul{1}}=&
  \sigma^{+}\otimes\mathbb{I}_{2}\otimes\mathbb{I}_{2},
  &\tilde{\gamma}^{\bar{\ul{1}}}=&
  \sigma^{-}\otimes\mathbb{I}_{2}\otimes\mathbb{I}_{2},\notag \\
  \tilde{\gamma}^{\ul{2}}=&
  \sigma_{3}\otimes\sigma^{+}\otimes\mathbb{I}_{2},
  &\tilde{\gamma}^{\bar{\ul{2}}}=&
  \sig_{3}\otimes\sigma^{-}\otimes\mathbb{I}_{2},\\
  \tilde{\gamma}^{\ul{3}}=&
  \sigma_{3}\otimes\sigma_{3}\otimes\sigma^{+},
  &\tilde{\gamma}^{\bar{\ul{2}}}=&
  \sig_{3}\otimes\sig_{3}\otimes\sigma^{-},\notag
\end{align}
with
\begin{equation}
  \sigma^{+}:=\begin{pmatrix}
    0 & 2 \\
    0 & 0
  \end{pmatrix},\quad
  \sigma^{-}:=\begin{pmatrix}
    0 & 0 \\ 2 & 0\end{pmatrix}.
\end{equation}
which satisfy
\begin{equation}
  \sigma^{\pm}\eta_{\pm}=0,\quad
  \sigma^{\pm}\eta_{\mp}=2\eta_{\pm}.
\end{equation}

\section{\label{app:bosonic_eom}Equations of motion from the Myers
  action}

In this appendix we check that the equations of motion for the
warped zero mode that were deduced by imposing the BPS
conditions~\eqref{eq:unmagnetized_bps_eom_2} imply those that follow
from the DBI and CS actions.  In the 10d Einstein frame, the
non-Abelian generalization of this action, which is appropriate for
describing multiple $\uD 7$-branes, is~\cite{Myers:1999ps}
\begin{subequations}
\label{eq:bosonic_action}
\begin{align}
  S_{\uD 7}=&S_{\uD 7}^{\mathrm{DBI}}+S_{\uD 7}^{\mathrm{CS}},\\
\label{eq:DBI}
  S_{\uD 7}^{\mathrm{DBI}}=&
  -\tau_{\uD 7}\int_{\cW}\ud^{8}x\,
  \Str\biggl\{\left(\im\tau\right)^{-1}
  \sqrt{\det M_{\alpha\beta}\det Q^{i}_{\, j}}\biggr\},\\
  S_{\uD 7}^{\mathrm{CS}}=&
  \tau_{\uD 7}\int_{\cW}\Str\biggl\{
  P\biggl[\ue^{\ui\lambda\iota_{\Phi}\iota_{\Phi}}
  \mathcal{C}\wedge \ue^{B_{2}}\biggr]
  \ue^{\lambda F_{2}}\biggr\},
\end{align}
\end{subequations}
where $\tau$ is the axio-dilaton, $\mathcal{C}$ is the formal sum of
all of the RR-potentials, $B_{2}$ is the NS-NS $2$-form potential, and
$F_{2}=\bigl(\ud-\ui A\wedge\bigr) A$ is the worldvolume field
strength and $\lambda=2\pi\alpha'$.  In terms of the deformation
moduli $\Phi^{i}=\lambda^{-1}X^{i}$, the tensor $Q$ is given by
\begin{equation}
  Q^{i}_{\, j}=\delta^{i}_{\, j}
  -\ui\lambda\bigl[\Phi^{i},\Phi^{k}\bigr]\bigl(\im\tau)^{-1/2}E_{kj},
\end{equation}
where the NS-NS rank $2$ tensor is the sum of the metric and the NS-NS
$2$-form potential
\begin{equation}
  E_{MN}=g_{MN}+\bigl(\im{\tau}\bigr)^{1/2}B_{MN},
\end{equation}
with $i, j$ transverse to the brane.  In terms of these
\begin{equation}
  M_{\alpha\beta}=P\bigl[E_{\alpha\beta}
  +\bigl(\im{\tau}\bigr)^{-1/2}E_{\alpha i}
  \bigl(Q^{-1}-\delta\bigr)^{ij}E_{j\beta}\bigr]
  +\lambda\bigl(\im\tau\bigr)^{1/2}F_{\alpha\beta},
\end{equation}
where the transverse indices are raised and lowered with $E^{-1}$ and
$E$.  $\iota_{\Phi}$ denotes the interior product which acts on a
$1$-form $\omega$ as
\begin{equation}
  \iota_{\Phi}\omega=\Phi^{i}\omega_{i}.
\end{equation}
Note that since generally the $\Phi^{i}$ are non-commuting,
$\iota_{\Phi}^{2}$ does not identically vanish.  On the worldvolume,
the closed-string fields are to be interpreted as a non-Abelian Taylor
expansion,
\begin{equation}
  \Psi=\sum_{n=0}^{\infty}\frac{\lambda^{n}}{n!}
  \Phi^{i_{1}}\cdots\Phi^{i_{n}}
  \bigl[\partial_{i_{1}}\cdots\partial_{i_{n}}\Psi\bigr]_{X^{i}=0},
\end{equation}
and pullbacks involve the covariant derivative
\begin{equation}
  P\bigl[v_{\alpha}\bigr]=v_{\alpha}+
  \lambda \bigl(D_{\alpha}\Phi^{i}\bigr)v_{i}.
\end{equation}
$\Str$ indicates that the trace is to be taken only after
symmetrization over $F_{\alpha\beta}$, $D_{\alpha}\Phi^{i}$,
$\bigl[\Phi^{i},\Phi^{j}\bigr]$, and the individual $\Phi^{i}$
appearing in non-Abelian Taylor expansions of closed string fields.
Finally, the $\uD 7$-brane tension is $\tau_{\uD
  7}^{-1}=8\pi^{3}\lambda^{4}g_{\mathrm{s}}^{-1}$.

Let us for simplicity now consider a warped
compactification~\eqref{eq:warpedmetric} such that the warping is
supported by~\eqref{eq:5form} and all other closed strings are trivial
(in particular we choose the dilaton to take a constant value). The
bosonic part of the Super Yang-Mills action is recovered after
expanding to leading non-trivial order in $\lambda$.  Using the
identity
\begin{equation}
  \det\left(1+\delta M_{1}+\delta^{2}M_{2}\right)
  =1+\delta\tr M_{1}+\delta^{2} M_{2}+
  \frac{\delta^{2}}{2}\left(\tr M_{1}\right)^{2}
  -\frac{\delta^{2}}{2}\tr\left(M_{1}^{2}\right)+
  \mathcal{O}\left(\delta^{3}\right),
\end{equation}
we get
\begin{align}
  S_{\mathrm{D7}}^{\mathrm{DBI}}
  =-\frac{1}{g_{8}^{2}}\int_{\cW}\ud^{8}x\,\sqrt{g}
  \tr\biggl\{&
  \frac{1}{\lambda^{2}}
  +\frac{1}{4}g^{\alpha\beta}g^{\gamma\delta}
  F_{\alpha\gamma}F_{\beta\delta}
  +\frac{1}{2}g^{\alpha\beta}g_{ij}D_{\alpha}\Phi^{i}
  D_{\beta}\Phi^{j}\notag \\
  &-\frac{1}{4}g_{ij}g_{kl}
  \bigl[\Phi^{i},\Phi^{k}\bigr]
  \bigl[\Phi^{j},\Phi^{l}\bigr]\biggr\},
\end{align}
where $g_{8}^{2}=\lambda^{2}\tau_{\uD 7}$. Here we have taken
$g_{\alpha\beta}$ to be the standard pull-back of the metric on ${\cal
  S}_4$, ignoring non-Abelian effects.  As argued in the main text,
this approximation is justified in the limit of small intersection
angles, and can be handled beyond this approximation as discussed in
Appendix~\ref{app:corrections}.  Note that then the metric carries no
additional factors of $\Phi^{i}$ and every object in the trace is
already symmetrized in the sense described above.

For the CS action, only $C_{4}$ is present in $\mathcal{C}$.  Since
the integral picks out $8$-forms and the interior derivative
decreases the rank of the form on which it acts, we have
\begin{equation}
  S_{\mathrm{\uD 7}}^{\mathrm{CS}}
  =\frac{1}{2g_{8}^{2}}\int_{\cW}
  \tr\biggl\{C_{4}\wedge F_{2}\wedge F_{2}\biggr\},
\end{equation}
where again $C_{4}$ is taken to be the standard pull-back on ${\cal
  S}_4$.  Note that in the case of a varying axio-dilaton, one would
have to include contributions from $C_{8}$, the magnetic dual of the
axion.

The equations of motion that lead to a stationary action are
\begin{subequations}
\begin{align}
  0=&\frac{1}{\sqrt{-g}}
  D_{\gamma}\bigl(\sqrt{-g}g^{\alpha\beta}g^{\gamma\delta}
  F_{\delta\beta}\bigr)
  +\ui
  g^{\alpha\beta}g_{ij}\bigl[\Phi^{i},D_{\beta}\Phi^{j}\bigr]
  +\frac{1}{2\cdot 4!\sqrt{-g}}
  \epsilon^{\alpha\beta\gamma\delta\epsilon\eta\zeta\theta}
  D_{\beta}\bigl(C_{\epsilon\eta\zeta\theta}F_{\gamma\delta}\bigr),\\
  0=&\frac{1}{\sqrt{-g}} D_{\alpha}\bigl[\sqrt{-g}
  g^{\alpha\beta}g_{ij}D_{\beta}\Phi^{j}\bigr]
  +g_{ij}g_{kl}\bigl[\Phi^{k},\bigl[\Phi^{j},\Phi^{l}\bigr]\bigr].
\end{align}
\end{subequations}
Specializing now to the case of local flat coordinates,
so that the warped K\"ahler form is given by~\eqref{eq:warped_kaehler_form}, 
the equations of motion read
\begin{subequations}
\label{eq:DBI_eom}
\begin{align}
  \label{eq:DBI_eom_maxwell}
  0=&\eta^{\rho\sigma}D_{\rho}F_{\sigma\mu}+\ue^{4a}
  \frac{2}{\alpha'}
  \sum_{m=1}^{2}
  \frac{1}{\left(2\pi R_{m}\right)^{2}}
  \bigl(D_{\bar{m}}F_{m\mu}+D_{m}F_{\bar{m}\mu}\bigr)\notag \\
  &+\frac{\ui\alpha'}{2}\left(2\pi R_{3}\right)^{2}
  \bigl(\bigl[\Phi,D_{\mu}\bar{\Phi}\bigr]+
  \bigl[\bar{\Phi},D_{\mu}\Phi\bigr]\bigr),\\
  \label{eq:DBI_eom_wilsonline1}
  0=&\eta^{\mu\nu}D_{\mu}F_{\nu 1}
  +\ue^{4a}\frac{2}{\alpha'}
  \biggl\{\frac{1}{\left(2\pi R_{1}\right)^{2}}D_{1}F_{\bar{1}1}+
  \frac{1}{\left(2\pi R_{2}\right)^{2}}
  \bigl(D_{2}F_{\bar{2}1}+D_{\bar{2}}F_{21}\bigr)\biggr\}\notag \\
  &+\ue^{4a}\frac{2}{\alpha'}
  \biggl\{4\partial_{1}a\biggl(
  \frac{1}{\left(2\pi R_{1}\right)^{2}}F_{\bar{1}1}+
  \frac{1}{\left(2\pi R_{2}\right)^{2}}F_{\bar{2}2}\biggr)
  +8\partial_{\overline{2}}a\frac{1}{\left(2\pi R_{2}\right)^{2}}
  F_{21}\biggr\}\notag \\
  &+\frac{\ui\alpha'}{2}
  \bigl(2\pi R_{3}\bigr)^{2}
  \bigl(\bigl[\Phi,D_{1}\bar{\Phi}\bigr]
  +\bigl[\bar{\Phi},D_{1}\Phi\bigr]\bigr),\\
  \label{eq:DBI_eom_wilsonline2}
  0=&\eta^{\mu\nu}D_{\mu}F_{\nu 2}
  +\ue^{4a}\frac{2}{\alpha'}
  \biggl\{\frac{1}{\left(2\pi R_{2}\right)^{2}}D_{2}F_{\bar{2}2}+
  \frac{1}{\left(2\pi R_{1}\right)^{2}}
  \bigl(D_{1}F_{\bar{1}2}+D_{\bar{1}}F_{12}\bigr)\biggr\}\notag \\
  &+\ue^{4a}\frac{2}{\alpha'}
  \biggl\{4\partial_{2}a\biggl(
  \frac{1}{\left(2\pi R_{1}\right)^{2}}F_{\bar{1}1}+
  \frac{1}{\left(2\pi R_{2}\right)^{2}}F_{\bar{2}2}\biggr)
  +8\partial_{\overline{1}}a\frac{1}{\left(2\pi R_{2}\right)^{2}}
  F_{12}\biggr\}\notag \\
  &+\frac{\ui\alpha'}{2}
  \bigl(2\pi R_{3}\bigr)^{2}
  \bigl(\bigl[\Phi,D_{2}\bar{\Phi}\bigr]
  +\bigl[\bar{\Phi},D_{2}\Phi\bigr]\bigr),\\
  \label{eq:DBI_eom_modulus}
  0=&\eta^{\mu\nu}D_{\mu}D_{\nu}\Phi
  +\ue^{4a}\frac{2}{\alpha'}
  \sum_{m=1}^{2}\frac{1}{\left(2\pi R_{m}\right)^{2}}
  \biggl\{D_{m},D_{\bar{m}}\biggr\}\Phi
  +\frac{\alpha'}{2}\left(2\pi R_{3}\right)^{2}
  \bigl[\Phi,\bigl[\Phi,\bar{\Phi}\bigr]\bigr].
\end{align}
\end{subequations}

In general, this set of coupled second order differential equations is
difficult to solve.  However, we can show that when the $F$-flatness
and $D$-flatness conditions are satisfied, the equations of motion are
satisfied as well.  Indeed, as we are interested in zero modes we set the
first term in each of~\eqref{eq:DBI_eom} to zero.  Furthermore, our
interest is only on the bifundamental fields and so after writing the
fluctuations as in~\eqref{eq:flucs}, we set the block-diagonal entries
to zero.  Since the non-trivial angle~\eqref{eq:angle} and magnetic
flux~\eqref{eq:magnetic_flux} Higgs the gauge group down, we do not expect
a massless bifundamental vector boson and set $A^{\mp}_{\mu}=0$.
Choosing again the background gauge field to take the
form~\eqref{eq:connection} and parameterizing the fluctuations
as~\eqref{eq:param_flucs}, from~\eqref{eq:DBI_eom_maxwell}, we get
\begin{align}
  \label{eq:DBI_eom_maxwell_2}
  0=&\partial_{\mu}\biggl\{\hat{D}_{1}^{\mp}\phi_{1}^{\mp}
  + \hat{D}_{2}^{\mp}\phi_{2}^{\mp}
  + 
  \ue^{-4a}\hat{D}_{3}^{\mp}\phi_{3}^{\mp}\biggr\}\notag \\
  &-\partial_{\mu}\biggl\{\bigl(\hat{D}^{\pm}_{1}\bigr)^{\dagger}
  \bigl(\phi_{1}^{\pm}\bigr)^{\dagger}+
  \bigl(\hat{D}^{\pm}_{2}\bigr)^{\dagger}
  \bigl(\phi_{2}^{\pm}\bigr)^{\dagger}+\ue^{-4a}
  \bigl(\hat{D}^{\pm}_{3}\bigr)^{\dagger}
  \bigl(\phi_{3}^{\pm}\bigr)^{\dagger}\biggr\},
\end{align}
where the covariant derivatives are defined
in~\eqref{eq:magnetized_covariant_derivs}.
Clearly~\eqref{eq:DBI_eom_maxwell_2} is satisfied in either the
unmagnetized or magnetized case whenever the
$D$-term~\eqref{eq:unmagnetized_d_term} is satisfied.

For the remaining equations, we first impose the $F$-term conditions.
Namely we take $F$ to be purely $\left(1,1\right)$, impose the
self-duality constraint~\eqref{eq:self-dual} and demand that $\Phi$ is
holomorphic in the sense that $D_{\bar{m}}\Phi=0$.  Then the remaining
equations can be cast in the form
\begin{align}
  0=&\hat{D}^{\pm}_{m}\biggl\{\hat{D}_{1}^{\mp}\phi_{1}^{\mp} +
  \hat{D}_{2}^{\mp}\phi_{2}^{\mp} +
  \ue^{-4a}\hat{D}_{3}^{\mp}\phi_{3}^{\mp}\biggr\}\notag \\
  &-\bigl(\hat{D}^{\mp}_{m}\bigr)^{\dagger}
  \biggl\{\bigl(\hat{D}^{\pm}_{1}\bigr)^{\dagger}
  \bigl(\phi_{1}^{\pm}\bigr)^{\dagger}+
  \bigl(\hat{D}^{\pm}_{2}\bigr)^{\dagger}
  \bigl(\phi_{2}^{\pm}\bigr)^{\dagger}+\ue^{-4a}
  \bigl(\hat{D}^{\pm}_{3}\bigr)^{\dagger}\bigl(\phi_{3}^{\pm}\bigr)^{\dagger}\biggr\},
\end{align}
where $m=1,2,3$ for~\eqref{eq:DBI_eom_wilsonline1},
\eqref{eq:DBI_eom_wilsonline2}, and~\eqref{eq:DBI_eom_modulus}
respectively.  Thus, when the $F$-flatness and $D$-flatness conditions
are satisfied so are the equations of motion~\eqref{eq:DBI_eom} for
the zero mode.

Note that for simplicity, in this section we have only considered the
constant dilaton case as was the case when we considered the equations
of motion~\eqref{eq:bosonic_zero_mode_eqs} following from the
fermionic part of the action.  However, we expect that even in the
case of a holomorphically varying axio-dilaton, the $D$-flatness and
$F$-flatness conditions will continue to imply the equations of motion
following from the bosonic action.

\section{\label{app:corrections}Large angle corrections}

While holomorphy forbids the existence of $\alpha'$-corrections to the
superpotential, the $D$-terms enjoy no such protection.  As a
consequence, the $D$-flatness conditions presented in
section~\ref{sec:setup}, and hence the equations of motion, receive
corrections when these effects are taken into account. In the T-dual
picture of magnetized $\uD 9$-branes the corrections are negligible in
the limit of diluted worldvolume fluxes, and here we find that the
corrections are suppressed by small angles between the $\uD 7$-branes.

The absence of corrections to the superpotential is expected on
general grounds but we can see it directly from~\eqref{eq:na_superpotential} as well.  
The only appearance of
$\alpha'$ is with the interior derivatives and the worldvolume field
strength $F_{2}$.  However, since we can choose $\gamma=z^{3}\ud z^{1}\wedge\ud z^{2}$ as
in the main text, $\iota_\Phi \gamma = 0$ and so 
$\ue^{\ui\lambda\iota_{\Phi}\iota_{\Phi}}\gamma=\gamma$ is a $2$-form.
Then  the integral is saturated by a single power of $F_{2}$ and
no further factors of $\alpha'$ follow from $\ue^{\lambda F_{2}}$.
Finally, since $\gamma$ has no legs transverse to $\cS_{4}$, the
pullback is trivial and no $\alpha'$ corrections follow from
$D_{\alpha}\Phi^{i}$ terms.  That is,~\eqref{eq:particular_W} is exact
to all orders in $\alpha'$.

The $D$-term, however, is not protected from $\alpha'$-corrections as is
immediate from~\eqref{eq:D_term_expanded}
\begin{equation}
  \label{eq:D_term_expanded_2}
  D=\int_{\mathcal{S}_{4}}\mathrm{S}\biggl\{
  \ue^{2\alpha}\biggl(\lambda P\bigl[J\bigr]\wedge F_{2}
  -\frac{i\lambda}{6}P\bigl[\iota_{\Phi}\iota_{\Phi} J^{3}\bigr]
  +\frac{i\lambda^{3}}{2}P\bigl[\iota_{\Phi}\iota_{\Phi} J\bigr]
  \wedge F_{2}\wedge F_{2}\biggr)\biggr\},
\end{equation}
where the warped K\"ahler form is~\eqref{eq:warped_kaehler_form} and
$\lambda=2\pi\alpha'$.  Consider the first term
of~\eqref{eq:D_term_expanded_2}.  In the analysis of the main text, we
dropped the derivative terms in the pull-back which are higher order
in $\alpha'$.  Incorporating them we find
\begin{multline}
  \ue^{2\alpha}P\bigl[J\bigr]\wedge F_{2} =\\\frac{\ui\alpha'}{2}
  \biggl\{\biggl[\bigl(2\pi R_{1}\bigr)^{2} +\lambda^{2}\bigl(2\pi
  R_{3}^{2}\bigr) D_{1}\bar{\Phi}\bar{D}_{\bar{1}}\Phi\biggr]
  F_{2\bar{2}}+ \biggl[\bigl(2\pi R_{2}\bigr)^{2}
  +\lambda^{2}\bigl(2\pi R_{3}^{2}\bigr)
  D_{2}\bar{\Phi}\bar{D}_{\bar{2}}\Phi\biggr]
  F_{1\bar{1}}\biggr\}\ud^{4}z.
\end{multline}
Note that in the $\uD 9$-picture, these additional terms are $F^{3}$
corrections. In writing this expression, we have imposed the $F$-term
condition $\bar{D}_{\bar{m}}\Phi=0$.  These $\alpha'$ corrections are
suppressed by the small angle between the branes.  Indeed,
taking~\eqref{eq:angle}, we have
\begin{equation}
  \mathrm{S}\biggl\{\ue^{2\alpha}P\bigl[J\bigr]\wedge F_{2}\biggr\}
  =\frac{\ui\alpha'}{2}
  \biggl\{\left(2\pi R_{1}\right)^{2}F_{2\bar{2}}
  +\left(2\pi R_{2}\right)^{2}\biggl[1+\frac{1}{3}
  \left(\frac{R_{3}M}{R_{2}}\right)^{2}\biggr]F_{1\bar{1}}\biggr\}\ud^{4}z,
\end{equation}
where for simplicity of presentation we have taken the special case
\begin{equation}
  \label{eq:simple_angle}
  M_{3}^{\left(a\right)}=-M_{3}^{\left(b\right)}=M.
\end{equation}
The factor of $\frac{1}{3}$ comes from the symmetrization procedure
which we discuss further below. The ratio $R_{3}M/R_{2}$ is the
tangent of half the angle between the intersecting $\uD 7$-branes and
when it is small, as was assumed throughout the analysis, it can be
neglected.  This suggests that the correction is a result of the fact
that the two stacks are not wrapping $\cS_{4}$, but are actually
wrapping different $4$-cycles.  At small angles, this correction can
be neglected as long as local questions are considered.

Considering now the second term in~\eqref{eq:D_term_expanded_2}, the
interior derivative acting on $J^{3}$ removes the legs transverse to
$\cS_{4}$ and so the pullback is trivial.  However, unlike the first
and third terms of~\eqref{eq:D_term_expanded_2}, the warp factor does
not cancel and further $\alpha'$ arise from the non-Abelian Taylor
expansion discussed below.

The final term in~\eqref{eq:D_term_expanded_2} is already explicitly
an $\alpha'$-correction to the $D$-flatness condition and in fact no
other powers of $\alpha'$ appear other than those appearing explicitly
in~\eqref{eq:D_term_expanded_2}.  Again the interior derivatives on
$J$ strip the legs from $J$ so the pullback is trivial.  Additionally,
the leading warp factor $\ue^{2\alpha}$ cancels the factor appearing
in $J$ so that there are no corrections from the non-Abelian Taylor
expansion.

Note that all three $\alpha'$-corrected terms appearing
in~\eqref{eq:D_term_expanded_2} contain additional open-string fields $\Phi$.
Thus, while at leading order in $\alpha'$ applying the symmetrization
procedure of~\cite{Myers:1999ps} to the $D$-term was trivial, at
sub-leading order the symmetrization must be taken into account.
Finally we note that much of the simplification of the
$\alpha'$-corrections was a result of the fact of that the K\"ahler
form~\eqref{eq:warped_kaehler_form} only depended on the transverse
coordinates through the warp factor.  For more general
compactifications, this will not be the case and further
$\alpha'$-corrections will result.

Among the $\alpha'$ corrections are the higher order terms in the
non-Abelian Taylor expansion.  From the point of view of the $\uD
7$-branes, $\ue^{-4\alpha}$ should be interpreted in terms of a
non-Abelian Taylor expansion~\eqref{eq:non_abelian_taylor_expansion}.
In particular, in the bulk the warp factor can be locally written as
\begin{equation}
  \ue^{-4\alpha}=\sum_{nm}c_{nm}\bigl(z^{3}\bigr)^{n}
  \bigl(\bar{z}^{\bar{3}}\bigr)^{\bar{m}},
\end{equation}
where $c_{nm}$ are functions of $z^{1}$, $z^{2}$, and their
conjugates.  Then on the worldvolume,
\begin{equation}
  \label{eq:warp_factor_non_abelian_taylor_expansion}
  \ue^{-4\alpha}=\sum_{nm}c_{nm}\lambda^{n+m}\Phi^{n}\bar{\Phi}^{m}.
\end{equation}

In the main text, we considered the small angle limit where
$M_{3}^{\left(a,b\right)}$ are small so that this sum can be truncated
after the zeroth order term.  However, taking into account these
higher order terms does not change the essential procedure.  The warp
factor appearing in the $D$-flatness
condition~\eqref{eq:unmagnetized_bps_eom_2d} comes from the second
term in~\eqref{eq:D_term_expanded_2}
\begin{equation}
  \frac{\lambda^{2}}{4}\left(2\pi R_{3}^{2}\right)\int_{\cS_{4}}
  \mathrm{S}\biggl\{\ue^{-4\alpha}\tilde{\mathfrak{J}}^{2}
  \bigl[\Phi,\bar{\Phi}\bigr]\biggr\},
\end{equation}
where $\mathfrak{\tilde{J}}$ is the unwarped version
of~\eqref{eq:warped_kahler_form_restricted}.  Thus, to take into account the
higher order terms
in~\eqref{eq:warp_factor_non_abelian_taylor_expansion}, we need
to consider
\begin{equation}
  \sum_{nm}c_{nm}\lambda^{n+m}
  \uS\biggl\{\Phi^{n}\bar{\Phi}^{m}\bigl[\Phi,\bar{\Phi}\bigr]\biggr\}.
\end{equation}
Expanding to linear order in fluctuations gives
\begin{equation}
  \label{eq:Dtermworking_2}
  \frac{2}{\sqrt{2\pi}R_{3}\lambda}\sum_{n,m}c_{nm}\lambda^{n+m}
  \uS\biggl\{\vev^{m}\bar{\vev}^{n}\bigl(
  \bigl[\vev,\bar{\phi}\bigr]+\bigl[\phi,\bar{\vev}\bigr]\bigr)\biggr\}.
\end{equation}
Writing
\begin{equation}
  \vev=\frac{z^{2}}{\lambda}N,\quad
  N=\begin{pmatrix}
    M_{3}^{\left(a\right)}\mathbb{I}_{N_{a}} & \\
    & M_{3}^{\left(b\right)}\mathbb{I}_{N_{b}},\end{pmatrix},\quad
  F=\frac{2}{\sqrt{2\pi}R_{3}\lambda}\bigl[\vev,\bar{\phi}\bigr],
\end{equation}
we have that~\eqref{eq:Dtermworking_2} can be written as
\begin{equation}
  \label{eq:Dtermworking_3}
  \sum_{nm}c_{nm}
  \bigl(z^{3}\bigr)^{n}\bigl(\bar{z}^{3}\bigr)^{m}
  \mathrm{S}\biggl\{N^{n+m}\bigl(F-\bar{F}\bigr)\biggr\}.
\end{equation}
The symmetrization procedure of~\cite{Myers:1999ps} requires that we
symmetrize over each factor of $N$ and $F$.  For example,
\begin{equation}
  \label{eq:symmetrization_ex}
  \mathrm{S}\biggl\{N^{4}F\biggr\}
  =\frac{4!}{5!}
  \biggl\{N^{4}F+N^{3}FN+N^{2}FN^{2}+NFN^{3}+FN^{4}\biggr\}.
\end{equation}
The factor of $5!$ accounts for each of the different ways to permute
each of the $5$ objects ($F$ and $4$ copies of $N$) and $4!$ counts
the number of ways that in each term in~\eqref{eq:symmetrization_ex}
the $N$s can be arranged.  Defining
\begin{equation}
  T=\begin{pmatrix}
    \mathbb{I}_{N_{a}} & 0 \\
    0 & -\mathbb{I}_{N_{b}}
  \end{pmatrix},
\end{equation}
we can write
\begin{equation}
  N=\frac{1}{2}\bigl(K_{3}^{\left(ab\right)}+I_{3}^{\left(ab\right)}T\bigr),
\end{equation}
where
\begin{equation}
  K_{3}^{\left(ab\right)}=M_{3}^{\left(a\right)}+M_{3}^{\left(b\right)},
\end{equation}
and $I_{3}^{\left(ab\right)}$ is again as in~\eqref{eq:define_I3}.
One can easily show that when only the bifundamental modes are
considered,
\begin{equation}
  S\bigl\{T^{a}F\bigr\}=
  \begin{cases}
    0 & a=2\ell+1,\\
    \frac{1}{a+1}F & a=2\ell
  \end{cases},\quad\ell\in\mathbb{Z}.
\end{equation}
Thus,~\eqref{eq:Dtermworking_3} takes the form
\begin{equation}
  \ue^{-4\hat{\beta}}\frac{2}{\sqrt{2\pi}R_{3}\lambda}
  \bigl(
  \bigl[\vev,\bar{\phi}\bigr]+\bigl[\phi,\bar{\vev}\bigr]\bigr),
\end{equation}
where
\begin{align}
  \label{eq:define_gamma}
  \ue^{-4\hat{\beta}}=\sum_{nm}
  \frac{c_{nm}\bigl(z^{2}\bigr)^{n}\bigl(\bar{z}^{\bar{2}}\bigr)^{m}}
  {2^{n+m}}
  \biggl\{&
  \bigl(K_{3}^{\left(ab\right)}\bigr)^{n+m}
  +\frac{1}{n+m+1}\bigl(I_{3}^{\left(ab\right)}\bigr)^{n+m}
  \delta_{n+m,2\ell}+\notag \\
   &+\frac{1}{m+1}\bigl(K_{3}^{\left(ab\right)}\bigr)^{n}
  \bigl(I_{3}^{\left(ab\right)}\bigr)^{m}\delta_{m,2\ell}
  + \frac{1}{n+1}\bigl(I_{3}^{\left(ab\right)}\bigr)^{n}
  \bigl(K_{3}^{\left(ab\right)}\bigr)^{m}\delta_{n,2\ell}\biggr\}.
\end{align}
The $z^{3}$ dependence of the warp factor can then be taken into
account by using the methods of section~\ref{sec:nonchiral}
and~\ref{sec:chiral} but with the substitution
$\ue^{-4\alpha}\to\ue^{-4\hat{\beta}}$.

This expression simplifies considerably in the special
case~\eqref{eq:simple_angle}.
Indeed, then $K_{3}^{\left(ab\right)}=0$ and~\eqref{eq:define_gamma}
becomes
\begin{equation}
  \label{eq:gamma_2}
  \ue^{-4\hat{\beta}}=\sum_{n+m=2k}\frac{c_{nm}M^{2k}}{2k+1}\left(z^{2}\right)^{n}
  \bigl(\bar{z}^{\bar{2}}\bigr)^{m}.
\end{equation}
The corrections considered here should arise when we move away from
the small angle approximation.  That is,
truncating~\eqref{eq:define_gamma} and~\eqref{eq:gamma_2} amounts to
neglecting some of the $z^{3}$ and $\bar{z}^{\bar{3}}$ dependence of
the warp factor in the equations of motion.  When the angle is small,
the branes are at approximately constant $z^{3}$ and this expansion is
justified, but when the angle is large, the $z^{3}$ dependence could
become important.  However, what apparently appears as the expansion
parameter in~\eqref{eq:gamma_2} is not quite the angle which should be
determined using the physical distances $\ud w^{m}=\left(2\pi
  R_{m}\right)\ud z^{m}$ and so is approximately $2R_{3}M/R_{2}$ when
it is small.  Instead, what explicitly appears as the expansion
parameter in~\eqref{eq:gamma_2} is $M$ without the accompanying radii.
Since we could change the physical angles by changing $R_{m}$ but
leaving $M$ fixed, there must be some hidden dependence appearing on
the radii in~\eqref{eq:gamma_2} as otherwise these corrections would
not depend on the physical distances between the branes.  Indeed, this
apparent confusion is simply an artifact of using the coordinates
$z^{m}$ throughout and the radii reappear if we use physical
distances.  The warp factor should be more naturally expressed in
terms of these physical scales and so one expects that $c_{nm}\sim
R_{3}^{n+m}$.  Then when~\eqref{eq:gamma_2} is expressed in terms of
$w^{m}$, we find
\begin{equation}
  \ue^{-4\hat{\beta}}=\sum_{n+m=2k}\frac{\tilde{c}_{nm}}{2k+1}
  \left(\frac{R_{3}M}{R_{2}}\right)^{2k}\bigl(w^{2}\bigr)^{n}
  \bigl(\bar{w}^{\bar{2}}\bigr)^{m},\quad
  \tilde{c}_{nm}=c_{nm}/R_{3}^{n+m},
\end{equation}
which makes manifest that the expansion truly is a small angle
expansion.  For comparison, note that the exact
solution~\eqref{eq:unwarped_solution} is expressed in terms of the
physical lengths as $\ue^{-q \absb{w^{2}}^{2}}$ where $q\sim
R_{3}M/R_{2}$ is proportional to the angle.

As a simple example, we consider a modification
of~\eqref{eq:quadratic_warp_factor}
\begin{equation}
  \ue^{-4\alpha}=1+\epsilon \ell^{-2}\bigl(R_{2}^{2}\abs{z^{2}}^{2}+
  R_{3}^{2}\abs{z^{3}}^{2}\bigr)
  =1+\epsilon L^{2}\left[\abs{z^{2}}^{2}+
  \left(\frac{R_{3}\abs{z^{3}}}{R_{2}}\right)^{2}\right].
\end{equation}
for which $c_{00}=1+\epsilon L^{-2}\abs{z^{2}}^{2}$ and
$c_{11}=\epsilon \left(LR_{3}/R_{2}\right)^{-2}$. For the simple
case~\eqref{eq:simple_angle}
we get
\begin{equation}
  \ue^{-4\hat{\beta}}=1+\epsilon L^{-2}\left[1+\frac{1}{3}
  \left(\frac{R_{3}M}{R_{2}}\right)^{2}\right]
  \abs{z^{2}}^{2}.
\end{equation}
The solutions~\eqref{eq:nonchiral_quadratic_solution}
and~\eqref{eq:chiral_quadratic_solution} still hold with the replacement
\begin{equation}
  L^{-2}\to L^{-2}\left[1+\frac{1}{3}
  \left(\frac{R_{3}M}{R_{2}}\right)^{2}\right].
\end{equation}

Note that if we are to take into account this correction, we must also
take into account the corrections to the first term
of~\eqref{eq:D_term_expanded_2}, while the third term has an
additional $\alpha'$ suppression and so can still be neglected.  This
modifies the D-term equation~\eqref{eq:unmagnetized_bps_eom_2d} by the
replacement $\hat{D}_{1}^{\mp}\to \left(1+t\right)D_{1}^{\mp}$ where
$t=\frac{1}{3}\left(R_{3}M/R_{2}\right)^{2}$.  This amounts to correcting the
same factor of $\hat{D}_{1}^{\mp}$ appearing in the first row
of~\eqref{eq:vector_diff_op}. For the first order corrections to the
warped zero mode, this can be accounted for by simply mapping
$\bs{\Phi}_{1;mnlp}^{\mp}\to \bigl(1+t\bigr)\bs{\Phi}_{1;mnlp}^{\mp}$
with an analogous statement in the magnetized case.

\section{\label{app:examples}Exact solutions for toy warp factors}

Although in general the equations of
motion~\eqref{eq:unmagnetized_bps_eom_2}
and~\eqref{eq:bosonic_zero_mode_eqs} for the warped zero mode cannot
be solved exactly, in special cases a simple analytic solution does
exist.  In this appendix we briefly present some of these solutions.
Setting $\phi_{0}^{\mp}=0$ and taking the
ansatz~\eqref{eq:f_term_solving_ansatz} gives the second order
equation~\eqref{eq:quadratic_eom_warped}. Taking the ansatz
$\psi^{\mp}=\kappa z^{2}f^{\mp}$ where $\kappa$ is the magnetized
width~\eqref{eq:magnetized_unwarped_width}, we consider the
special case where the warp factor is a function of only
$w=\abs{z^{2}}^{2}$. Then, $\psi^{\mp}$, which is proportional to
$\phi_{3}^{\mp}$, will also depend on $z^{2}$ and $\bar{z}^{\bar{2}}$
only through $w$.  Further, taking $f^{\mp}$ to be in the kernel of
$\bigl(\hat{D}_{1}^{\pm}\bigr)^{\dagger}$, $\psi^{\mp}$ satisfies
\begin{equation}
  \frac{\partial^{2}\psi^{\mp}}{\partial w^{2}}-
  \kappa^{2}\ue^{-4\alpha}\psi^{\mp}=0.
\end{equation}
In what follows, we will focus only on the $w$ dependence of
$\psi^{\mp}$; $\psi^{\mp}$ will be independent of $z^{1}$ and
$\bar{z}^{1}$ in the unmagnetized case and will depend on these
coordinates through theta functions as detailed in
section~\ref{sec:chiral}.  Additionally, we will suppress family indices
and the $\mp$ superscript in this appendix.

As one would expect and was already demonstrated in the weak-warping limit,
the wavefunctions remain highly localized along the intersection of
the $\uD 7$-branes in the presence of warping.  However, for certain
special warp factors, the solutions to the equations of motion diverge
along the matter curve and the field theory treatment breaks down.

We first consider warp factors of the form
\begin{equation}
  \ue^{-4\alpha}=1+L^{-2n}w^{n}.
\end{equation}
Such warp factors were considered in the main text, but here do not
make the assumption of weak warping.  A simple analytic solution is
not available for general $n$.  However, for $n=1$ the solution is
given in Airy functions as in~\eqref{eq:quadratic_exact}.  Similarly,
for $n=2$, the solution satisfying the boundary condition $\psi\to 0$
as $w\to\infty$ is, up to an overall normalization constant
\begin{equation}
  \psi=D_{\nu}\bigl(\frac{\sqrt{2\kappa}w}{L}\bigr),\quad
  \nu=-\frac{1}{2}\bigl(L^{2}\kappa-1\bigr),
\end{equation}
where $D_{\nu}$ is a parabolic cylinder function.  The function is
peaked at $w=0$ and is normalizable in both the warped
($\int\ud^{4}z\ue^{-4\alpha}$) and unwarped ($\int\ud^{4}z$) sense.

Another special case is the $n=-2$.  Although the $n\ge 0$ case was
easily addressable for weak warping using the massive mode analysis
and ladder operators, because the annihilation operator is not
invertible, negative $n$ is not as easily handled.  However, in this
case there is an exact solution given as a modified Bessel function of
the second kind
\begin{equation}
  \psi=\sqrt{w}K_{\nu}\bigl(\kappa w\bigr),\quad
  \nu=\frac{1}{2}\sqrt{1+L^{4}\kappa^{2}}.
\end{equation}
The solution is again peaked at $w=0$ and diverges such that the
function is not normalizable in either the warped or unwarped sense.

Finally, we can consider the class of warp factors
\begin{equation}
  \ue^{-4\alpha}=L^{-2n}w^{n}.
\end{equation}
Such warp factors do not have a weak-warping limit.  For $n>0$, it is
convenient to define a new variable
\begin{equation}
  x=\left(\frac{\kappa}{L^{n}}\right)^{2/\left(n+2\right)}w,
\end{equation}
the equation of motion becomes
\begin{equation}
  \psi''-x^{n}\psi=0,
\end{equation}
and so the solution is given in terms of modified Bessel functions of the
second kind
\begin{equation}
  \psi=\sqrt{x}K_{\nu}\bigl(2\nu x^{1/2\nu}\bigr),\quad
  \nu=\frac{1}{n+2}.
\end{equation}
The solutions are peaked at $w=0$ and are normalizable.

For the case $n=-2$, the solution that vanishes at large $w$ is given by
\begin{equation}
  \psi=w^{\left(1-\sqrt{1+4L^{4}\kappa^{2}}\right)/2},
\end{equation}
which, while again localized at the intersection, diverges for small
$w$ and is not normalizable.

\section{\label{app:fourier_theta_overlap}Overlap of Fourier modes and
 theta functions}

In this appendix, we consider again the
expansion~\eqref{eq:fourier_theta}.  Using the orthonormality of the
massive modes, we have
\begin{align}
  B_{mnq}^{kj,-}=&\bigl\langle
  \vp_{q00}^{k,-},h_{mn}\vp_{0}^{j,-}\bigr\rangle\notag \\
  =&\frac{1}{\sqrt{\hat{M}_{1}^{q}q!}}
  \bigl\langle\bigl(\ui \hat{D}'^{-}_{1}\bigr)^{q}
  \vp_{0}^{k,-},h_{mn}\vp_{0}^{j,\mp}\bigr\rangle \notag \\
  =&\frac{1}{\sqrt{\hat{M}_{1}^{q}q!}}
  \bigl\langle \vp_{0}^{k,-},
  \bigl[\ui\bigl(\hat{D}'^{+}_{1}\bigr)^{\ast}\bigr]^{q}\bigl(h_{mn}
  \vp_{0}^{j,-}\bigr)\bigr\rangle.
\end{align}
Now since $\ui\bigl(\hat{D}'^{+}_{1}\bigr)^{\ast}$ is a $-$-sector
lowering operator, we have
\begin{equation}
  \bigl(\hat{D}'^{+}_{1}\bigr)^{\ast}
  \bigl(h_{mn}\vp_{0}^{j,-}\bigr)
  =\bigl(\hat{\partial}_{1}^{\ast}h_{mn}\bigr)\vp_{0}^{j,-}
  =t_{mn}h_{mn}\vp_{0}^{j,-},
\end{equation}
where $t_{mn}$ is defined in~\eqref{eq:define_t_mn}.  Thus,
\begin{equation}
  B^{kj,-}_{mnq}=\frac{\bigl(\ui t_{mn}\bigr)^{q}}{\sqrt{\hat{M}_{1}^{q}q!}}
  \bigl\langle\vp_{0}^{k,-},h_{mn}\vp_{0}^{j,-}\bigr\rangle.
\end{equation}
A similar expression holds in the $+$-sector.

In this case, the inner product can be calculated explicitly.  In
fact, since the integrals over $z^{1}$ and $z^{2}$ factor, we
have
\begin{align}
  \bigl\langle \vp_{0lp}^{j,\mp},&h_{mn}\vp_{0l'p'}^{j',\mp}\bigr\rangle
  =\delta_{ll'}\delta_{pp'}
  \sqrt{\frac{\pm 2I^{\left(ab\right)}_{1}}{\mathrm{Im}\, \tau_{1}}}
  \sum_{r\,r'}
  \exp\biggl\{\pm\pi\ui I^{\left(ab\right)}_{1}
  \biggl[\biggl(\frac{\pm j}{I^{\left(ab\right)}_{1}}
  +r\biggr)^{2}\tau_{1}-
  \biggl(\frac{\pm j'}{I^{\left(ab\right)}_{1}}+r'\biggr)^{2}
  \bar{\tau}_{1}\biggr]\biggr\}\notag \\
  &\times\int_{0}^{\mathrm{Im}\,\tau_{1}}
  \ud\bigl[\mathrm{Im}\, z^{1}\bigr]\,
  \exp\biggl\{\mp 2\pi I^{\left(ab\right)}_{1}
  \frac{\bigl(\mathrm{Im}\, z^{1}\bigr)^{2}}
  {\mathrm{Im}\, \tau_{1}}
  \mp 2\pi I^{\left(ab\right)}_{1}
  \biggl(\pm\frac{j+j'}{I^{\left(ab\right)}_{1}}
  +r+r'\biggr)\im{z^{1}}\notag \\
  &\phantom{\times\int_{0}^{\mathrm{Im}\,\tau_{1}}
    \ud\bigl[\mathrm{Im}\, z^{1}\bigr]\,
    \exp\biggl\{}
  +2\pi\ui\frac{\left(m-n\mathrm{Re}\,\tau_{1}\right)}
  {\mathrm{Im}\left(\tau_{1}\right)}\im{z^{1}}\biggr\}\notag\\
  &\times\int_{0}^{1}\ud\bigl[\re{z^{1}}\bigr]\,
  \exp\biggl\{\pm 2\pi\ui I^{\left(ab\right)}_{1}
  \biggl[\pm\frac{j-j'}{I^{\left(ab\right)}_{1}}
  +r-r'\biggr]\re{z^{1}}+2\pi\ui n\re{z^{1}}\biggr\}.
\end{align}
Writing $n=k\pm sI^{\left(ab\right)}_{1}$, where
$k<\absb{I^{\left(ab\right)}_{1}}$, the integral over the real part of
$z^{1}$ vanishes unless $s=r-r'$ and $k=j-j'$.  The latter condition
is the statement that $n=j-j'\mod \pm I^{\left(ab\right)}_{1}$.
Defining the symbol
\begin{equation}
  \delta^{a}_{bc}=\begin{cases}
    1 & b=c\mod a,\\
    0 & b\neq c\mod a,
  \end{cases},
\end{equation}
we get
\begin{align}
  \bigl\langle\vp_{0lp}^{j,\mp},&h_{mn}\vp_{0l'p'}^{j',\mp}\bigr\rangle\notag \\
  =&\delta_{ll'}\delta_{pp'}\delta_{n,j-j'}^{I^{\left(ab\right)}_{1}}
  \sqrt{\frac{\pm 2I^{\left(ab\right)}_{1}}{\mathrm{Im}\,\tau_{1}}}
  \sum_{r}
  \exp\biggl\{\pm \pi\ui I^{\left(ab\right)}_{1}
  \biggl[\biggl(\frac{\pm j}{I^{\left(ab\right)}_{1}}
  +r\biggr)^{2}\tau_{1}
  -\biggl(\frac{\pm\left(j-k\right)}{I^{\left(ab\right)}_{1}}
  +r-s\biggr)^{2}\bar{\tau}_{1}\biggr]\biggr\} \notag \\
  &\times\int_{0}^{\mathrm{Im}\, \tau_{1}}\ud\bigl[\mathrm{Im}\, z^{1}\bigr]\,
  \exp\biggl\{
  \mp 2\pi I^{\left(ab\right)}_{1}
  \frac{\left(\mathrm{Im}\, z^{1}\right)^{2}}{\mathrm{Im}\,\tau_{1}}-
  2\pi I^{\left(ab\right)}_{1}
  \biggl(\frac{\pm \bigl(2j-k\bigr)}{I^{\left(ab\right)}_{1}}
  +2r-s\biggr)\im{z^{1}}\notag \\
  &\phantom{\times
    \int_{0}^{\mathrm{Im}\, \tau_{1}}\ud\bigl[\mathrm{Im}\, z^{1}\bigr]\,
    \exp\biggl\{}
  +2\pi\ui\frac{m-n\mathrm{Re}\, \tau_{1}}{\mathrm{Im}\, \tau_{1}}
  \im z^{1}\biggr\}.
\end{align}
Completing the square gives
\begin{align}
  \bigl\langle\vp_{0lp}^{j,\mp},h_{mn}\vp_{0l'p'}^{j',\mp}\bigr\rangle
  =&\delta_{ll'}\delta_{pp'}\delta_{n,j-j'}^{I^{\left(ab\right)}_{1}}
    \sqrt{\pm 2I^{\left(ab\right)}_{1}\mathrm{Im}\,\tau_{1}}
  \sum_{r}
  \exp\biggl\{\mp \pi\mathrm{Im}\left(\tau_{1}\right)n^{2}/2
  I^{\left(ab\right)}_{1}\biggr\}\notag \\
  &\times\int_{0}^{1}\ud\xi\,
  \exp\biggl\{\mp 2\pi I^{\left(ab\right)}_{1}
  \mathrm{Im}\,\tau_{1}\left(\xi\pm\frac{j-n/2}
    {I^{\left(ab\right)}_{1}}+r\right)^{2}\biggr\}
  \exp\biggl\{2\pi\ui m\xi\biggr\}\notag \\
  &\phantom{\times\int_{0}^{1}\ud\xi\,}
  \times\exp\biggl\{- 2\pi\ui n\ \mathrm{Re}\,\tau_{1}
  \left(\xi\pm\frac{j-n/2}{I^{\left(ab\right)}_{1}}
    +r\right)\biggr\},
\end{align}
where $\xi=\im{z^{1}}/\im{\tau_{1}}$.  If we make the substitution
$\xi\to\xi+r$, then sum over $r$ turns this into an integral of $\xi$
over all $\mathbb{R}$.  The rest follows straightforwardly,
\begin{equation}
  \bigl\langle\vp_{0lp}^{j,\mp},h_{mn}\vp_{0l'p'}^{j',\mp}\bigr\rangle=
  \delta_{ll'}\delta_{pp'}\delta_{n,j-j'}^{I^{\left(ab\right)}_{1}}
  \ue^{\mp\hat{m}_{mn}^{2}\cV_{1}/4\pi^{2}I^{\left(ab\right)}_{1}}
  \ue^{\mp 2\pi\ui m\left(j+j'\right)/2I^{\left(ab\right)}_{1}},
\end{equation}
giving
\begin{equation}
  B_{mnq}^{kj,\mp}=\delta^{I^{\left(ab\right)}_{1}}_{n,k-j}
  \frac{\bigl(\ui t_{mn}\bigr)^{q}}{\sqrt{\hat{M}_{1}^{q}q!}}
  \ue^{\mp\hat{m}_{mn}^{2}\cV_{1}/4\pi^{2}I^{\left(ab\right)}_{1}}
  \ue^{\mp 2\pi\ui m\left(k+j\right)/2I^{\left(ab\right)}_{1}}.
\end{equation}
Note that the higher Fourier modes have an exponentially suppressed overlap.
 
\section{\label{app:hermite}Massive modes and Hermite functions}

The $\triangle^{\mp}$ eigenstates $\vp_{00lp}^{\mp}$ in the
unmagnetized case~\eqref{eq:unmagnetized_component_MMs} and
$\vp_{0lp}^{j,\mp}$ in the magnetized
case~\eqref{eq:magnetized_plus_sector_component_MMs}
and~\eqref{eq:magnetized_minus_sector_component_MMs} are defined the
same way that the excited mode of a quantum harmonic oscillator are,
and thus should be expressible in terms of Hermite functions.

The Hermite functions are defined by
\begin{equation}
  \tilde{H}_{n}\left(u\right)=\sqrt{\frac{1}{2^{n}n!\sqrt{\pi}}}
  \biggl(\frac{\ud}{\ud u}-u\biggr)^{n}\ue^{-u^{2}/2},
\end{equation}
and are normalized such that
\begin{equation}
  \int_{-\infty}^{\infty}\ud u\,
  \tilde{H}_{n}\left(u\right)\tilde{H}_{n'}\left(u\right)
  =\delta_{nn'}.
\end{equation}
Defining
\begin{equation}
  u=\sqrt{2\kappa}\,\re{z^{2}},\qquad
  v=\sqrt{2\kappa}\,\im{z^{2}},
\end{equation}
the ladder operators can be written
\begin{align}
  \hat{D}'^{\mp}_{2}=&\frac{c}{\sqrt{2\pi}R_{2}}
  \sqrt{\frac{\kappa}{2}}
  \biggl\{\biggl(\frac{\partial}{\partial u}\pm u\biggr)
  -\ui\biggl(\frac{\partial}{\partial v}\pm v\biggr)\biggr\},\\
  \hat{D}'^{\mp}_{3}=&\frac{-\ui s}{\sqrt{2\pi}R_{2}}
  \sqrt{\frac{\kappa}{2}}
  \biggl\{\biggl(\frac{\partial}{\partial u}\mp u\biggr)
  -\ui\biggl(\frac{\partial}{\partial v}\mp v\biggr)\biggr\}.
\end{align}
Then, in the magnetized case
\begin{multline}
  \vp_{nlp}^{j,\mp}
  =\bigl(\mp 1\bigr)^{l}
  \ui^{p}
  2^{-\left(l+p\right)}
  \left(\frac{2\kappa
    \sqrt{2I^{\left(ab\right)}_{1}\mathrm{Im}\,\tau_{1}}}
  {\cV_{1}\cV_{2}}\right)^{1/2}
  \Omega_{n}^{j,\mp}\bigl(z^{1},\bar{z}^{\bar{1}}\bigr)\\
  \times\sum_{r=0}^{l}\sum_{s=0}^{p}
  \bigl(\mp\ui\bigr)^{r-s}
  \binom{l}{r}
  \binom{p}{s}
  \sqrt{\frac{\left(r+s\right)!\left(l+p-s-r\right)!}
    {l!p!}}
  \tilde{H}_{r+s}\left(u\right)
  \tilde{H}_{l+p-r-s}\left(v\right),
\end{multline}
where
\begin{equation}
  \Omega_{n}^{j,\mp}=
  \sqrt{\frac{1}{n!\left(\pm\hat{M}_{1}\right)^{n}}}
  \ue^{\pm\pi\ui I^{\left(ab\right)}_{1}z^{1}
    \mathrm{Im}\, z^{1}/\mathrm{Im}\, \tau_{1}}
  \vartheta\begin{bmatrix}
    \pm j/I^{\left(ab\right)}_{1} \\ 0\end{bmatrix}
  \bigl(\pm I^{\left(ab\right)}_{1}z^{1},
  \pm I^{\left(ab\right)}_{1}\tau_{1}\bigr).
\end{equation}

The same relations apply in the unmagnetized case except that
$\Omega_{n}^{j,\mp}$ is replaced with the appropriate Fourier modes.

\bibliography{chiral_wfs}

\end{document}